\documentclass[12pt]{revtex4-1}
\usepackage{graphicx}
\usepackage{amssymb}
\usepackage{amsmath}
\usepackage{bm}

\newcommand{\bu}{\bm{u}} 
\newcommand{\bj}{\bm{j}} \newcommand{\bi}{\bm{i}}

\newcommand{\br}{\bm{r}} \newcommand{\bv}{\bm{v}}
\newcommand{\bg}{\bm{g}}

 \newcommand{\bn}{\bm{n}}
\newcommand{\bk}{\bm{k}} 
\newcommand{\bF}{\bm{F}} 
\newcommand{\bs}{\bm{s}} 
\newcommand{\bc}{\bm{c}}

\newcommand{\etext}[1]{\quad\mbox{#1}\quad}

\newcommand{\oder}[2]{\frac{d #1}{d #2}}

\newcommand{\sub}[1]{_{\mbox{\small #1}}}
\newcommand{\beq}{\begin{equation}}
\newcommand{\eeq}{\end{equation}}
\newcommand{\fracp}[2]{\left(\frac{#1}{#2}\right)}
\newcommand{\Kn}{\mbox{K}}

\begin{document}
\begin{center}
.
\vskip 4cm
{\bf Modelling  of carving turns in alpine skiing}
\end{center}

\vskip 1cm
{\bf Authors:} {\it Serguei Komissarov}, School of Mathematics, University of Leeds, Leeds, LS29JT, UK. 
{\it email}: s.s.komissarov@leeds.ac.uk.

{\bf Status}: This is a preprint of the paper which has not been accepted for publication because of its length. A shorter version may be prepared later.  

{\bf Preprint DOI}: 10.31236/osf.io/u4ryc 

{\bf Cite as}: Komissarov, Serguei, 2018. ÒModelling of Carving Turns in Alpine Skiing.Ó SportRxiv. November 22. doi:10.31236/osf.io/u4ryc.

\newpage
\setcounter{page}{0}
\title{Modelling  of carving turns in alpine skiing}
\author{ Serguei S.\surname{Komissarov}}
\email{s.s.komissarov@leeds.ac.uk}
\affiliation{Department of Applied Mathematics\\
The University of Leeds\\ Leeds, LS2 9JT, UK}
\date{Received/Accepted}

\begin{abstract}  
In this paper we present a simple mathematical theory of carving turns in alpine skiing and snowboarding. The theory captures the  basic dynamics of carving runs and thus provides a useful tool for assessing the potential and limitations of the carving technique.  We also apply the model to simulate runs on slopes of constant gradient and describe the results.  We find that pure carving is possible only on relatively flat slopes, with the critical slope angle in the range of $8^\circ-20^\circ$.  The exact value depends mostly on the coefficient of snow friction and to a lesser degree on the sidecut radius of the skis.  In wiggly carving runs on slopes of subcritical gradient, the aerodynamic drag force remains  relatively unimportant and the speed stops growing well below the one achieved in fall-line gliding. This is because the increased g-force of carving turns leads to enhanced snow friction which can significantly exceed the one found in the gliding.  At the critical gradient, the g-force of carving turns becomes excessive.   For such and even steeper slopes only hybrid turns, where at least some part of the turn is skidded, are possible. The carving turns of all alpine racing  disciplines are approximately similar. This is because in the dimensionless equations of the model the sidecut radius of skis enters only via the coefficient of the aerodynamic drag term, which always remains relatively small.  Simple modifications to the model are made to probe the roles of skier's angulation and skidding at the transition phase of hybrid turns.  As expected, the angulation gives certain control over the turn trajectory but confines pure carving to even flatter slopes.         
\end{abstract}

\maketitle

\section{Introduction}
\label{introduction}

When making their way down the hill, expert skiers execute  complex coordinated body movements, often within a fraction of a second, which
allow them to ski at speed and yet to remain in control. Their decisions are dictated by many factors, such as terrain, snow condition, equipment etc.  
A ski racer faces additional challenges as a race course significantly reduces the freedom of choosing trajectory.  There is a great deal of qualitative 
understanding of skiing techniques and race tactics based of the personal experiences of ski professionals, coaches and instructors.  
When designing a race course, the setter has a pretty good idea of how the turns have to be executed  in order to excel.  All this knowledge 
is based on countless previous runs in similar circumstances and trial-and-error attempts to figure out what works and what does not.  
Understandably, this empiric knowledge is very subjective, often imprecise and sometimes even subconscious and this keeps the door open 
to misunderstanding,  misconceptions  and controversies. 

Science can help to solidify this knowledge and to raise the level of understanding.  This road has already been followed by many 
scientifically-minded skiing experts as well as  academics.    For example, in the book ``Ultimate Skiing'' by Ron LeMaster \cite{LM10}, the mechanical 
and bio-mechanical principles are used to elucidate the reasons behind skiing techniques 
and in ``The Physics of Skiing'' by Lind and Sanders \cite{LS04} a rudimentary mathematical theory of skiing is developed.        
Few studies have tried  to identify quantitative parameters of skiing turns which could tell to racers how to ski in order to minimise 
the run time  \cite{S08,S11,S13}.   First attempts have been made to mathematically model ski runs. These studies were  limited to the so-called steered (or skidded) turns where skis are pivoted to an angle with the direction of motion, thus introducing enhanced
dissipation of the skier's kinetic energy via snow-ploughing and ice-scraping  \cite{NSK99,TH02,H06}.    

The advance of modern shaped skis has moved the focus of competitive skiing from skidded turns to carving turns. The main feature of a 
carving turn is a near-perfect alignment of skis with their trajectory, which significantly reduces the energy dissipation and increases the speed.  Nowadays even mass-produced skis are shaped, thus giving the opportunity to enjoy carving runs to all skiing enthusiasts. This has made an impact on the way the alpine skiing is taught by ski instructors (e.g. \cite{HH06}).  The key property of sharped skis is their ability to bend into an arc when put on edge and pressed down in the middle, the higher the inclination angle of the ski the tighter is the arc.  When pressed against hard snow (or ice ), such a ski cuts in it an arc-like platform which has the following two main outcomes:  First this helps to inhibit ski's lateral displacement and hence skidding. Second, the shape of the bended ski (or rather of its inside edge) becomes the local shape of its trail in the snow.                    

The dynamics of carving turns has already received a lot of attention in the theory of skiing. In particular, significant insights have been gained from the studies of a specialised sledge whose skis could be made to run on edge \cite{M06,M08,M09}.  This approach allowed to compare the results of computer simulations with the data of sufficiently controlled simple field experiments. A particular attention was paid to ski-snow interaction. In the simulations, the skis were modelled as a collection of rigid segments connected via revolute joints with prescribed spring and damping parameters. Eventually, the sledge had evolved towards a more complex Hanavan-like model of a skier, where the human body is also represented by several rigid segments connected by mechanical joints \cite{M10}.       

Naively, one may think that skiers are free to change the inclination angle of their skis as they like and hence fully control the local shape (curvature) of their  trajectory. However, this angle in largely dictated by the inclination angle of the skier, and  an arbitrary change of this angle may upset skier's lateral balance.  Indeed, as we all know a stationary skier must stay more-or-less vertical to avoid falling to the one side or the other.  The theoretical analysis of the lateral balance in a carving turn allowed Jentshura and Fahrbach \cite{JF04} to derive the so-called Ideal Carving Equation (ICE), which shows that the inclination angle, and hence the local radius of curvature of skier's trajectory, are completely determined by his speed and direction of motion relative to the fall line. To a mathematician, this immediately tells that the speed and direction of motion at the initiation of a carving turn (initial conditions) should determine the rest of the turn.  The skier is almost  like a passenger.  Although the analysis is restricted to the simplified case of a mono-skier, this model allows to develop the understanding of the  key properties of carving turns, which is needed before we proceed towards more comprehensive and hence less transparent models.  In our paper we continue this analysis and formulate the key dynamic equations of carving turns. We also present a number of solutions describing simple runs down a slope of constant gradient, which reveal some intriguing properties. 

The inclination is only one of several body manoeuvres executed by advanced skiers during their runs in order to maintain their balance.  Another one is the  lateral angulation, mostly at the hip but also at the knees.  The angulation  breaks the rigid relationship between the ski's and skier's inclination angles imposed by ICE and allows skiers to regain some control over their motion in pure carving turns. We generalise ICE by taking the angulation into account and analyse its role in some details.      
      
Because the carving technique minimises the losses of kinetic energy and allows to ski faster, one could naively expect that ski racers would execute pure carving turns from start to finish, but this is not the case.  The deviations from pure carving often manifest themselves in the form of snow plumes, ejected from under their skis. In many cases, this is the result of mistakes made by skiers while deciding when to terminate one turn and to start next one, mistakes which are corrected by braking. However, in this paper we argue that it may not be possible in principle to get rid of the skidding completely.   In fact, ICE implies an  existence of maximum speed consistent with the lateral balance in carving turns -- no mater how strong and skilful the athlete is,  she/he would not be able to perform a carving turn at a speed exceeding this limit \cite{JF04}. In our paper, we presents a somewhat different analysis of this issue and extend it to the case of angulated skier. We find that the speed limit is robust and then show that, unless the slope is sufficiently flat, some amount of skidding has to be introduced just to stay on the run.   We modify our mathematical model of skiing accordingly and present the results for runs with a certain degree of skidding during transitions between the carving phases of turns.

\section{Aerodynamics drag and characteristic scales of alpine skiing}
\label{estimates}

\subsection{Fall-line gliding}
\label{flg}

Although recreational skiing can be very relaxed and performance skiing physically most demanding,  the dominant source of energy in both cases is the Earth's gravity.  The total available gravitational energy  is  
\beq
    U=mgh \,,  
\label{i1}
\eeq   
 where $m$ is skier's mass, $g$ is the gravitational acceleration and $h$ is the total vertical drop of the slope.   
 If all this energy was converted into the kinetic energy of the skier, $K=mv^2/2$, then at the bottom of the slope the speed would reach 
\beq
    v=\sqrt{2gh} \approx  227 \fracp{h}{200\,\mbox{m}}^{1/2} \mbox{km/h} \,.  
\label{i2}
\eeq   
In fact, this not far from what is achieved in the speed skiing competitions, where skiers glide straight down the fall line.   
For example, the current speed record of the Riberal slope in 
Grandvalira (Andorra), which has exactly 200m vertical drop and is often used for 
speed skiing competitions, is just short of $200\mbox{km/h}$\cite{SR18}. 
However, the typical speeds in other alpine disciplines are significantly lower, indicating that only a fraction of the available gravitational energy is 
converted into the kinetic energy  of the skiers. In other words, there are some forces working against gravity. Two of the candidates are the dynamic 
snow friction and the aerodynamic drag \cite{LS04}. The friction force is antiparallel to the skier velocity vector $\bv$ and its magnitude relates 
to the normal reaction force $\bF\sub{n}$ via
\beq
    F\sub{f} = \mu F\sub{n} \,,  
\label{i3}
\eeq   
where $\mu$ is the dynamic coefficient of friction.  For a brick of mass $m$ residing on a firm surface, the normal reaction force is normal to 
this surface and balances the normal component of the gravity force:
\beq
    F\sub{n} = mg\cos\alpha \,,  
\label{i4}
\eeq   
where $\alpha$ is the angle between the surface and the horizontal plane. This result also applies to a skier gliding down the fall line, like in the 
speed skiing competitions.  Because the friction force determined by equations (\ref{i3}) and (\ref{i4}) does not depend on the skier's speed, 
it cannot limit its growth but only to reduce the growth rate.

The aerodynamic drag force is also antiparallel to the velocity vector and has the magnitude  
\beq
    F\sub{d} =  \kappa v^2  \etext{where} \kappa=\frac{C\sub{d}A\rho}{2} \,,
\label{i5}
\eeq   
where $C\sub{d}$ is the drag coefficient, $A$ is the cross-section area of the skier normal to the direction of motion and $\rho$
is the mass density of the air. The fact that the drag force grows with speed without limit implies that at some speed it will balance the gravity 
along the fall line. Below this speed, the gravity wins and skier accelerates. Above this speed, the drag wins and the skier decelerates. Hence the equilibrium 
speed is also a growth saturation speed (an attractor).  The value of the  saturation speed $v\sub{s}$ can be easily found from the energy principle.  

In the saturation regime, the work carried out by the drag and friction forces over the distance $L$ along the fall line is   
\beq
    W = (F\sub{f}+F\sub{d}) L  \,.
\label{i6}
\eeq   
This must be equal to the gravitational energy $U=mgh$ released over the same distance. This leads to    
\beq
    v\sub{s}^2 =  \frac{mg}{\kappa} \sin\alpha(1-\mu\cot\alpha)
\label{i7}
\eeq   
 (cf. \cite{LS04}). Incidentally, the result shows that the slope angle has to exceed $\alpha\sub{min}=\arctan(\mu)$, as otherwise $v\sub{s}^2<0$, indicating    
 that skiing is impossible.  For the realistic value $\mu=0.04$ \cite{LS04} this gives $\alpha\sub{min}=2.3^o$. Usually ski 
 slopes are significantly steeper than this and the snow friction contribution is small. In this case, the saturation speed is determined mostly by the 
 balance between the gravity and aerodynamic drag, which  yields the speed
\beq
       V\sub{g} = \sqrt{\frac{mg}{\kappa} \sin\alpha} \,. 
    \label{Vg}
\eeq   
From this analysis it follows that $V\sub{g}$ is a characteristic speed for the problem of fall-line gliding and presumably for alpine skiing 
in general.  We note here that it depends  on the slope gradient -- the steeper is the slope, the higher is the speed.  
The time required to reach this speed under the action of gravity,
\beq
       T\sub{g} = \frac{V\sub{g}}{g\sin\alpha} = \sqrt{\frac{m}{\kappa g \sin\alpha}} \,,
    \label{Tg}
\eeq   
is the corresponding characteristic time scale.  This scale is shorter for steeper slopes.  However,  the corresponding length scale 
\beq
       L\sub{g} = V\sub{g} T\sub{g} = \frac{m}{\kappa} 
    \label{Lg}
\eeq   
does not depend of the slope gradient, which is not very intuitive ($L\sub{g}$ is twice the distance required to reach $V\sub{g}$ under the action of gravity 
alone.).     
Lind \& Sanders \cite{LS04} state the values $C\sub{d}=0.5$, $m=80\,$kg, $A=0.4\,\mbox{m}^2$, 
and $\rho=1.2\,\mbox{kg}/\mbox{m}^3$  as typical for downhill (DH) competitions.  
For these parameters 
\beq
      V\sub{g} = 148 \, \sqrt{\frac{\sin\alpha}{\sin 15^\circ} }\mbox{km}/\mbox{h} \,,\quad
      T\sub{g} = 16 \sqrt{\frac{\sin 15^\circ}{\sin\alpha} }\mbox{s} \,,\quad
      L\sub{g} =  0.667 \mbox{km} \,.
\label{i8}
\eeq
Though rather large, $L\sub{g}$ is definitely smaller than the typical length of downhill tracks, which are usually few kilometres long.   
Similarly, $T\sub{g}$ is significantly shorter compared to the typical duration of competition runs.  This indicates that $V_g$  must be well within reach during 
the runs and indeed it is quite close to the typical speeds recorded in downhill  -- the fastest speed ever recorded with a speed gun in downhill is 
$v\approx162\,\mbox{km}/\mbox{h}$ (skier Johan Clarey, Wengen, 2013).  Although downhill courses are not entirely straight, they still include a number of 
long gliding sections and this is why the gliding model works quite well here. (The significantly higher speed of speed skiing is due to the extreme measures 
undertaken by its participants in order to reduce the aerodynamic drag, and hence the lower typical values of $\kappa$.)

\subsection{Skiing with turns}
\label{swt}

Slalom (SL) race tracks are much shorter than the downhill ones -- for example the Ganslern track in Kitzbuhel  
is only about 590 meters long which is about the same as the value $L\sub{g}$ estimated above. Moreover, for a lighter slalom skier 
(smaller $m$) in a less compact position (higher $\kappa$ ), $m/\kappa$  can be about twice as low, 
leading to  
\beq
      V\sub{g} = 104 \, \sqrt{\frac{\sin\alpha}{\sin 15^\circ} }\mbox{km}/\mbox{h} \,,\quad
      T\sub{g} = 11 \sqrt{\frac{\sin 15^\circ}{\sin\alpha} }\mbox{s} \,,\quad
      L\sub{g} =  0.333 \mbox{km} \,.
\label{i9}
\eeq

Thus even in slalom, tracks are few times longer than the distance required to hit the drag-imposed limit.  However, the typical speeds of slalom runs  
stay well below this limit, near or below  $40\,\mbox{km}/\mbox{h}$ (e.g. www.scholastic.com).    
For such low speeds, the aerodynamic drag is about six times below of what is required to balance gravity (see also \cite{S13}) and hence skiers must be loosing a lot of kinetic energy in a way not accounted for in equation (\ref{i6}). 
 
On average, slalom slopes are steeper compared to the downhill ones. For example, the mean inclination angle of the Ganslern track is 
$\left<\alpha\right>\approx 19^\circ$.  This makes the discrepancy slightly more pronounced as according to equation (\ref{i8}) $V\sub{g}$ increases 
with $\alpha$ (by approximately 12\% in the case of the Ganslern).  The fact that a slalom course is not a straight line, contrary to what is assumed
in the derivation of equation (\ref{i7}),  but wanders from side to side does not make much difference either.  The wandering makes the skier path longer 
compared to the length $L$ of the slope.  However, even when each turn is a full half-circle, the length of 
skier's path $L'=(\pi/2)L$ is only about 50\% longer. Replacing $L$ with $L'$ in equation (\ref{i6}) one finds that this results in a decrease of the saturation speed 
by only about 20\%, compensating the increase due to steeper gradients.

Equation (\ref{i7}) assumes that no snow-ploughing or ice-scraping by the ski edges is involved.  In the case of skiing with turns, this implies ideal carving 
where the skis are bend to adopt the local shape of their trajectory and leave on the snow two very narrow tracks.  Moreover, it assumes that the skier's weight 
is just $mg$, which is not necessarily true.  Indeed, turning introduces the centrifugal force, which leads to an increased effective weight 
of the skier and hence higher normal reaction and friction forces compared to the case of fall-line gliding.  As the centrifugal force depends on the speed and local 
radius of skier's trajectory, it varies during the run and its overall effect cannot be estimated easily.  A detailed mathematical model of carving turns is 
required for this purpose.  

When watching alpine skiing competitions or instruction videos, on can see that pure carving turns are executed mostly on flat slopes, whereas on steeper 
slopes skiers use hybrid turns which are partly carved and partly skidded.  The skidding introduces additional losses kinetic energy and could be another 
reason behind the fact that the typical speeds in slalom are significantly below $V\sub{g}$. 
 
\section{The mathematical model}
\label{mmodel}
  
Most phenomena surrounding us are usually complicated with many connections and numerous factors, whereas our mathematical models of them 
are normally quite simplified.  The simplifications are usually driven by the understandable desire to keep the resultant mathematical equations 
treatable by known mathematical techniques.  The typical evolution of a mathematical model starts from the most basic form that still captures 
some key factors and progresses to a more complex form where some additional factors, recognised as potentially important, are included as well.        
Since the theory of carving is in a rather immature state,  the model which we present here is rather simplistic and ignores a number of factors that 
may end up being important. Moreover, the accounted for factors might have been included in a rather simplistic way.   Yet we believe
 the model is useful as it captures the nature of carving rather well and shows the way forward. 

\begin{figure}
\begin{center}
\includegraphics[width=0.49\textwidth]{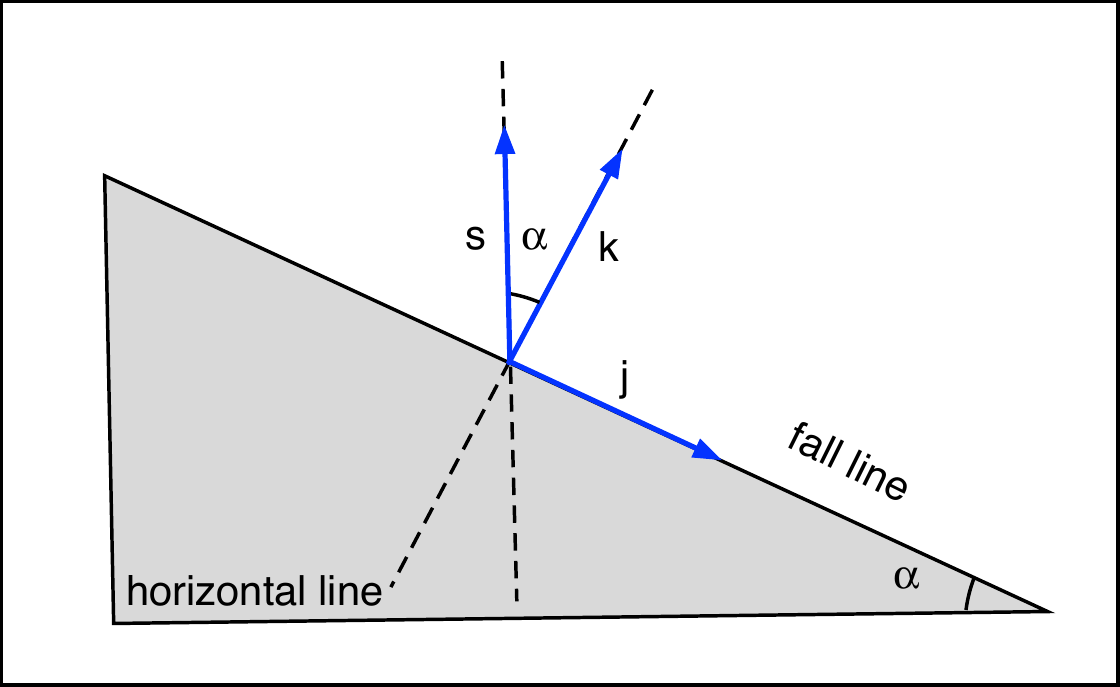}
\includegraphics[width=0.49\textwidth]{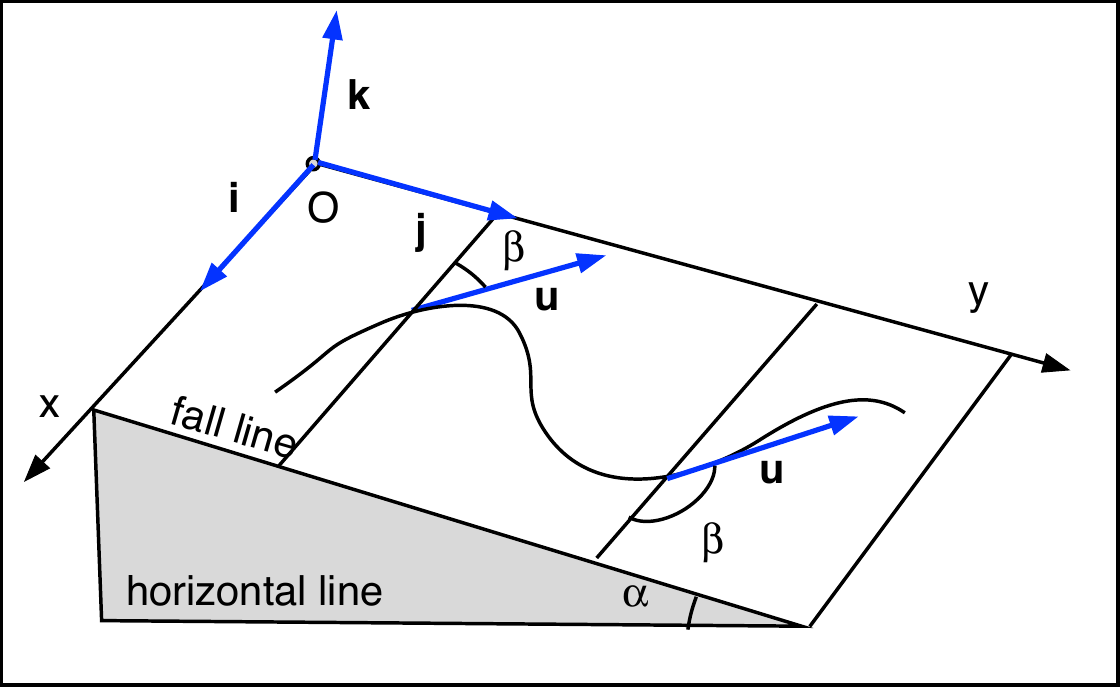}
\caption{ Geometry of the slope. {\it Left panel:} The vertical section of the slope along the fall line. 
{\it Right panel:} The slope as seen at an angle from above. The curved line in the middle is the skier trajectory. }
\label{figure1}
\end{center}
\end{figure}

Instead of listing all the simplifications from the start, we rather describe them as they come into force.     
The first one concerns the geometry of the slope.  Here we limit ourselves to the idealised case of a plane slope with constant gradient 
and introduce such system of Cartesian coordinates $\{x,y,z\}$ associated with the slope that on its surface $z=0$. The unit vectors parallel to the coordinate 
axes will be denoted as $\{\bi,\bj,\bk\}$  respectively. For convenience, we direct the y-axis along the fall line, pointing downwards.    
We also introduce the vertical unit vector $\bs=-\sin(\alpha)\bj+\cos(\alpha)\bk$, so that the gravitational acceleration 
$ \bg=-g \bs $ (see figure \ref{figure1}).

In the model, we will focus mostly of the motion of skier's centre of mass (CM).  We will also consider the torques that may force the skier to rotate  
about CM, focusing on prevention of falling to the side.     
 
\subsection{Basic dynamics of alpine skiing}
\label{bdc}

When only the gravity, normal reaction, dynamic friction and aerodynamic drag forces are taken into account, the second law of Newtonian mechanics 
governing the motion skier CM reads 
\beq
    m\oder{\bv}{t} = \bF\sub{g} + \bF\sub{f}+\bF\sub{d}  +\bF\sub{n}\,,
\label{m2}
\eeq   
where 
\beq
     \bF\sub{g} = m\bg \,, \quad  \bF\sub{f}  = -  \mu F\sub{n} \bu\,, \quad  \bF\sub{d} = -kv^2 \bu \,
\label{m3}
\eeq   
are the total gravity, friction, aerodynamic drag, and snow reaction force acting on the skier respectively and $\bF\sub{n}$ is the normal reaction.  
In these expressions,  $\bu$ is the unit vector in the direction of motion. It is convenient to introduce 
the {\it angle of traverse} $\beta$ as the angle between $-\bi$ and $\bu$ for the right turns and between $\bi$ and $\bu$ for the left turns 
of a run (see figure \ref{figure1}).  With this definition, 
\beq
   \bu= \mp\cos(\beta)\bi + \sin(\beta) \bj  
  \label{m4}
\eeq   
where the upper sign of $\cos\beta$ corresponds to the right turns and the lower sign to the left turns (we will use this convention throughout the paper).

The normal reaction force $\bF\sub{n}$ is not as easy to describe as the other forces.  First, it is not normal to the slope surface but to the surface of contact between the snow and and the skis. When skis are put on edge  they carve a platform (or a step) in the snow and the normal reaction force is normal to the surface (surfaces) of this platform \cite{LM10}.  The orientation of this platform depends on a number of factors including the state of skier motion.  Second, this force equals the total effective weight of the skier, which is determined not only by the gravity but also by the centrifugal force. The latter depends not only on the skier speed but also on their acceleration and hence the local curvature of their trajectory.  Thus the system of equations (\ref{m2}) is not closed.  We will discuss the ways of closing the system later in the paper but in this section we focus on the results that do not depend on how this is done.

\begin{figure}
\begin{center}
\includegraphics[width=0.35\textwidth]{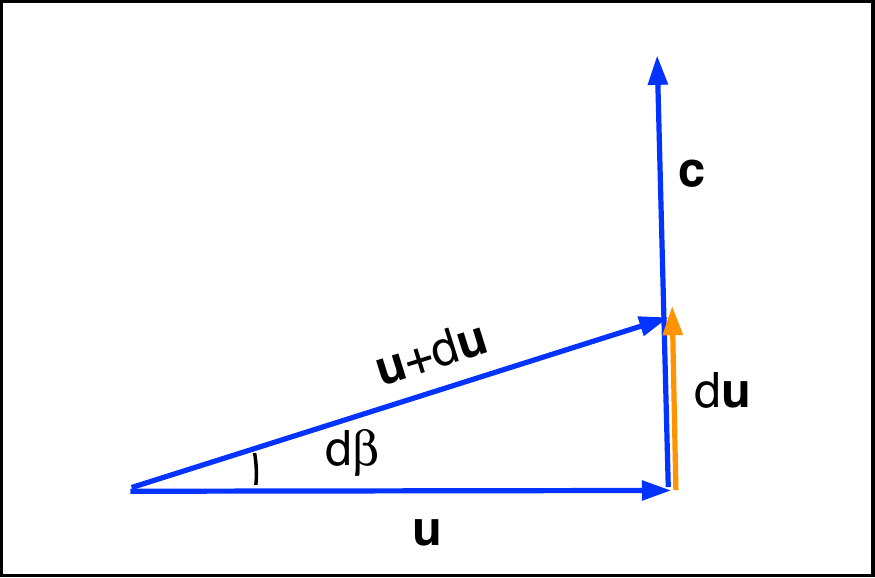}
\caption{The unit velocity vector $\bu$, its infinitesimal variation $d\bu$ and the local centripetal unit vector $\bc$.}
\label{figure2}
\end{center}
\end{figure}

Since the velocity vector $\bv=v\bu$ we have 
\beq
   \oder{\bv}{t}=  \bu \oder{v}{t} + v\oder{\bu}{t}  \,.
  \label{m5a}
\eeq   
Ignoring the up and down motion of CM, we can write $d\bu/dt = \bc |d\beta/dt|$, where  the unit vector $\bc$ is parallel to the slope plane, 
perpendicular to the instantaneous velocity direction $\bu$ and points in the same direction as  $d\bu$  (see figure \ref{figure2}).  
$\bc$ is called the centripetal unit vector as it points towards the local centre of curvature.  Hence 
\beq
   \oder{\bv}{t}=   \bu \oder{v}{t} + v \bc \left|\oder{\beta}{t} \right| \,.
  \label{m5}
\eeq   
Since $dt=dl/v$, where $l$ is the distance measured along the skier trajectory, the last equation can also be written as 
\beq
   \oder{\bv}{t}=  \bu \oder{v}{t} + \frac{v^2}{R} \bc  \,,
  \label{m6}
\eeq   
where
\beq
   R =  \left|\oder{l}{\beta}\right| \,.
  \label{curve-radius}
\eeq   
$R$ is called as the ( local ) radius of curvature of the trajectory. Hence we can rewrite equation (\ref{m2}) as 
\beq
    m\bu \oder{v}{t} + \frac{mv^2}{R} \bc = \bF\sub{g} + \bF\sub{f}+\bF\sub{d}  +\bF\sub{n}\,.
\label{newton}
\eeq   
%

Scalar multiplication of equation (\ref{newton}) with  $\bu$ delivers the equation governing the evolution of skier's speed.  Since 
$$
     \bu\cdot\bc=0 \,, \quad \bu\cdot\bu=1 \,,\quad \bu\cdot\bs = -\sin\alpha\sin\beta \,,
$$
this equation reads 
\beq
   \oder{v}{t}=g  \sin\alpha\sin\beta -\mu \frac{F\sub{n}}{m}  -\frac{k}{m} v^2 \,.
  \label{eq-speed}
\eeq   
Like in the case of fall-line gliding the speed is govern the gravity, friction and aerodynamic drag but now it also depends on the traverse angle. 
Notice that we do not utilise  the expression (\ref{i4}) for $F\sub{n}$ here. This is because, as we are about to show, its area of application is limited to 
the case of fall-line gliding only. 

The normal reaction force can be decomposed into components parallel to $\bc$ and $\bk$: 
\beq
   \bF\sub{n} =  F\sub{n,c} \bc + F\sub{n,k} \bk  \,.
  \label{m7}
\eeq   
The scalar multiplication of equation (\ref{newton}) and $\bk$ immediately yields
\beq
  F\sub{n,k} =mg\cos\alpha  \,.
  \label{m8}
\eeq   
Thus, the normal to the slope component of the snow reaction force $\bF\sub{n}$ is the same as in the case of the fall-line gliding. This is what is needed  
to match exactly the normal to the slope component of the gravity force and hence to keep the skier on the slope. 

\begin{figure}
\begin{center}
\includegraphics[width=0.49\textwidth]{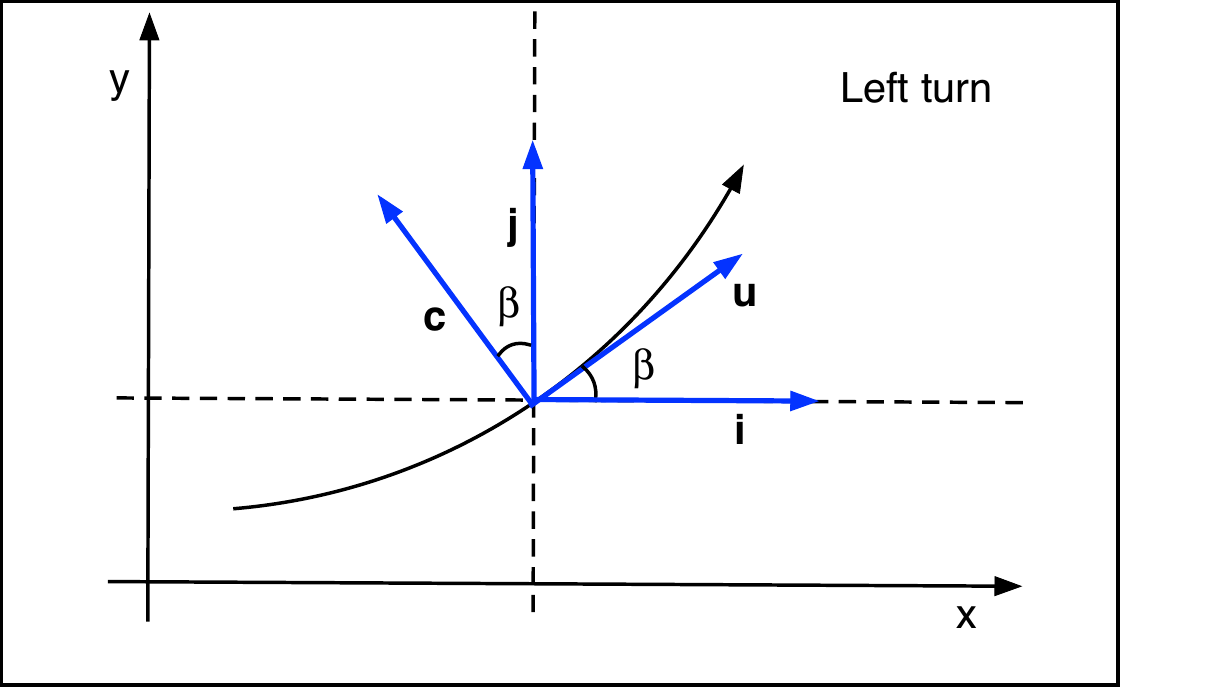}
\includegraphics[width=0.49\textwidth]{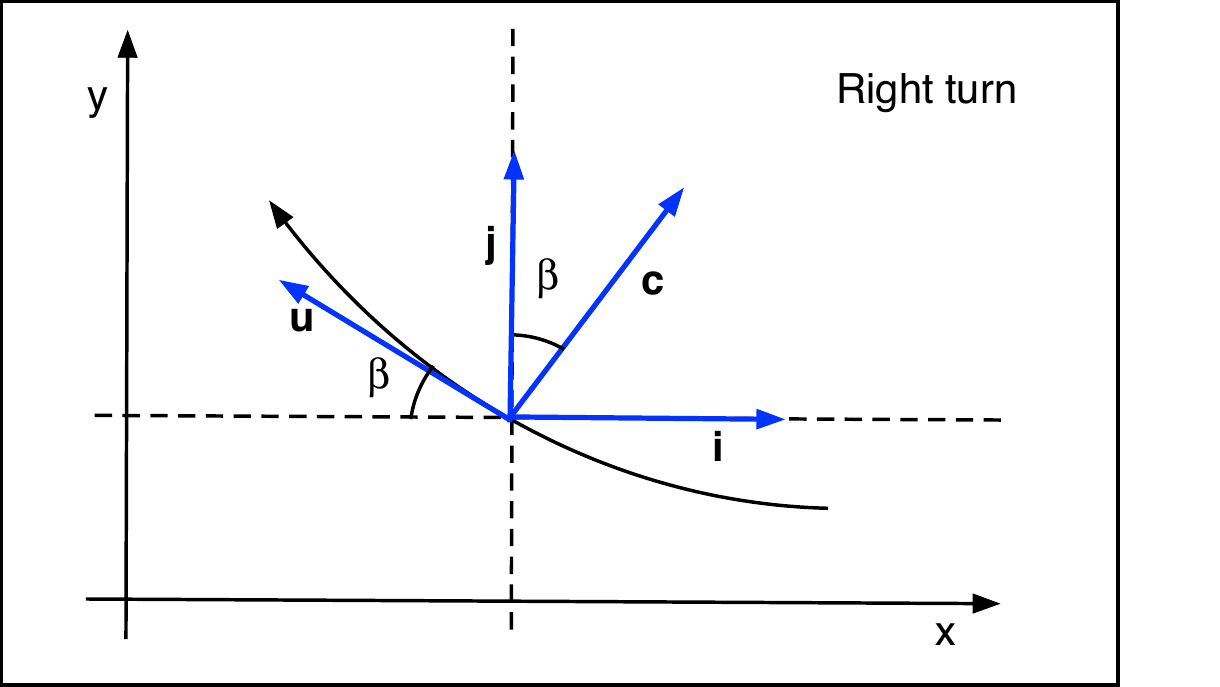}
\caption{ The centripetal unit vector $\bc$ for left and right turns.  }
\label{figure3}
\end{center}
\end{figure}

\begin{figure}
\begin{center}
\includegraphics[width=0.7\textwidth]{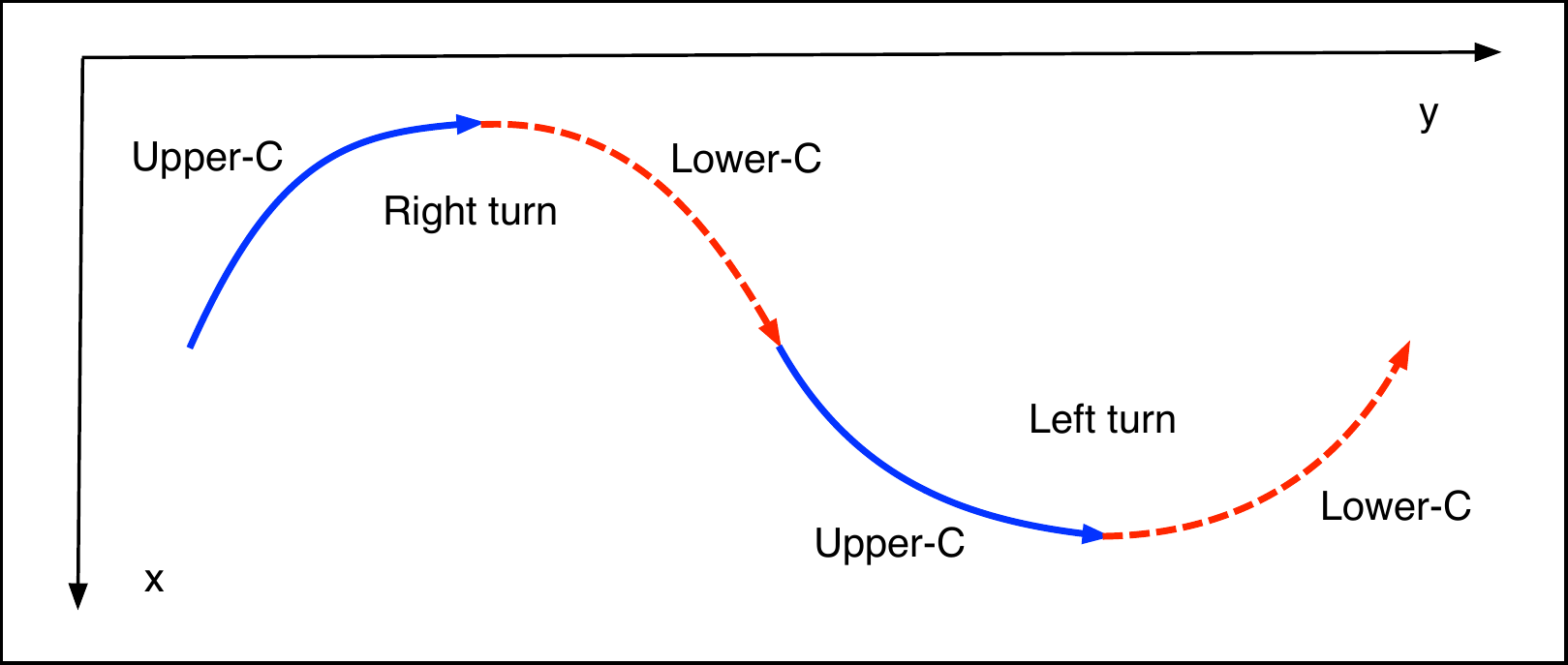}
\caption{ Upper-C and Lower-C sections of the right and left turns.  }
\label{figure4}
\end{center}
\end{figure}

Since 
\beq
  \bc =\pm \sin(\beta)\bi+\cos(\beta)\bj \,
  \label{m9}
\eeq   
(see figure \ref{figure3}) and hence $(\bs\cdot\bc)=-\sin\alpha\cos\beta$,   the scalar multiplication of equation (\ref{newton}) with $\bc$ yields  
\beq
   F\sub{n,c} = \frac{mv^2}{R}  -mg\sin\alpha\cos\beta \,.
  \label{m10}
\eeq   
Now we can find the angle between $\bF\sub{n}$ and the normal direction to the slope $\bk$ 
\beq
   \tan\Phi = \frac{F\sub{n,c}}{F\sub{n,k}} = \frac{v^2}{gR} \frac{1}{\cos\alpha} -  \tan\alpha\cos\beta 
  \label{turn-phi}
\eeq   
and its total strength (and hence the effective weight of the skier) 
\beq
   F\sub{n} = \frac{mg\cos\alpha}{\cos\Phi}  \,.
  \label{f-normal}
\eeq   
In the case of a fall-line glide $\Phi=0^\circ$ and this equation reduces to the familiar $F\sub{n}=mg\cos\alpha$, 
as expected.  It is also clear that the normal reaction force is stronger in the 
Lower-C part of the turn ($90^\circ<\beta<180^\circ$), where the gravity force and the centrifugal force act in the same direction, 
and weaker in the Upper-C ($0^\circ<\beta<90^\circ$), where they act in opposite directions (see figure \ref{figure4}. The terminology is acquired from \cite{HH06}. ).

Equation (\ref{m10}) has the appearance of a force balance.  However this not the case as $mv^2/R$ is not a force but the skier mass multiplied by its radial 
acceleration. Strictly speaking (\ref{m10}) is a differential equation, whose time derivative is hidden away through the definition of $R$.   
It defines the component of normal reaction along the slope plane which is needed to force a curvilinear motion with the local radius of curvature $R$.

Formally, equations (\ref{m8},\ref{m10}) can be written as one vector equation 
\beq
   \bF\sub{n} + \bF\sub{c} + \bF\sub{g,lat} = 0 \,,   
  \label{m11}
\eeq   
where 
\beq
\bF\sub{g,lat}=  -(mg\cos\alpha) \bk + (mg\sin\alpha\cos\beta) \bc 
\eeq  
is the lateral (normal to $\bu$) component of the gravity force and
\beq
\bF\sub{c} = -(mv^2/R)\bc \,.
\eeq  

Equation (\ref{m11}) also holds in the accelerated (non-inertial) frame of the skier. In this frame, the skier is at 
rest and equation  (\ref{m11})  is a balance law between the normal reaction, gravity and the inertial force $\bF\sub{c}$ called the centrifugal force.   
This inertial force has the same properties as the gravity force and can be combined with the lateral gravity into the effective gravity force
\beq
\bF\sub{g,eff} = \bF\sub{g,lat} + \bF\sub{c}  \,,
\eeq         
turning  equation  (\ref{m11}) into the balance between the normal reaction force and the effective gravity 
\beq
   \bF\sub{n} +\bF\sub{g,eff} = 0 \,.   
  \label{m12}
\eeq   
%

The weight is often measured in the units of the standard weight $mg$, in which case it is called a g-force. Hence we have got      
\beq
  \mbox{g-force} = \frac{\cos\alpha}{\cos\Phi} \,.
  \label{g-force}
\eeq

The expressions (\ref{f-normal},\ref{turn-phi}) for the strength and direction of $\bF\sub{n}$ do not allow to close the system. Indeed, they involve the turn radius $R$ which remains undetermined yet. It has entered the model via equations (\ref{m6},\ref{curve-radius}) which cannot be used to calculate it as this requires to know the trajectory, which is not known until the whole solution to the problem is found.   In order to specify the normal reaction force we need to 
consider the interaction of skis with the snow as well as the torques acting on the skier.   

\subsection{Turn radius}
\label{Rsc}

\begin{figure}
\begin{center}
\includegraphics[width=0.3\textwidth]{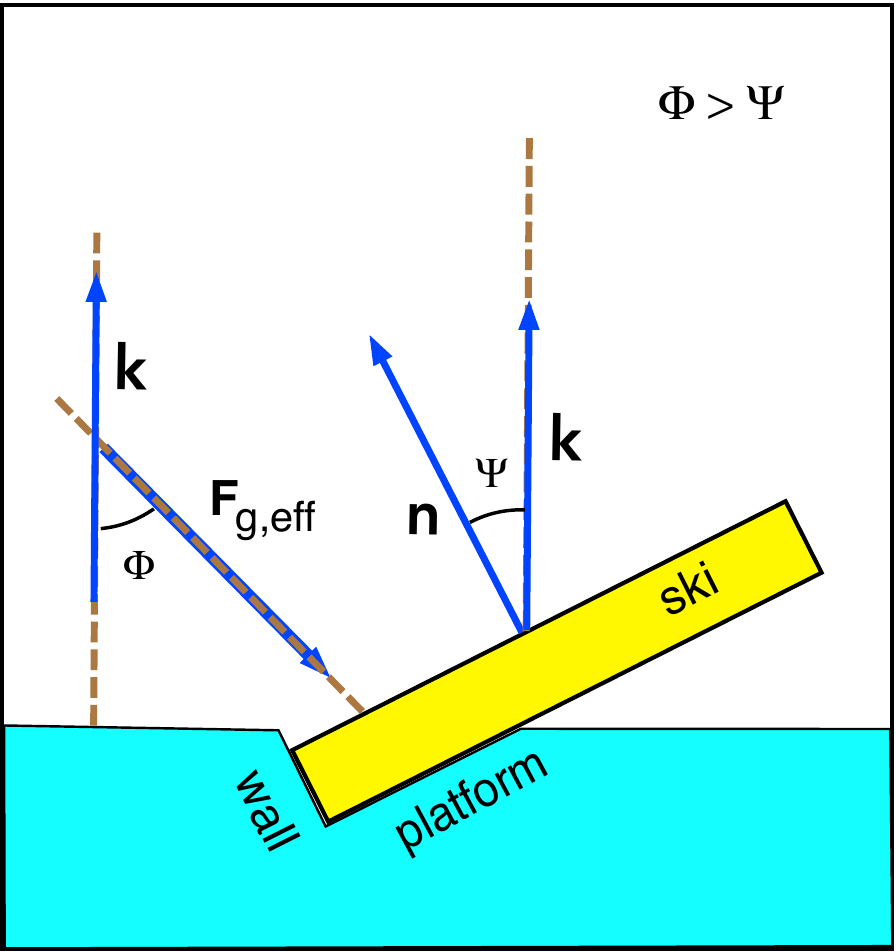}
\includegraphics[width=0.3\textwidth]{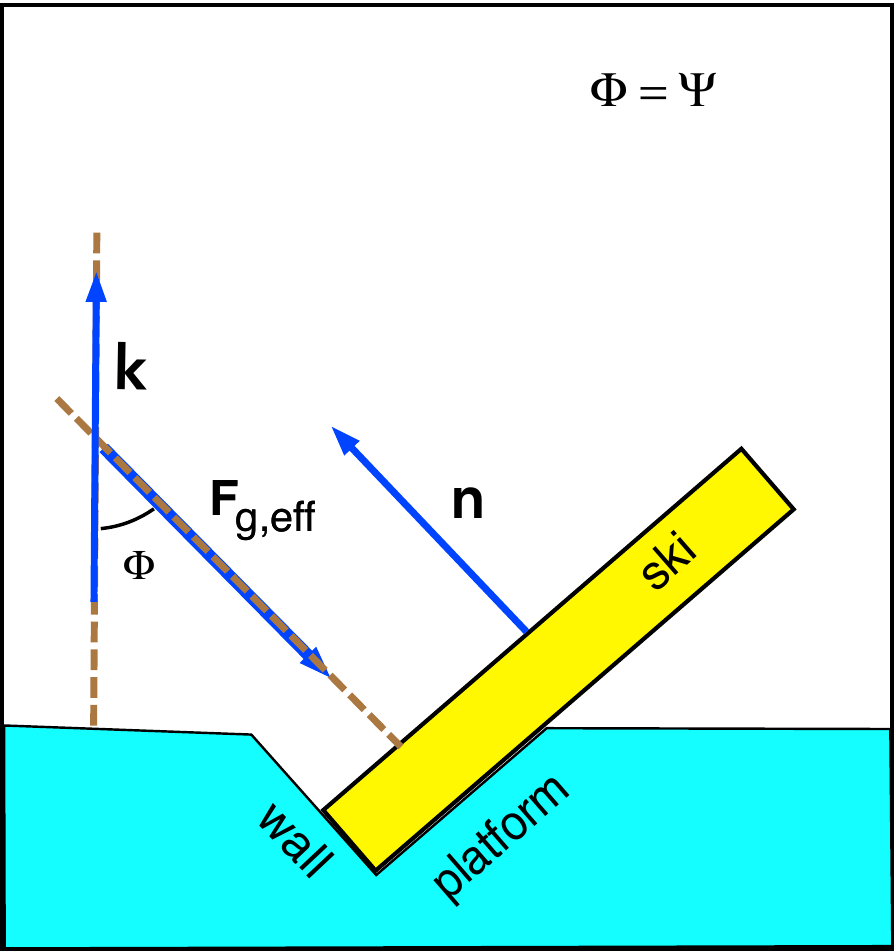}
\includegraphics[width=0.3\textwidth]{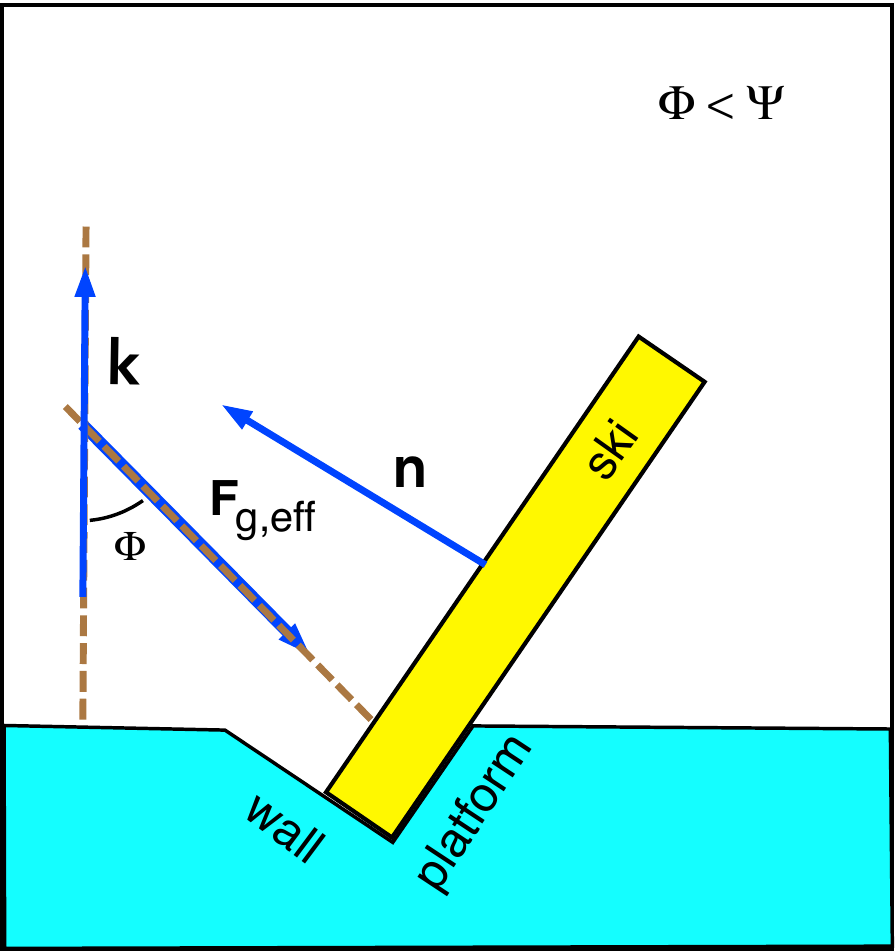}
\caption{Ski-snow contact and the force balance. {\it Left:} The inclination angle of the ski is smaller than the inclination angle of the effective gravity. In this case the effective gravity is pushing against the platform only. Hence the normal reaction force from the snow is normal to the platform and cannot balance the effective gravity.   
{\it Centre: } The inclination angle of the ski is the same as the inclination angle of the effective gravity. In this case the effective gravity is still pushing against the platform only but the normal reaction force is now aligned with the effective gravity and hence can balance it.  {\it Right: } The inclination angle of the ski is higher than the inclination angle of the effective gravity. In this case the effective gravity is pushing against the platform and its wall.   The combined normal reaction force from the platform and the wall can balance the effective gravity.  
}
\label{canting}
\end{center}
\end{figure}

By its nature, $\bF\sub{n}$ is normal to the contact surface of skis with the snow and hence depends on how they are edged. 
Let us assume that the effective gravity force $\bF\sub{g,eff}$ is known and consider the possible orientations of the normal $\bn$ to 
the skis surface relative to $\bF\sub{g,eff}$ (see figure \ref{canting}). We denote the angle between $\bn$ and the slope normal $\bk$ 
as $\Psi$ and call it the ski inclination angle. As the ski carves a platform in the snow it necessary makes an inside wall of this platform and hence in general there are two contact surfaces involved in the problem -- in addition to the contact between the ski base and the platform we also have the contact between the sidewall of the ski and the wall of the platform. Both contacts can give rise to the snow reaction forces.

If $\Psi<\Phi$ then $\bF\sub{g,eff}$ is pushing only against the platform and hence the combined normal reaction force must be parallel to $\bn$. Obviously, it cannot balance $\bF\sub{g,eff}$, which is not aligned with $\bn$.  The unbalanced component of 
$\bF\sub{g,eff}$ results in side-slipping of the skis and the associated with this motion friction force has to be taken into account.

If $\Psi=\Phi$ then $\bF\sub{g,eff}$ is parallel to $\bn$ and in this marginal case the normal reaction originated at the platform contact can balance $\bF\sub{g,eff}$.  Since the side-slipping is not promoted, this case is consistent with carving turns. This case is relevant 
to the particular carving technique where the key parts of the skier body are well aligned with each other, the so-called stacked configuration.  

If $\Psi>\Phi$ then $\bF\sub{g,eff}$ is pushing the skis not only against the platform but also against its wall. However, the wall effectively prevents side-slipping and allows the force balance without invoking the friction force. Hence this case is also consistent
with carving turns. Here the snow reaction forces originate from both contact surfaces and their sum  is no longer parallel to $\bn$. 
This case is most common in carving as skiers tend to angulate their body in such a way that $\Psi>\Phi$. The angulation will be considered in some detail in Sec.\ref{sec:angulation}   

In the mechanics of solid bodies, by a balance we understand not only vanishing combined (total) force but also vanishing combined torque \cite{LL69}. Although skiers are not exactly solid bodies,  torques are still important in their dynamics.  Indeed, the combined force governs the motion of skier's CM only.  The combined torque determines if the skier stays in control or ends up over-rotated and even crashing.   When discussing torques it is important to know where each of the involved forces are applied.  Both the centrifugal and the gravity force (and hence the effective gravity) per unit mass are uniform and hence one may think of these forces as being applied directly at the CM. This is not true for the snow friction and normal reaction forces, which are applied at the skis, and for the aerodynamic force which is distributed over the skiers body in a rather complicated way.  

When discussing torques acting on a skier it is convenient to use the skier's frame of reference and introduce the three mutually orthogonal fundamental planes passing through skier's CM : the frontal, sagittal and transverse planes\cite{LM10}.  The frontal plane is normal to $\bu$, the sagittal plane is parallel to $\bu$ and $\bF\sub{g,eff}$,  the transverse plane is defined as the orthogonal to the other two.  In the transverse plane, the total torque does not have to vanish as skiers often rotate in this plane in the same sense as the skis (though a bit of counter-rotation can be used as well).  In the other two planes, the total torque must be close to zero. If the torque in the sagittal plane did not vanish, the skier would fall forward (or backward) and if the torque in the frontal plane did not not vanish the skier would fall to the inside (or outside) of the turn.    

In our model, we will assume that everything is fine with the skier balance in the sagittal and transverse planes and consider in details only  the balance in the frontal plane as it is the one which determining the turn radius.  How we analyse torques in this plane depends on the nature of skier's contact with the snow. It can be either (1) a sliding contact, in which case we have to deal with the torques about skier's CM  or (2) the skier can be regarded as pinned to the slope at the inside edge of the outside ski, in which case we should consider torque about the edge. The former is relevant for skidded turns and the latter  for carving turns.  Hence in carving turns the torque balance requires that effective gravity is directed along the line connecting CM with the inside edge of the outside ski.  

\begin{figure}
\begin{center}
\includegraphics[width=0.7\textwidth]{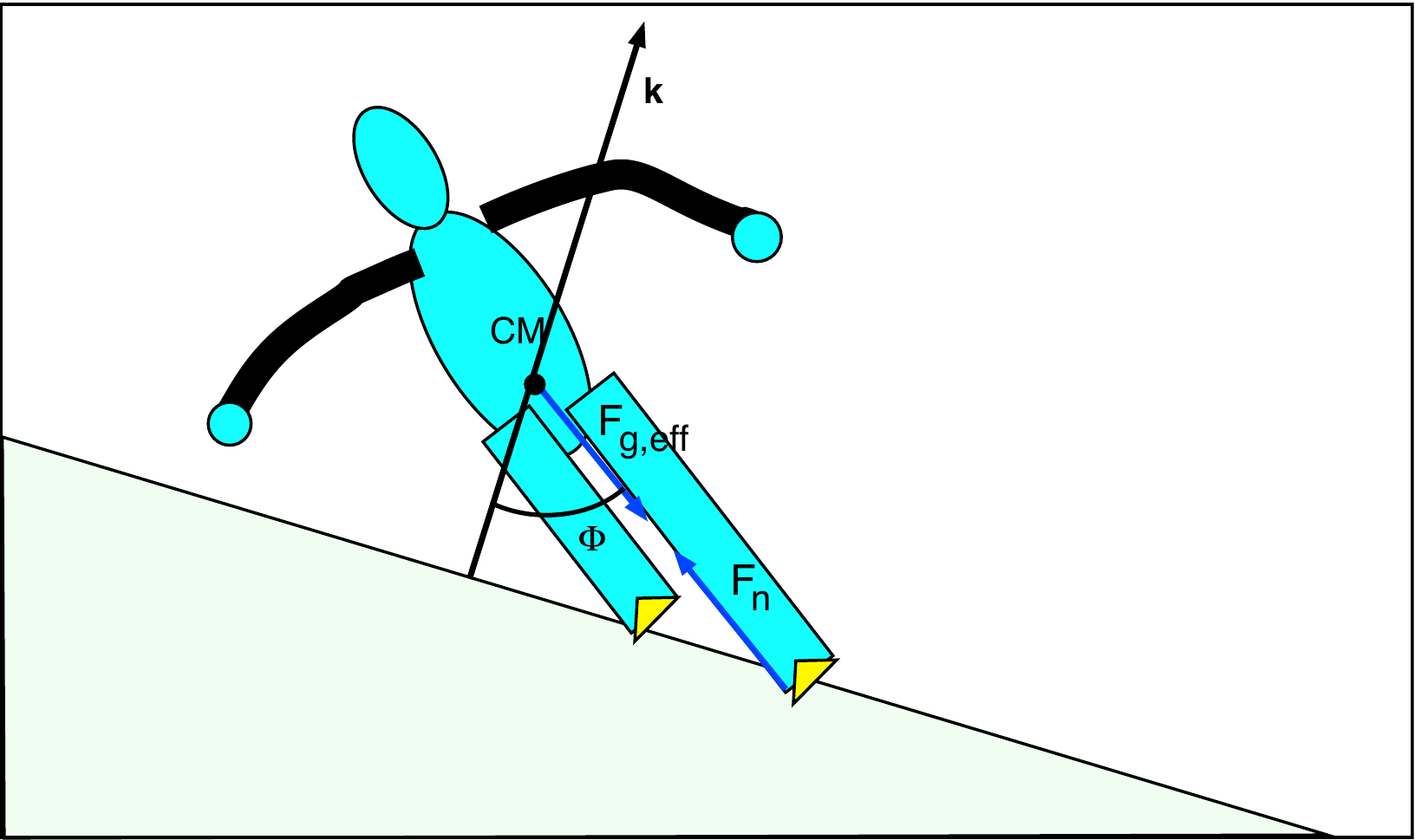}
\caption{Balance of a stacked (un-angulated skier) in the frontal plane. The normal reaction force balances the effective gravity force 
and introduces zero torque because it points directly at the skier's centre of mass (CM).  
This configuration corresponds to the Lower-C part of the turn.}
\label{inclination}
\end{center}
\end{figure}

Combining the torque condition with the condition of vanishing total force, we arrive to the balanced configuration where 1) $\bF\sub{g,eff}=-\bF\sub{n}$, 2)  $\bF\sub{g,eff}$ is applied at CM, 3) $\bF\sub{n}$ is applied at inside edge of the outside ski and 4) both these forces act along the line connecting CM with the inside edge of the outside ski (see figure \ref{inclination}). 

If the skier is stacked (their legs are aligned with their torso in the frontal plane) then their CM is located about their belle button and in the balanced position their whole body is also aligned with  $\bF\sub{g,eff}$. Hence in order to allow the force balance one has to ensure that in this position the ski surface is perpendicular to the line connecting CM with the ski edge. 
Obviously, this requirement is reflected in the overall design of the standard skiing equipment (boots, skis, bindings).    
However, the exact outcome also depends on the personal physiological features of individual skiers and a fine tuning 
is normally carried out via the so-called canting of ski boots  \cite{LM10}.  Ideally, when a skier is standing in full gear on a flat horizontal surface with all their weigh on one ski, the plumb-line attached to their belle button should lead to the inside edge of the ski.  If instead the line points to the outside of the edge then side-slipping will occur when trying to carve in a stacked position, as this results in $\Phi>\Psi$ during the turn (see figure \ref{canting}). If it points to the inside of the edge then carving entirely on the outside ski will be possible, as this implies  $\Phi<\Psi$ for the outside ski. However in this case the inside ski may slip when loaded and eventually collide with the outside ski, resulting in a loss of balance.    

If the skier body is not stacked but angulated at the hip than their CM becomes displaced away from their belle button. As the result the condition $\Phi=\Psi$ is no longer satisfied even if the boots are canted perfectly. The angulation at the skier knee leads to the same result.  Beginners often angle their body in such a way that $\Phi > \Psi$. This causes side-slipping of the skis even if the ski boots are perfectly canted.  Technically adept skiers angle their body in the opposite way so that $\Phi<\Psi$.  In this case, carving is possible even if the boot canting is not precise.    

In our analysis we assumed that the inside ski is not loaded. If it is then in the balanced position $\bF\sub{g,eff}$ does no longer point to the inner edge of the outside ski but to a point between the running edges of the inner and outer skis (See Appendix \ref{LIS}).  This leads to $\Phi<\Psi$ and hence the effect is similar to that of the angulation.  

From this point we focus mostly on the case of a stacked skier whose weight is loaded fully on the outside ski.   
The effect of the angulation on the turn dynamics will be considered near the end of the paper in Sec. \ref{sec:angulation}.

The running edge of a flatten carving ski is close to an arc of a circle. The radius of this circle is called the ski sidecut radius, $R\sub{sc}$.
When the ski is placed at the inclination angle  $\Psi$ to a flat surface and then pressed in the middle until it comes to contact with the surface, it edge can still be approximated as an arc but of a smaller radius  

\beq
   {R} = {R\sub{sc}} \cos\Psi \,.
  \label{turn-r}
\eeq   
(See \cite{LS04} and Appendix \ref{RBS}.) 
In ideal carving turns, there is no side-slippage and the ski is transported along its curved loaded edge. This makes the ski curvature radius $R$  the same as the local curvature radius of the ski trajectory.  This trajectory may not be exactly the same as the CM trajectory but here we ignore this difference. 

Combining equations (\ref{turn-r}) and (\ref{turn-phi})  with the balance condition  $\Phi=\Psi$ one finds that 
\beq
   \left( \fracp{R\sub{sc}}{R}^2 -1\right) ^{1/2} = \frac{v^2}{gR} \frac{1}{\cos\alpha} -  \tan\alpha\cos\beta \,.
  \label{ICE}
\eeq   
This equation, called by Jentschura and Fahrbach the Ideal-Carving Equation \cite{JF04},  defines $R$ as a function of skier's velocity and therefore allows to close the CM equations of motion.  Hence it allows deterministic mathematical modelling of carving runs using ordinary differential equations.  In practical terms this means that in a carved turn the skier has no control over its local curvature and hence trajectory!  
What is up to the skier is the decision when to switch from one turn to another and carve the next arc.

The equilibrium of a stacked skier who keeps all the load on the outside ski is similar to that of an inverted pendulum and hence unstable (\cite{LS04}, Appendix \ref{stability}).  However, skiers have ways of controlling this instability.  Lind and Sanders \cite{LS04} discuss the stabilising arm moment, similar to what is used by  trapeze artists.  Ski poles can provide additional points of support when planted into or dragged against the snow. Nowadays, the inclination of top racers in GS and super-G is so dramatic that they can use their arms to provide an additional sliding contact with the slope. A similar technique, but with fully extended arms, is applied by SL racers on steep slopes.   Some control can be provided by the body angulation (Appendix \ref{stability}).    Finally and presumably most importantly, when both skis are sufficiently wide apart and loaded, the skier is more like a flexible table than an inverted pendulum.  Indeed, in this case instead of the unique balanced inclination angle we have a continuum of balanced positions and so a small perturbation just shifts the skier to a nearby equilibrium. The skier has a simple way of controlling the inclination -- via relaxing and extending legs.          
Yet most of the load is still put on the outside ski and hence the mono-ski assumption underpinning  ICE is quite reasonable.

\subsection{Key limitations on the inclination angle}
\label{top-angle}

According to equation (\ref{turn-r}),  $R$ is a monotonically decreasing function of $\Psi$.   It is easy to see that  $R\to R\sub{sc}$ as $\Psi\to 0$ and hence the turn radius of marginally edged ski is exactly $R\sub{sc}$. At the other end of the range, $R\to 0$ as $\Psi\to 90^\circ$.  Obviously, $R=0$ corresponds to a ski curled into a point, which does not make any sense. This result reflects the fact that the approximation (\ref{r3})  breaks down for $\delta = l/R \ge 1$.  
Although we are not aware of any actual experiments, it feels all but impossible to bend a ski into an arc of the radius as small as the ski length without actually breaking it.  For $R=l\sub{ski}$ the exact equations (\ref{r2},\ref{r2a}) yield the inclination angle $\Phi\sub{s,max}=80^\circ$, which can be considered as an upper limit on $\Psi$. 

Another upper limit on $\Phi$ can be set by the athlete's ability to sustain the high g-force which builds up via the centrifugal acceleration.  Using equation (\ref{g-force}) we find that for $\alpha=15^\circ$ and $\Phi=80^\circ$ the g-force is almost as high as six.  According to LeMaster\cite{LM10} the best athletes can only sustain the g-force up to about three, which corresponds to the less extreme inclination angle $\Phi=70^\circ$. 

Finally, as $\Phi$ increases so does the tangential to the slope component of the effective gravity. This is effectively a shearing 
force acting on the snow. Above certain level it will cause snow fracturing and hence a loss of grip. As the result, the carving turn will 
turn into a skidded one \cite{M09,M13}.  The snow shear stress $S$ relates to $F\sub{n,c}$ via 
\beq
    F\sub{n,c} = S l\sub{ski} e \,, 
    \label{shear}
\eeq
where $e$ is the snow penetration depth in the direction normal to the slope. The penetration is related to the show 
hardness $H$ and $F\sub{n,k}$ via
\beq
    F\sub{n,k}= H V \,,
    \label{hardness}
\eeq
where $V=l\sub{ski} e^2/2\tan\Psi$ is the volume of the imprint made by the ski in the snow \cite{M13}.  The snow fractures when 
the shear stress exceeds the critical value $S\sub{c}$.  Based on equation (\ref{shear}) alone one would naively expect that carving requires
deep snow penetration. However, this ignores the fact that the penetration is dictated by the normal component of the same force. 
Combining equations  (\ref{shear}) and (\ref{hardness})  one finds the carving condition 
\beq
    e< \frac{2S\sub{c}}{H} \,, 
    \label{e-cond}
\eeq
which put an upper limit on the penetration, contrary to the naive expectation. 
This can be turned into the condition on the skier inclination angle. Indeed, using equation (\ref{m8}) one finds
$$
      e^2= \frac{2mg\cos\alpha\tan\Psi}{Hl\sub{ski}} 
$$
and, upon substitution of this result into equation (\ref{e-cond}), the condition
\beq 
\tan\Phi <\frac{2S^2\sub{c} l\sub{ski}}{mgH\cos\alpha} \,. 
\label{Phi-cond}
\eeq
According to the analysis in \cite{M13}, a well prepared race track  can be attributed with $S\sub{cr} \approx 0.52\,$N$\,$mm$^{-2}$ and 
$H\approx 0.21\,$N$\,$mm$^{-3}$. Using these values along with $m=70\,$kg, $l\sub{ski}=165\,$cm and $\alpha=15^o$ we 
obtain  $e<5\,$mm and $\Phi < 81^\circ$.  Thus all three conditions yield approximately the same upper limit on $\Phi$.    
( Looser snow will fracture under much lower shear stress allowing only small-inclination carving. )

\section{Speed limits imposed by the Ideal Carving Equation}
\label{SL}

Equation (\ref{ICE}) can be written in the form 
\beq
   \left( \xi^2 -1\right) ^{1/2} = a\xi +b \,,
  \label{s2}
\eeq   
where $\xi=R\sub{sc}/R$,  $a=v^2/gR\sub{sc}\cos\alpha$ and $b=-\cos\beta \tan\alpha$. From the definitions it is clear that 
$$
\xi \geq 1\,,\quad a>0 \etext{and}   -\tan\alpha <b<+\tan\alpha \,.
$$  
When squared, equation (\ref{s2}) becomes quadratic, which 
is easy to analyse,  but this operation may introduce an additional root and for this reason we will deal with this equation as it is.  

The right-hand side of (\ref{s2}) is the linear function $g(\xi)\equiv \tan\Psi(\xi)=a\xi+b$. Since its gradient $g'(\xi)=a>0$ this function increases with $\xi$. The left-hand side of (\ref{s2}) is the radical function $f(\xi)\equiv\tan\Phi(\xi)=\sqrt{\xi^2-1}$.  Since $d(\tan\Phi(\xi))/d\xi=\xi/\sqrt{\xi^2-1}>0$ it also grows with $\xi$ . Moreover,  $d(\tan\Phi(\xi)/d\xi))\to+\infty$ as $\xi \to 1$ and  $d(\tan\Phi(\xi))/d\xi \to 1$ as $\xi \to +\infty$. Finally, $d^2(\tan\Phi(\xi))/d\xi^2=-1/\sqrt{\xi^2-1}<0$ and hence $d(\tan\Phi(\xi))/d\xi$ decreases monotonically from $+\infty$ to 1 for $\xi\in[1,+\infty)$.

\begin{figure}
\begin{center}
\includegraphics[width=0.49\textwidth]{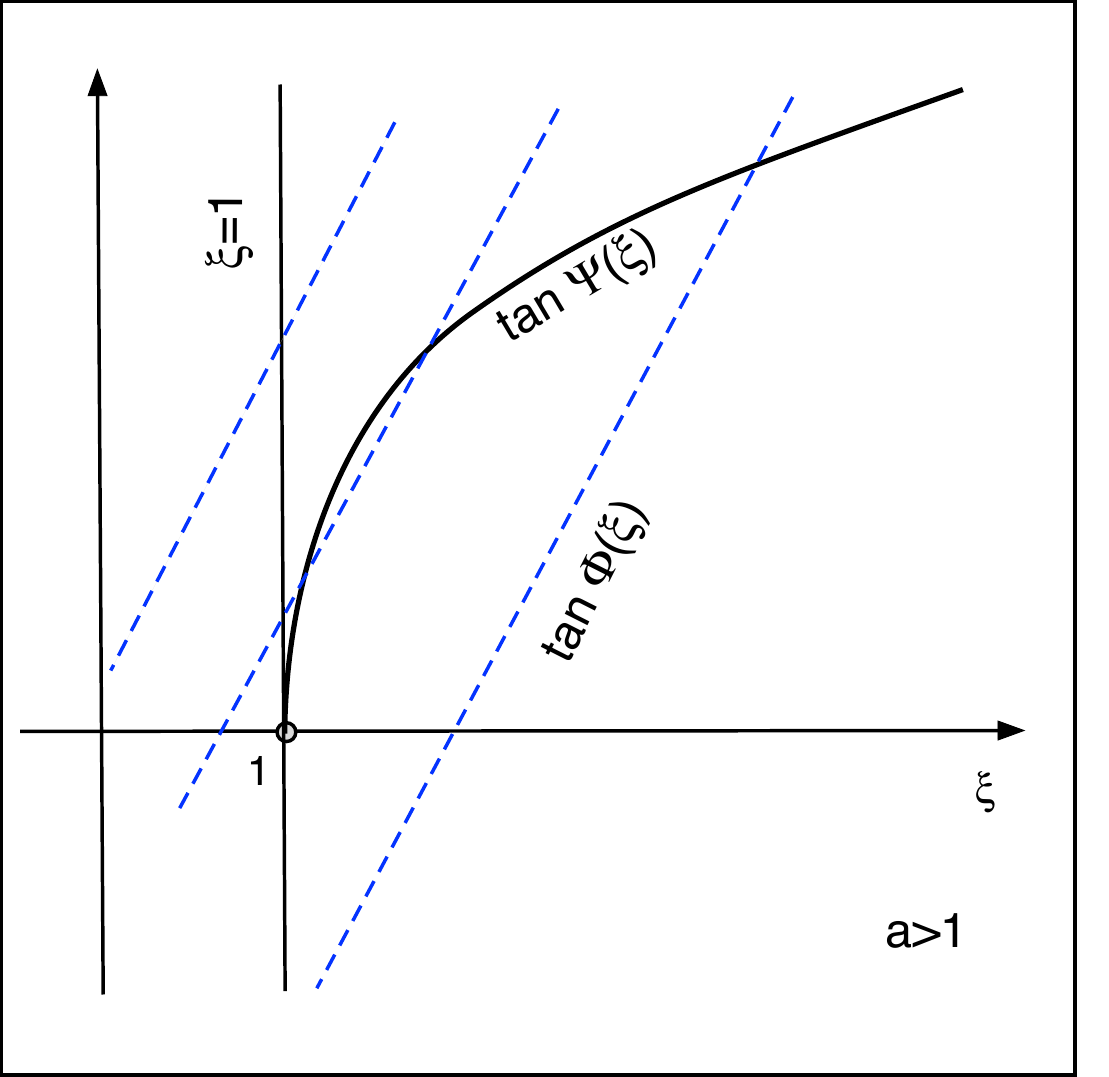}
\includegraphics[width=0.49\textwidth]{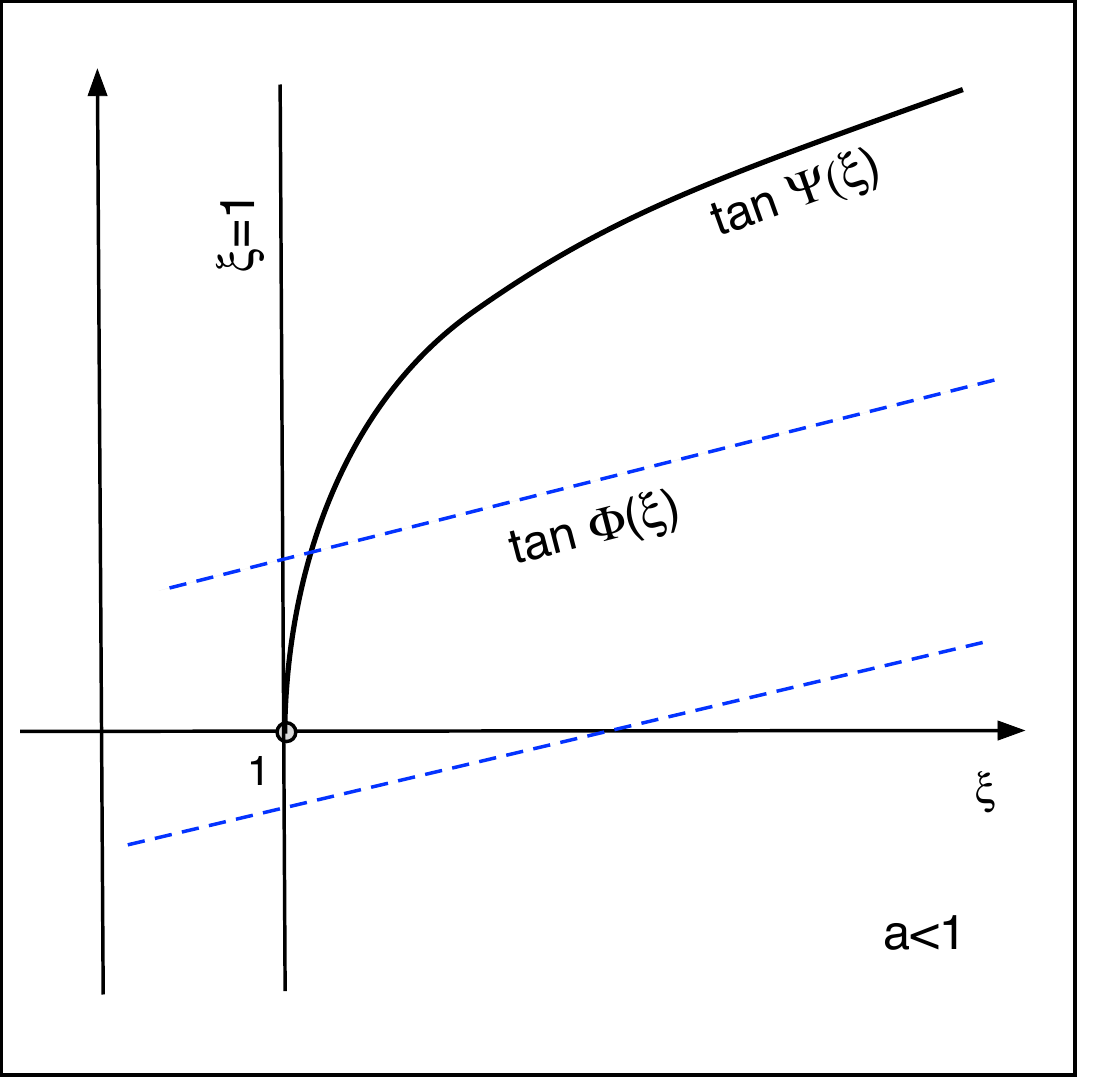}
\caption{ Finding roots of the ideal carving equation (ICE).  The solid line shows $\tan\Phi(\xi)=\sqrt{\xi^2-1}$ and 
the dashed lines $\tan\Psi(\xi)=a\xi+b$. The roots of ICE are given by the intersection points of the curves. 
{\it Left panel}:   The case $a>1$.  Depending on the value of $b$, one can have none, one or two roots.   
{\it Right panel}: The case $0<a<1$. Now one can have either one root or none. }
\label{figure6}
\end{center}
\end{figure}

\subsection{The case $a>1$}
\label{lover-c-limit}

In this case, the gradient of $\tan\Psi(\xi)$ is higher than the asymptotic gradient of $\tan\Phi(\xi)$ and 
as one can see in the left panel of figure \ref{figure6} there are three distinguish possibilities with either none, one or two roots. 
First, if $\tan\Psi(1)=a+b<0$ then there exists one and only one root. Indeed, in this case  $\tan\Psi(1)<\tan\Phi(1)$ but the faster growth of $\tan\Psi(\xi)$ for 
large $\xi$ ensures that it will eventually overcome $\tan\Phi(\xi)$ (see the lower dashed line in the left panel of figure \ref{figure6}).  

As we increase $b$ above $-a$, initially there are two roots but eventually they merge and disappear.  The bifurcation occurs at the point $\xi\sub{c}$ where 
$$
   \tan\Psi(\xi\sub{c}) = \tan\Psi(\xi\sub{c}) \etext{and}   \oder{\tan\Psi}{\xi}(\xi\sub{c}) = \oder{\tan\Phi}{\xi}(\xi\sub{c}) \,. 
$$   
Solving these two simultaneous equations for $\xi\sub{c}$ and $b\sub{c}$, we find that 
\beq
       b\sub{c}=-(a^2-1)^{1/2}  \
\label{b-dissapear}
\eeq
and
\beq
       \xi\sub{c}=\frac{a}{(a^2-1)^{1/2}}  \,.
\label{xi-dissapear}
\eeq

There no solutions when for $b>b\sub{c}$,  which includes all positive values of  $b$ and hence all $\beta>90^\circ$ (Lower-C).    
  
\subsection{The case $a<1$}

In this case, the gradient of $\tan\Psi(\xi)$ is lower than the asymptotic gradient of $\tan\Phi(\xi)$ and 
as illustrated in the right panel of figure \ref{figure6} there is either one solution or none.  
The corresponding condition for existence  is now 
\beq
   a + b \ge 0 \,. 
  \label{s3}
\eeq   
This includes the case of $b>0$ and hence there is always a unique solution for the Lower-C. 

\begin{figure}
\begin{center}
\includegraphics[width=0.7\textwidth]{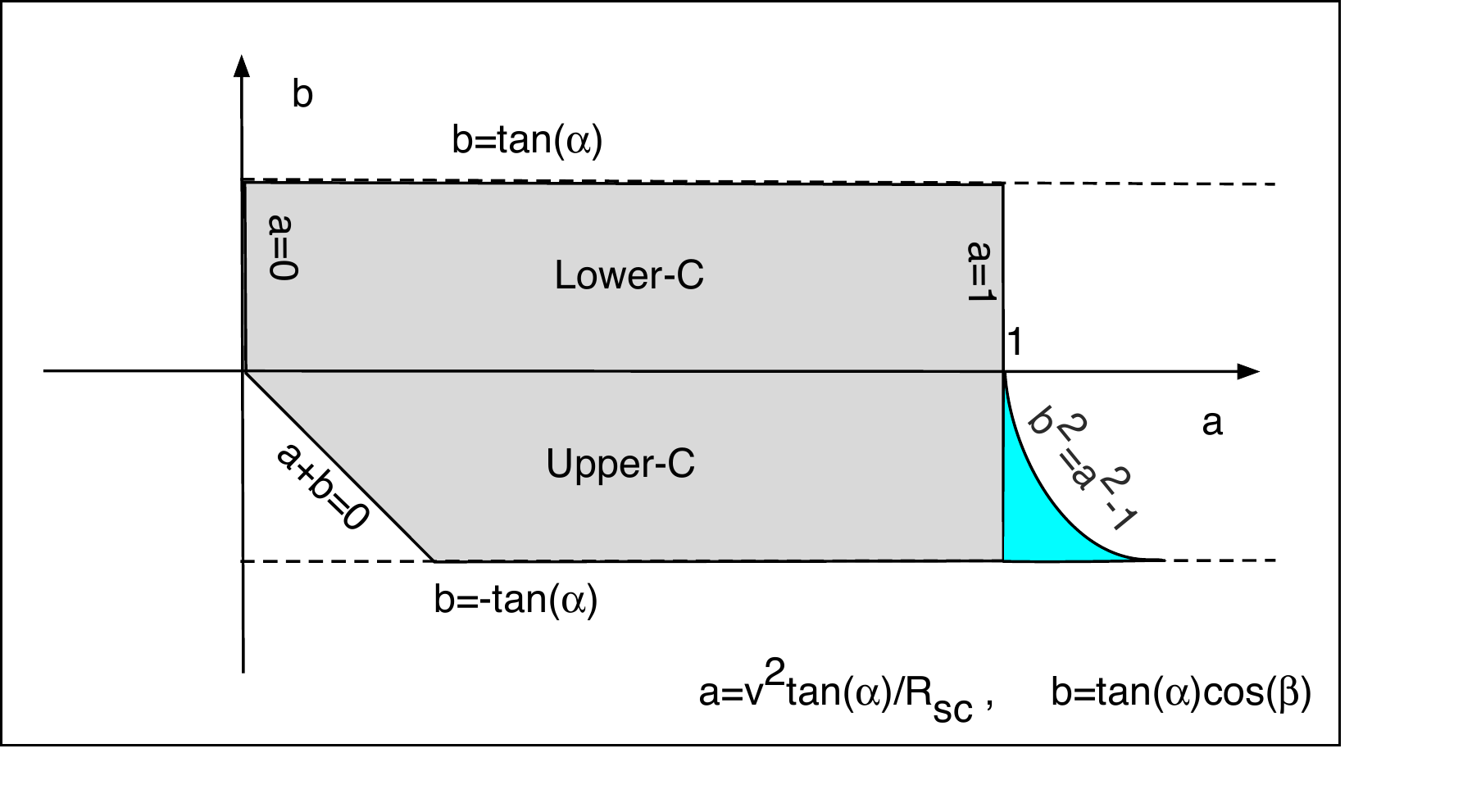}
\caption{ The parameter space of carving turn.  The regions where ICE has solutions are shaded, 
in grey where the solution is unique and in cyan where there are two solutions.}
\label{figure7}
\end{center}
\end{figure}

If $a+b < 0$ then there is no solutions. This condition implies negative $b$ and hence the Upper-C part of the turn.   
The results for both cases are illustrated in figure \ref{figure7}. 
  
\subsection{Interpretation}
\label{interpretation}

In terms of the usual physical parameters,  the condition $a<1$ reads  
\beq
      v <  V\sub{sc} \etext{where} V\sub{sc} = \sqrt{g R\sub{sc} \cos\alpha}  
  \label{cc1d}
\eeq   
This is the same as equation (36) in \cite{JF04}.  For $v>V\sub{sc}$, there are no solutions for sufficiently large angle of traverse, 
including the whole of the Lower-C.  Thus, the speed should not be too high or the skier will not be able to carve Lower-C. 

What are the indicators that a skier is about to hit the speed limit? Consider the entrance point to the Lower-C where   
$\beta=90^\circ$, $b=0$ and hence equation (\ref{s2}) reads  
$$
   \left( \xi^2 -1\right) ^{1/2} = a\xi  \,.
$$
Its solution 
$$
   \xi=\frac{1}{\sqrt{1-a^2}} \to\infty \etext{as} a\to 1 \,. 
$$
Thus the turn becomes very tight (formally $R\to 0$) and the skier's body close to horizontal  (formally $\Phi\to 90^\circ$). 
On approach to this point, something will give up. As we discussed earlier this will be either the skis, the skier's legs or the snow resistance to the applied shearing force (see Sec.\ref{top-angle}).   

However, when the speed limit is exceeded in the Upper-C there may not be such a clear indicator.  In fact,  after this the turn can be continued 
for a while until it reaches the critical traverse angle $\beta\sub{c}$ where 
$$
    \cos\beta\sub{c}= \cot\alpha\left( \fracp{v}{V\sub{sc}}^4-1 \right)^{1/2} 
$$   
and
\beq
   \tan\Phi\sub{c}=  \left(\fracp{v}{V\sub{sc}}^4 -1 \right)^{-1/2} 
  \label{cc2da}
\eeq   
(see equations \ref{b-dissapear} and \ref{xi-dissapear}). After this point the lateral balance can no longer be sustained.
Figure \ref{figure6} shows that in this case $\Phi(\xi)>\Psi(\xi)$ for any $\xi\ge1$. Hence the inclination angle of the effective gravity is higher than the inclination angle of the skis for any turn radius, making side-slipping unavoidable.

When $v$ grows slowly at the point of going over the upper speed limit and stays only just above $V\sub{sc}$, the loss of balance occurs close to the fall line at  extreme inclination angles.   If however the growth is fast, $v$ may significantly overshoot $V\sub{sc}$ quite early in the High-C. In this case the loss of balance may also occur early at $\Phi$ significantly below  $90^\circ$. As we shall see in Sec.\ref{sslope} this can occur on very steep slopes.  

When $v<V\sub{sc}$ and the carving turn is possible in the Lower-C, it may still not be possible in the Upper-C. 
This is what is governed by the condition (\ref{s3}).     In terms of the turn speed and the angle of traverse this condition reads   
\beq
   \cos\beta \le \frac{v^2}{gR\sub{sc}\sin\alpha} \,.
  \label{cc2d}
\eeq   
It is convenient to rewrite this as a condition on the angle between the skier velocity and the fall line, $\gamma=|\beta-90^\circ |$,
\beq
   \sin\gamma  \le \frac{v^2}{gR\sub{sc}\sin\alpha} \,.
  \label{cc3d}
\eeq   
Thus, for sufficiently low speeds, namely $v^2<gR\sub{sc}\sin\alpha$, carving is possible only close to the fall line. 
The physical nature of this condition is easy to understand by considering the limit $v\to 0$. In this case the centrifugal force is 
small and in order to keep their balance in a traverse perpendicular to the fall line the skier must be alined with the vertical 
direction (vector $\bs$).  This implies ski edging which is consistent with the Lower-C only.  Carving of the Upper-C becomes possible for 
the first time when the skis can run flat and hence $R=R\sub{sc}$. This requires the tangential component of the gravity force to balance the centrifugal force with leads to the condition 
$$
   v^2 = g R\sub{sc} \sin\alpha\cos\beta \,. 
$$     

To allow fully rounded turns,  with $\gamma$ approaching $90^\circ$ in the transition between the Lower-C and Upper-C parts of the turn, the speed must satisfy the condition
\beq
   v > \sqrt{gR\sub{sc}\sin\alpha} \,.
  \label{cc4d}
\eeq   

For a half-circle turn, which starts and finishes perpendicular to the fall line  ($\beta=0,\pi$)  the skier speed has to be within 
the range 
\beq
   \sqrt{gR\sub{sc}\sin\alpha} < v <  \sqrt{gR\sub{sc} \cos\alpha} \,,
  \label{cc5d}
\eeq   
which can be quite narrow for sufficiently large $\alpha$ and collapses completely for $\alpha\ge45^\circ$, making such turns impossible.   

It is easy to see that 
\beq
     V\sub{sc} = V\sub{g} \fracp{R\sub{sc}}{L\sub{g}\tan\alpha} ^{1/2}\,.
  \label{cc6d}
\eeq   
For the typical parameters  of slalom competitions, this gives
\beq
   V\sub{sc} = 0.3 V\sub{g}  
     \fracp{R\sub{sc}}{13\,\mbox{m}}^{1/2} 
     \fracp{L\sub{g}}{400\,\mbox{m}}^{-1/2} 
      \fracp{\tan\alpha}{\tan 20^\circ}^{-1/2}   \,.
  \label{s4}
\eeq   
Thus, for carving turns in slalom  $V\sub{sc}$ is significantly smaller than the characteristic speed speed $V\sub{g}$ (see equation \ref{Vg}) set 
by the aerodynamic drag.   This suggests that pure ideal carving turns are normally impossible for the typical parameters of slalom competitions and the racers have to shave their speed via skidding on a regular basis. The only exceptions are probably 
(i) very flat sections of the track where the last factor of equation (\ref{s4}) can be sufficiently large,
(ii) the first few turns right after the start, where the speed has not yet approached $V\sub{sc}$   
and (iii) the first turns after transitions from steep to flat. In the last case, $V\sub{sc}$ significantly increases at the transition, creating the 
opportunity for accelerating pure carving turns.  It is easy to verify that the situation is quite similar in other race disciplines. 
For example, for the downhill parameters 
\beq
   V\sub{sc} = 0.52 V\sub{g}  
     \fracp{R\sub{sc}}{50\,\mbox{m}}^{1/2} 
     \fracp{L\sub{g}}{700\,\mbox{m}}^{-1/2} 
      \fracp{\tan\alpha}{\tan 15^\circ}^{-1/2}   \,.
  \label{s4a}
\eeq   
%

\section{Dimensionless equations of carving turn}
\label{d-eqs}

It is a common practice of mathematical modelling to operate with dimensionless equations, which are derived using a set of scales characteristic to the problem under consideration, instead of standard of units. This leads to simpler equations which are easier to interpret and to the results which are at least partly scale-independent. The analysis of fall-line gliding given in  Section \ref{estimates}  suggests  that $V\sub{g}$, $T\sub{g}$  and $L\sub{g}$ are good candidates for the characteristic speed, time, and length scales of alpine skiing.  However, the analysis of ICE given in Sec.\ref{SL} shows that in ideal carving the speed remains significantly below 
$V\sub{g}$  and suggests to use $V\sub{sc}$ as a speed scale instead. Moreover, the turn radius is bound to remain below $R\sub{sc}$ which suggest that this is a more suitable length scale. The corresponding time scale is $T\sub{sc}=R\sub{sc}/V\sub{sc}$.   These are the scales we adopt here.  

In order to derive the dimensionless equations we now write 
$$
    v= V\sub{sc} \tilde{v} \,,\quad t = T\sub{sc} \tilde{t} \,,\quad  R = L\sub{sc} \tilde{R}   \,,
$$  
substitute these into the dimensional equations and where possible remove common dimensional factors. Finally, we ignore tilde in the 
notation. In other words, we do the substitutions $v \to V\sub{sc} v$,   $t \to T\sub{sc} t$, $R \to R\sub{sc} R$ and then simplify the results.  For example, substituting $v \to V\sub{sc} v$ into the equation (\ref{i7}) gives the dimensionless equation for the saturation speed of fall-line gliding as 
\beq
   v\sub{s}= \mbox{Sr} \sqrt{1-\mu\cot\alpha} \,, 
  \label{saturation-speed}
\eeq   
which includes the dimensionless speed parameter $\mbox{Sr}=V\sub{g}/V\sub{sc}$.
Similarly we deal with other dimensional variables should they appear in the equations, e.g.   $x \to R\sub{sc} x$ and 
$y \to R\sub{sc} y$.  In particularly, the application of this procedure to (\ref{eq-speed}) gives the dimensionless speed equation 
\beq
   \oder{v}{t}= \sin\beta\tan\alpha  -\frac{\mu}{R} -  \Kn v^2 \,,
  \label{deq-speed}
\eeq   
where 
\beq
\Kn=\frac{R\sub{sc}}{L\sub{g}}\,,
  \label{K-number}
\eeq   
is a dimensionless parameter which we will call the dynamic sidecut parameter.  
The g-force can be written as 
\beq
  \mbox{g-force} = \frac{\cos(\alpha)}{R} \,.
  \label{Gforce-Fn}
\eeq   

The dimensionless ideal-carving equation (\ref{ICE})  reads 
\beq
   \left( \frac{1}{R^2} -1\right) ^{1/2} = \frac{v^2}{R}   - \cos\beta \tan\alpha \,.
  \label{dice}
\eeq   
The equation governing the evolution of $\beta$ follows from equation (\ref{curve-radius}) upon the substitution $dl=vdt$. It reads
\beq
   \oder{\beta}{t} =  \frac{v}{R} 
     \label{dbdt}
\eeq   
and the dimensionless skier coordinates can be found via integrating 
\beq
   \oder{x}{t} =\mp v\cos\beta \,,  
     \label{dxdt}
\eeq   
where we use the sign minus for right turns and the sign plus for left turns, and 
\beq
   \oder{y}{t} = v \sin\beta \,.   
     \label{dydt}
\eeq    
 
Equations (\ref{deq-speed}-\ref{dydt}) determine the arc of a carving turn and the skier motion along the arc. What they do not tell us is when one turn ends and the next one begins.  These transitions have to be introduced independently. In this regard the angle of traverse is a more suitable independent variable than the time because its value is a better indicator of how far the turn has progressed.  Replacing $t$ with $\beta$ via equation (\ref{dbdt}) we finally obtain the complete system of equations which we use to simulate carving runs in this study. It includes three ordinary differential equations     

\beq
   \oder{v}{\beta}= \frac{R}{v} \left(   \sin\beta\tan\alpha  -\frac{\mu}{R}  -  \Kn v^2     \right) \,,
  \label{DVDB}
\eeq   
\beq
   \oder{x}{\beta} = \mp R \cos\beta  \,, 
     \label{DXDB}
\eeq   
\beq
   \oder{y}{\beta} = R \sin\beta \,, 
     \label{DYDB}
\eeq    
 %
%
%
and the constitutive  equation
\beq
  \left( \frac{1}{R^2} -1\right) ^{1/2} = \frac{v^2}{R}   - \cos\beta \tan\alpha  \,.
  \label{DICE}
\eeq

The definition of  $\beta$ implies that it increases both during the left and the right turns, but not in the transition between turns. 
In the ideal transition the direction of motion $\bu$ remains unchanged and hence the angle of traverse has to change from  $\beta$ to $180^\circ-\beta$.  According to equation (\ref{DICE}) this implies a jump in the local turn radius and hence the skier inclination. 
 Hence during the whole run, which may consist of many turns, $\beta$ remains confined between  $0^\circ$ and  $180^\circ$, 
provided each turn terminates before going uphill.

Finally, we observe that equations (\ref{DVDB}) and (\ref{DICE})  do not involve $x$ and $y$ and hence can be solved  independently from equations (\ref{DXDB}-\ref{DYDB}). However, all these equations should be integrated simultaneously when we are interested in skier's trajectory.  Equation (\ref{dbdt}) should be added as well when the time characteristics are important.

\section{Ideal carving solutions}

 Here we present the results of our study of ideal carving runs as described by equations (\ref{DVDB}-\ref{DICE}) with instantaneous 
 ideal transitions between turns. This means the jump $\beta \to 180^\circ-\beta$ at a single point of skier's trajectory, which is the termination point of the previous turn and at the same time the initiation point of the next one.  At the start we specify the initial angle of traverse $\beta\sub{ini}$ and speed  $v\sub{ini}$, achieved during the run-up phase. Each turn is terminated when the traverse angle reaches a given value, denoted as $\beta\sub{fin}$.  Hence, beginning from second one, all turns are initiated at $\beta\sub{ini}=180^\circ-\beta\sub{fin}$.  In all the runs we set the coefficient of friction to  $\mu=0.04$.  In this section we  focus on the effect of slope steepness and fix the sidecut parameter to $K=0.0325$. For the aerodynamic length scale $L\sub{g}=400\,$m this corresponds to SL skis with the dimensional radius of $R\sub{sc}=13\,$m.  Later we discuss the effect of sidecut radius as well. 
  
\subsection{Flat slope}

First we consider a slope with $\alpha=5^\circ$.  For such a flat slope, the friction plays an important role even in the case of fall-line gliding where 
it keeps the speed well below  $V\sub{g}$. Indeed, for $\beta=90^\circ$ equation (\ref{saturation-speed})  implies that  
the gliding speed stops growing after reaching $v\sub{s}=0.737 V\sub{g}$.  Secondly, equation (\ref{s4})  gives the speed limit $V\sub{sc}=0.612 V\sub{g}$, indicating that in carving runs the speed may eventually get over the speed limit.

\begin{figure}
\begin{center}
\includegraphics[width=1.0\textwidth]{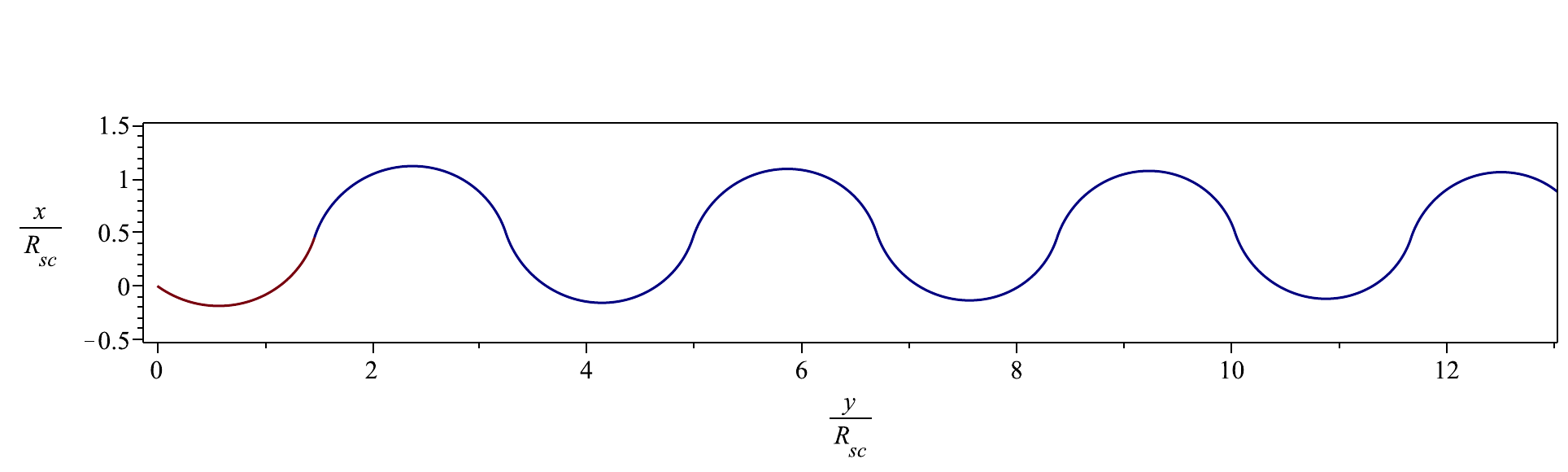}
\caption{ The first few turns of the flat-slope run ($\alpha=5^\circ$).}
\label{fig-tr-r1}
\end{center}
\end{figure}

\begin{figure}
\begin{center}
\includegraphics[width=0.4\textwidth]{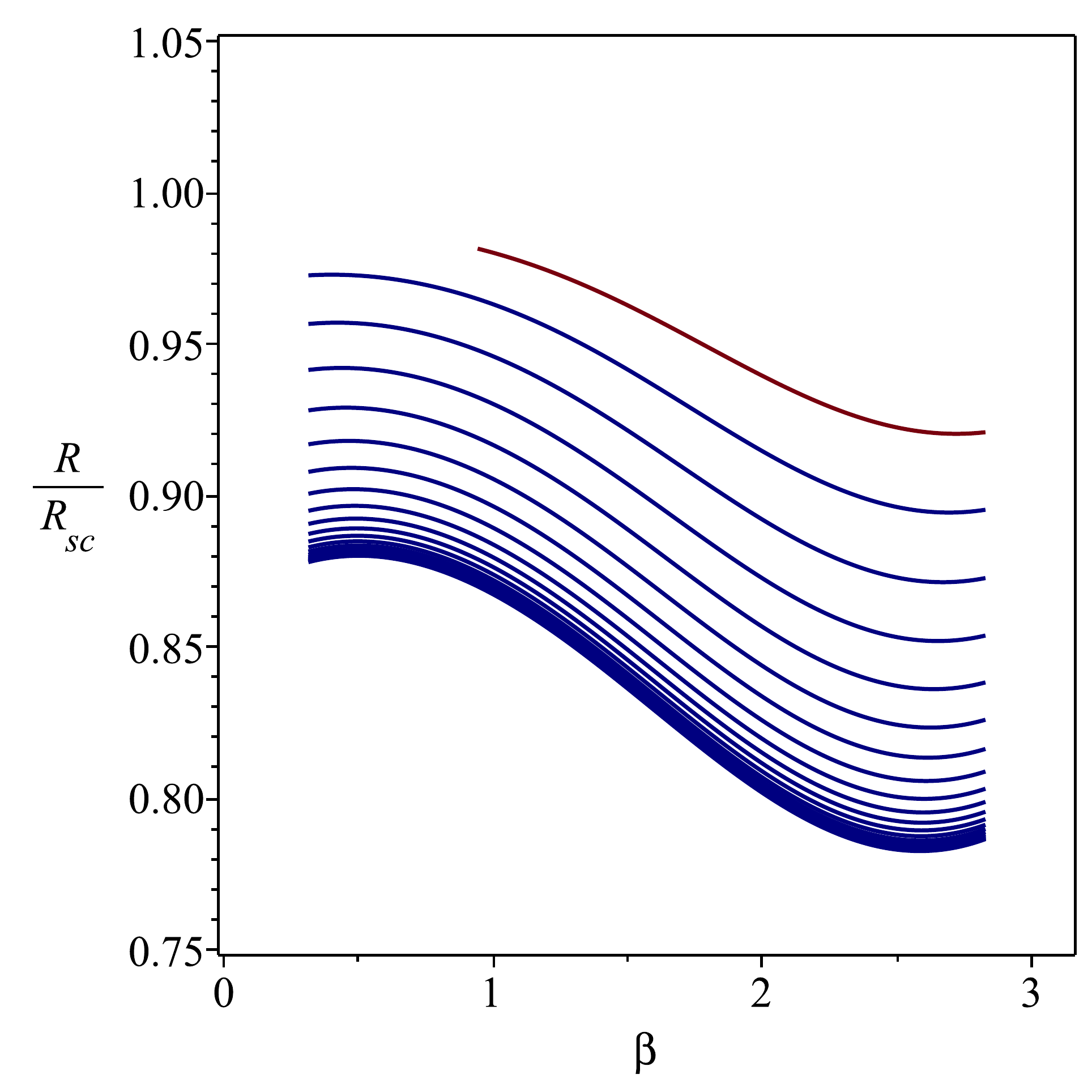}
\includegraphics[width=0.4\textwidth]{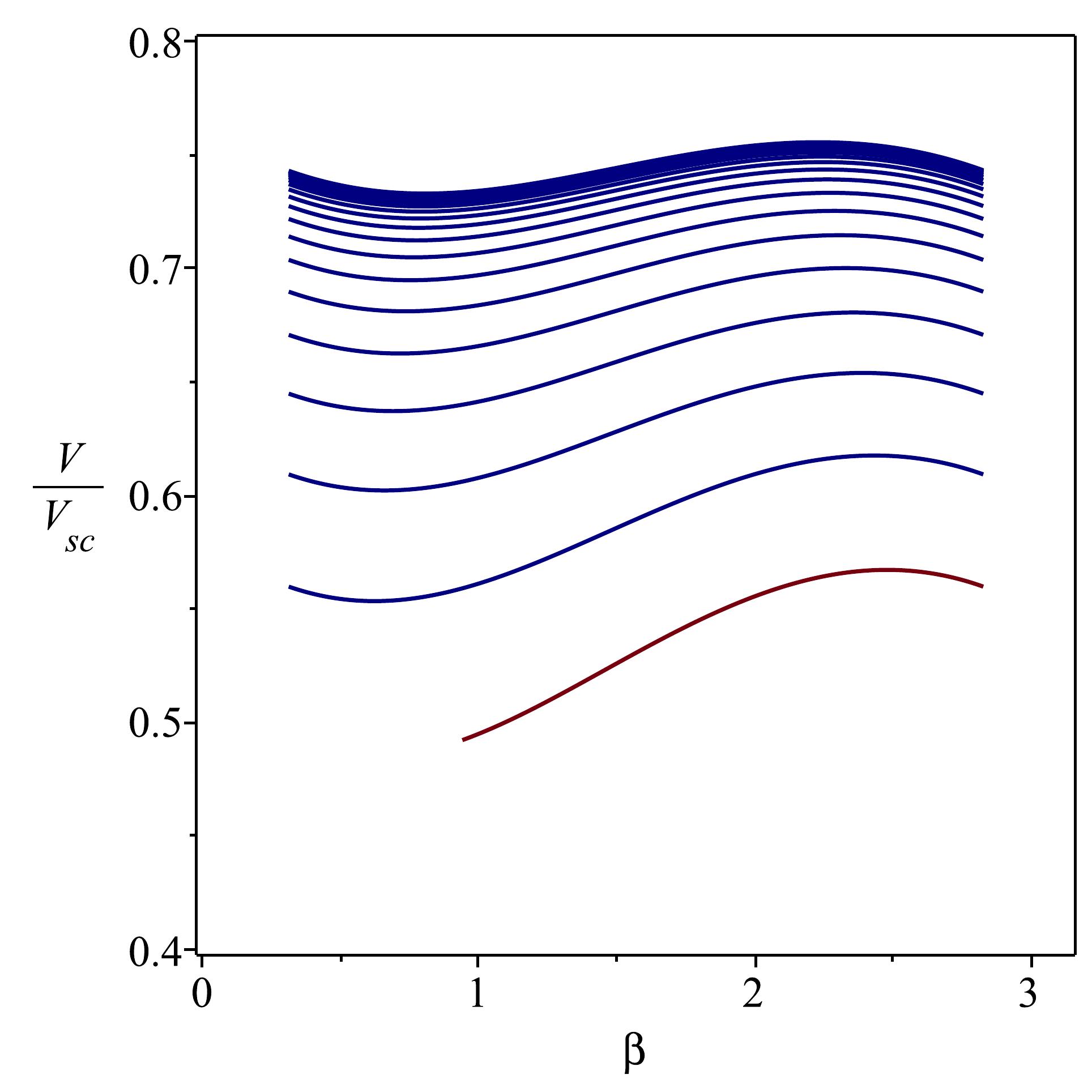}
\includegraphics[width=0.4\textwidth]{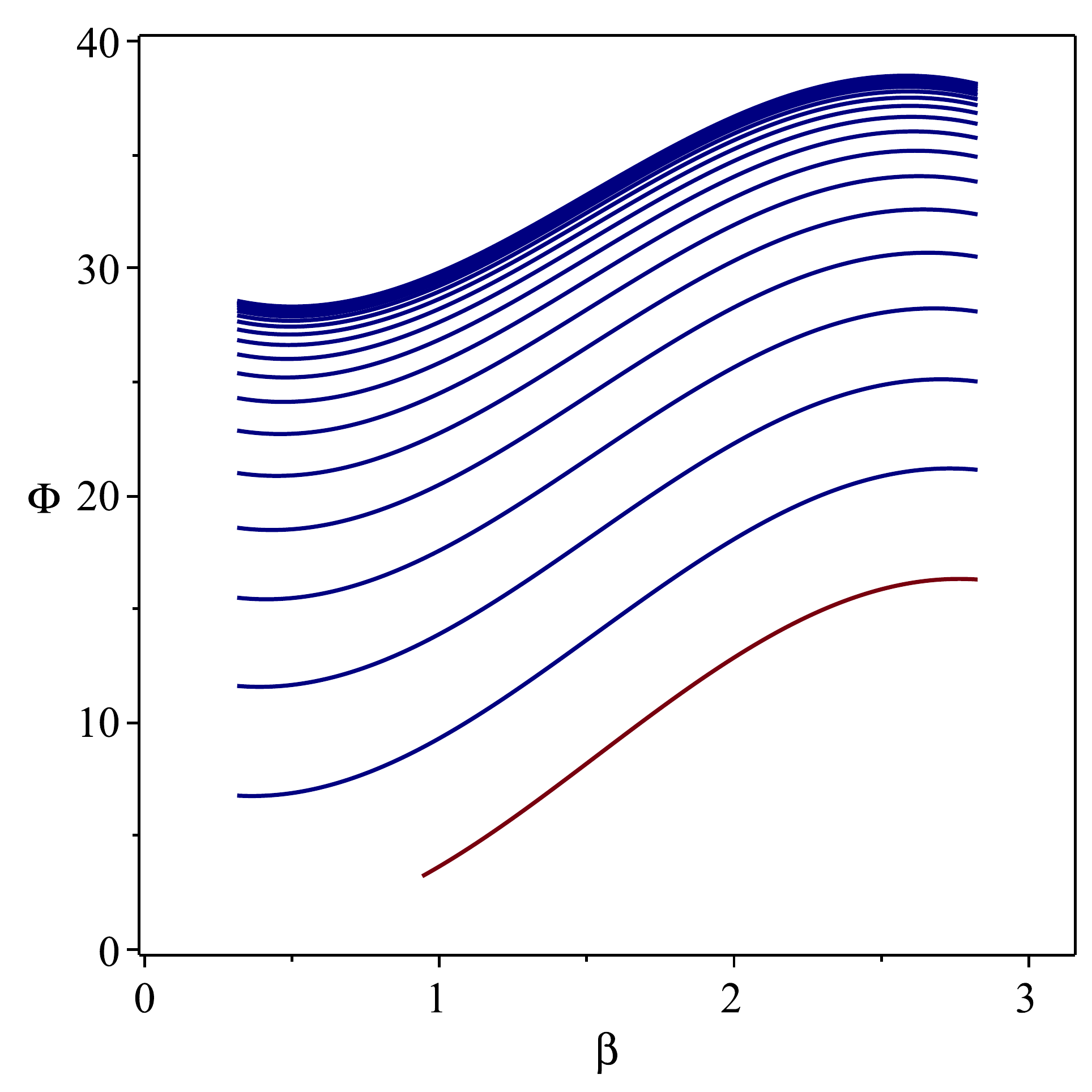}
\includegraphics[width=0.4\textwidth]{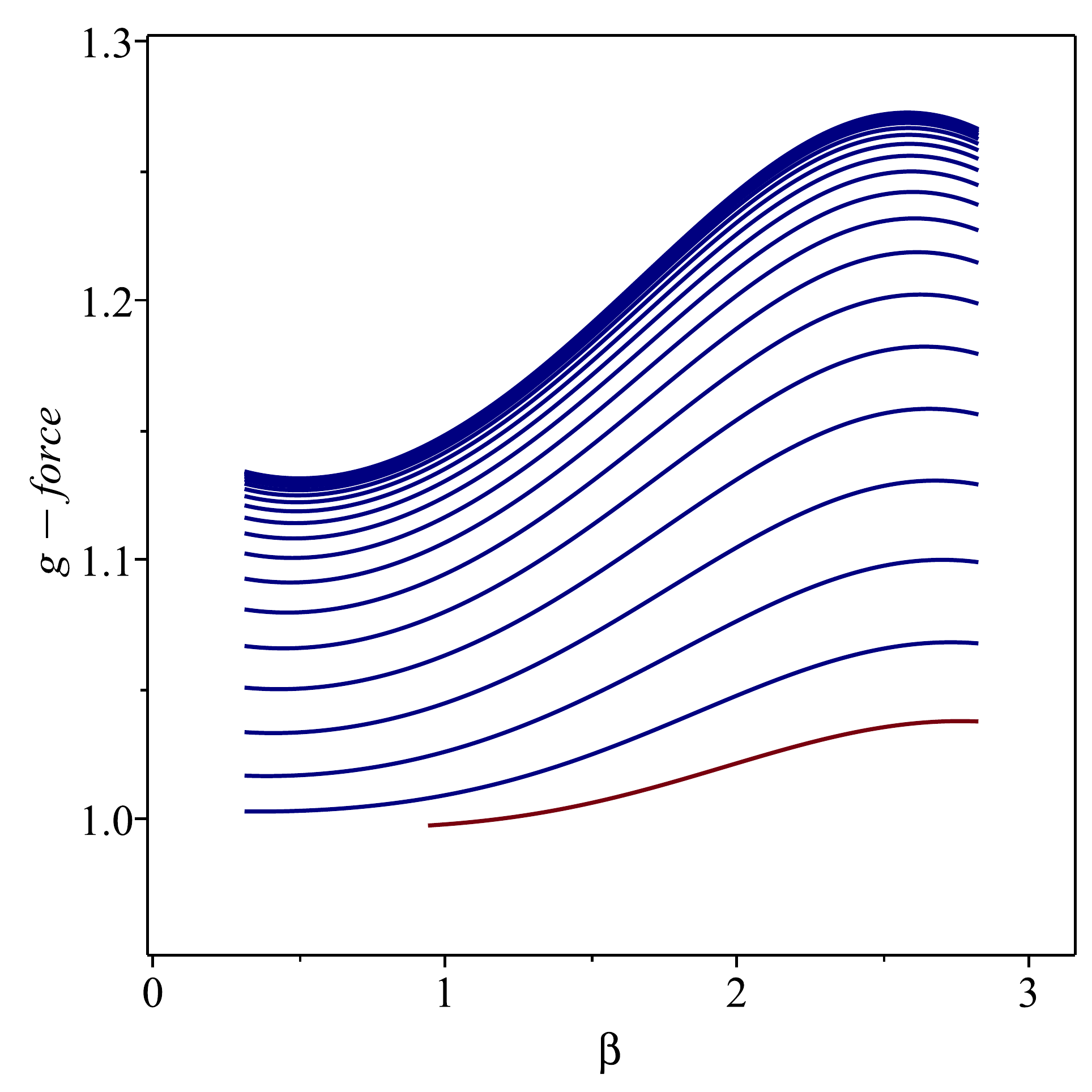}
\caption{Evolution of $R(\beta)$, $v(\beta)$, $\Phi(\beta)$ and $F\sub{n}$ for the flat-slope run ($\alpha=5^\circ$) during the first 20 turns.
In each of these plots, there are 20 curves which show the evolution of these variables during each turn. Each such curve (except the one corresponding to the first turn which originates at $\beta=0.3\pi$) originates at $\beta=0.1\pi$ (the turn initiation point)  and terminated at $\beta=0.9 \pi$ (the turn completion point).  The transition between turns is a jump from the termination point of the previous turn to the initiation point of the next turn. 
In the $R$-panel this transition is a jump to the lower curve and in the $v$-, $\Phi$- and $F\sub{n}$-panels to the upper curve.}
\label{fig-r1}
\end{center}
\end{figure}

Here we present the results for the initial data $x\sub{ini}=y\sub{ini}=0$, $\beta\sub{ini}=0.3\pi$ ($54^\circ$), and  $v\sub{ini}=0.49 V\sub{sc}$ ($0.3V\sub{g}$). Each turn terminates when the angle of traverse reaches $\beta\sub{fin}=0.9\pi$ ($162^\circ$). 
Figure~\ref{fig-tr-r1} shows the trajectory for the few first turns of the run, which indicates a rather slow evolution of the turn shape. This is confirmed  by the data presented in Figure~\ref{fig-r1}, which shows the evolution of $R$, $v$, $\Phi$ and the g-force for the first 20 turns. One can that the turn radius does not vary much along the track.  Moreover,  although each next turn is not an exact copy of the previous turn, for each of the variables we observe a convergence to some limiting curve, which will will refer to as the asymptotic turn solution. (This is reminiscent of the so-called limit cycle solutions in the theory of dynamical systems.)  In practical terms, well down the slope each next turn becomes indistinguishable from the previous one.

Interestingly, the  speed of the asymptotic solution remains everywhere below $V\sub{sc}$, with its mean value $\langle v \rangle \approx 0.75 V\sub{sc}$. As we mentioned in Section \ref{estimates}, this can be down to the longer length of our wiggly trajectory compared to the  length of 
fall line and the higher friction force due to the centrifugally-increased skier weight.  In order to test this explanation one can use the equilibrium version of the speed equation (\ref{DVDB})   

\beq
  \langle \sin\beta\rangle \tan\alpha-\mu \langle \frac{1}{R} \rangle  - \Kn \langle v^2 \rangle =0 \,,
  \label{s5}
\eeq   
where $\langle A \rangle$ stands for the mean turn value of the quantity $A$.  One can estimate $\langle \sin\beta\rangle$ via  
$$
\langle \sin\beta\rangle = \frac{1}{0.8\pi} \int\limits_{0.1\pi}^{0.9\pi} \sin(\beta) d\beta \approx 0.796 \,. 
$$ 
According to figure \ref{fig-r1},  $\langle 1/R \rangle = \langle \mbox{g-force}\rangle/\cos\alpha \approx 1.2$. 
Substituting the estimates into equation (\ref{s5}) we find  
$\langle v \rangle \approx 0.8V\sub{sc}$, which is quite close to what we have actually see in the numerical data.  
Using  equations (\ref{i9},\ref{cc6d}), we find the corresponding dimensional value  $ \langle v \rangle  \approx 27\,$km/h.

\subsection{Moderately steep slope}
\label{mss}
 
The setup of this run differs with that of the flat slope run only by the slope gradient,  $\alpha=15^\circ$.  In this case, equation (\ref{saturation-speed})  gives the saturation speed of a fall-line run $v\sub{s}=0.922 V\sub{g}$.  The fact that it is very close to $V\sub{g}$ tells us that for this gradient the role of the friction force in determining $v\sub{s}$ is substantially smaller than in the flat case.  On the other hand, equation (\ref{s4})  gives $V\sub{sc}=0.35 V\sub{g}$. This is significantly below $v\sub{s}$, suggesting that the skier speed will eventually exceed the carving upper limit. However, this is not what we find in the simulations. 
 
\begin{figure}
\begin{center}
\includegraphics[width=0.8\textwidth]{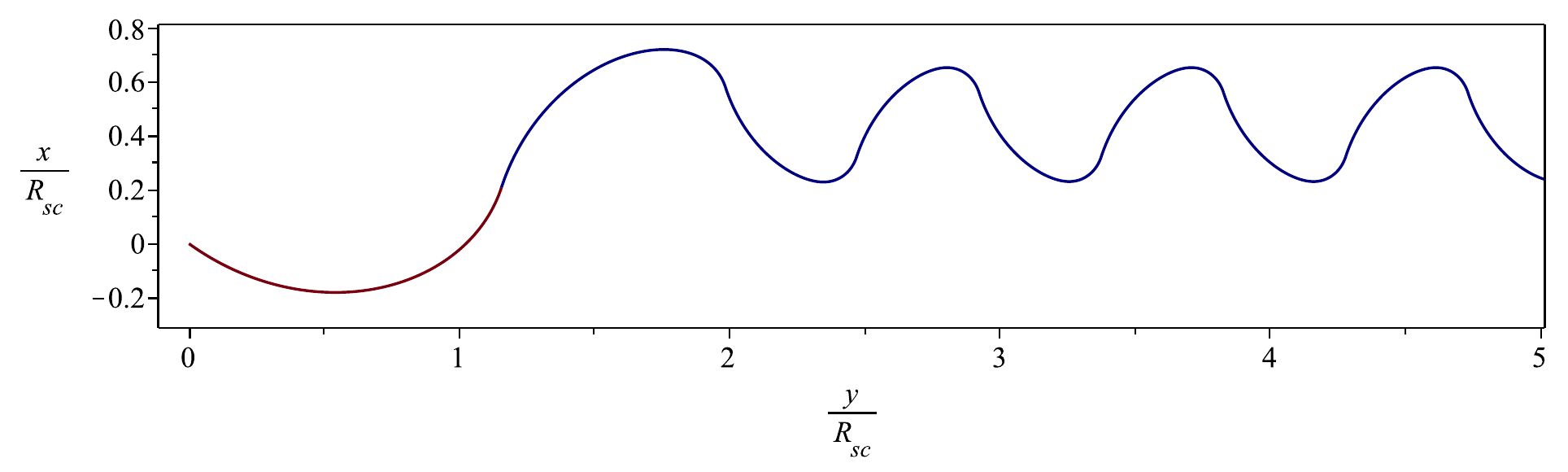}
\caption{ The first few turns for the moderately steep slope ($\alpha=15^\circ$).}
\label{fig-tr-r2}
\end{center}
\end{figure}

\begin{figure}
\begin{center}
\includegraphics[width=0.4\textwidth]{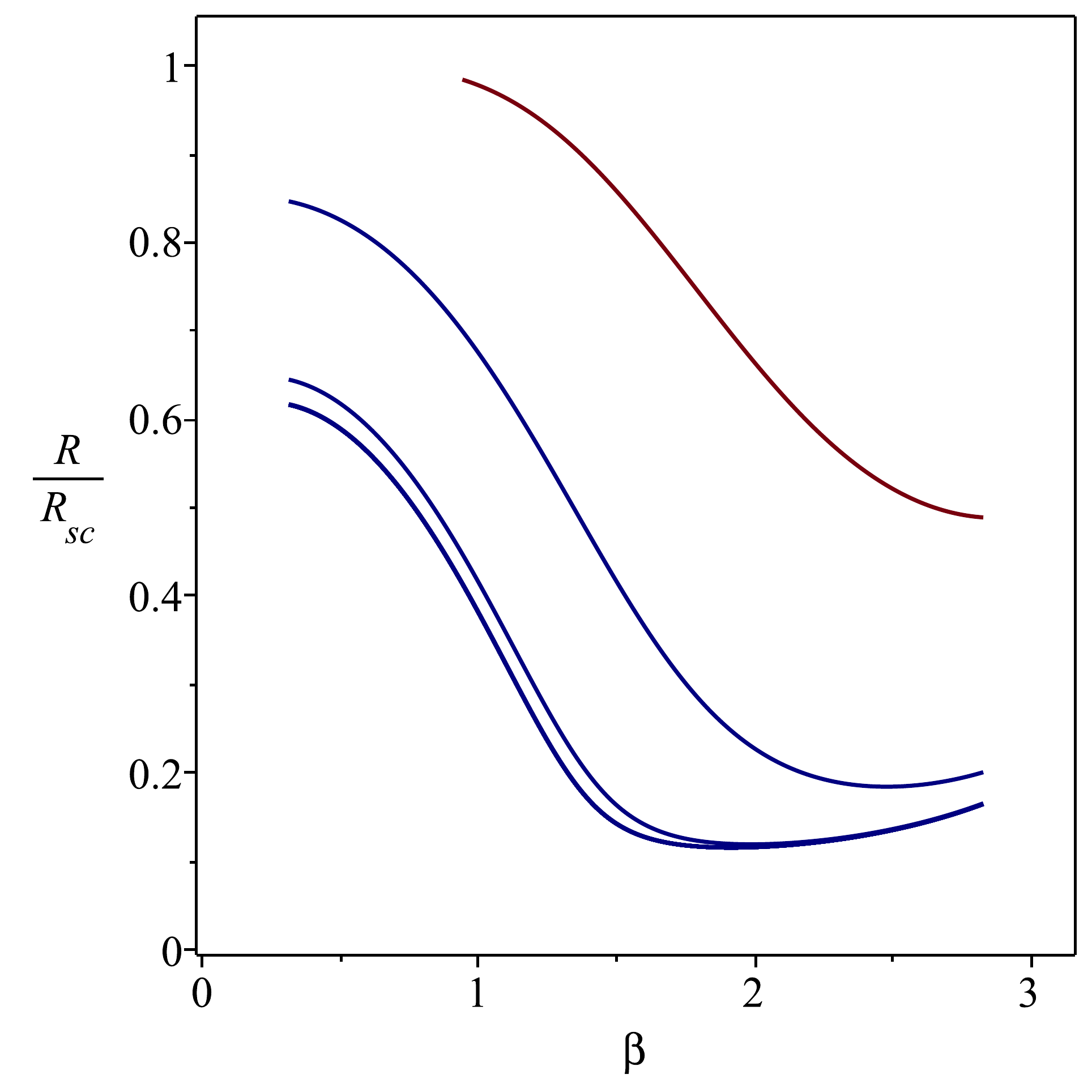}
\includegraphics[width=0.4\textwidth]{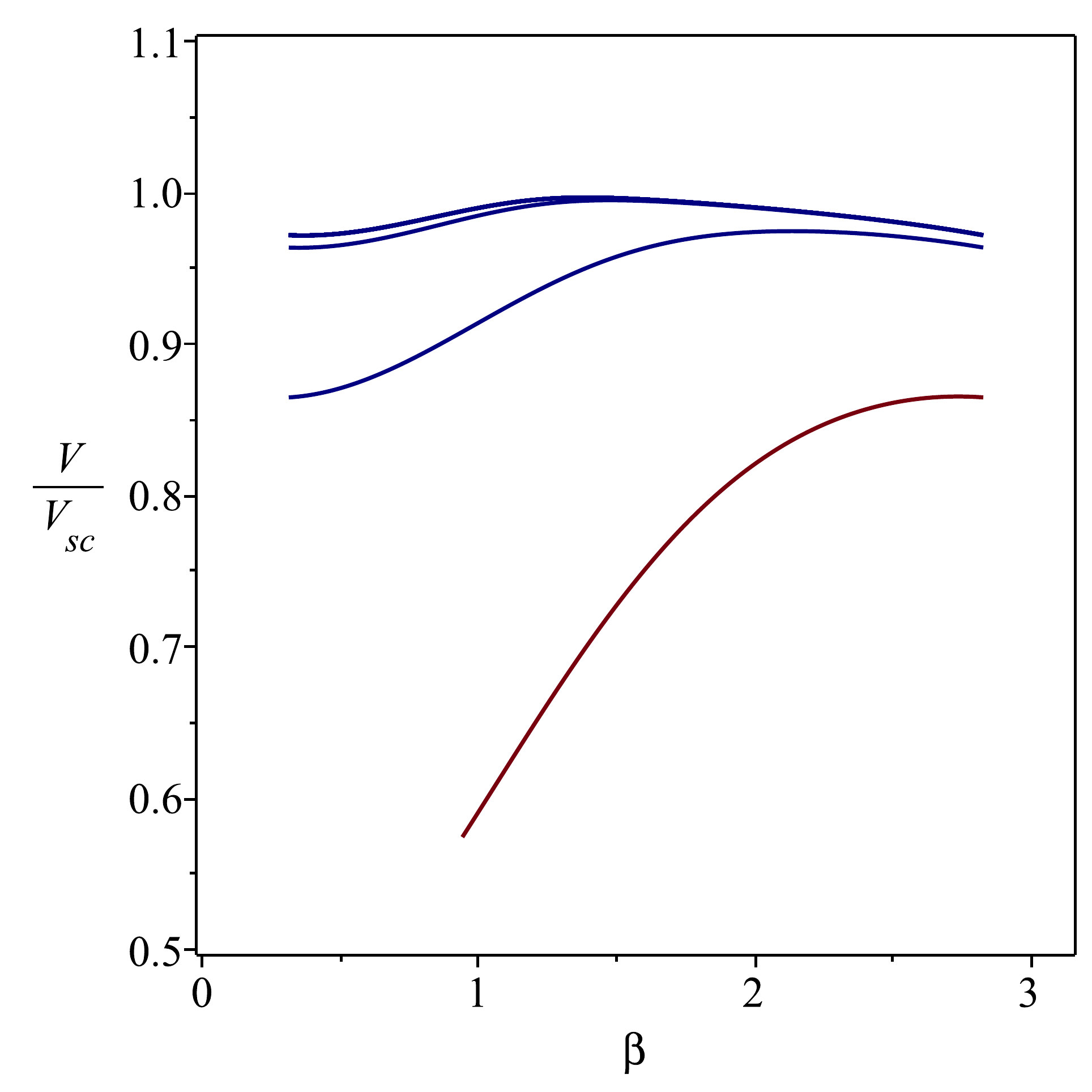}
\includegraphics[width=0.4\textwidth]{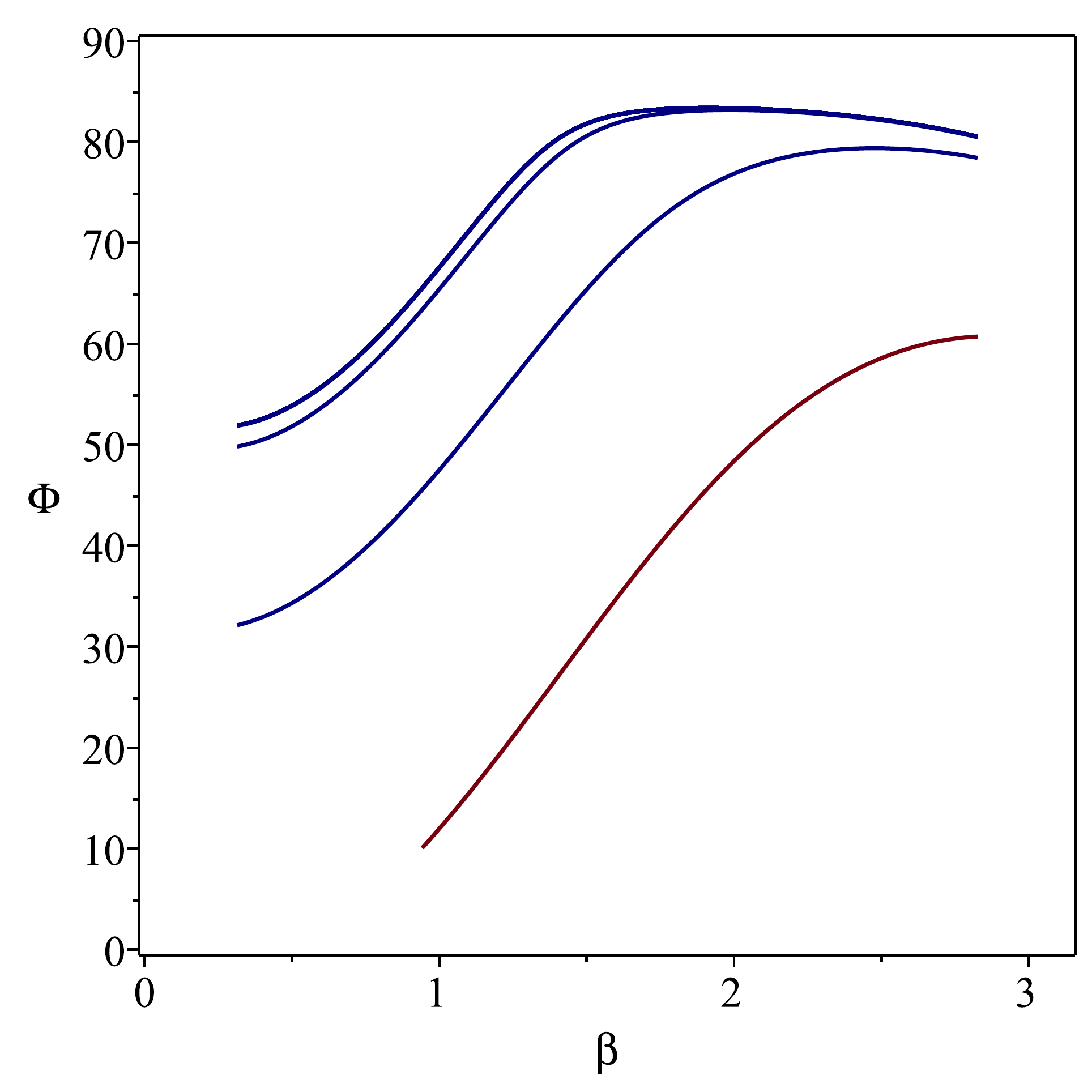}
\includegraphics[width=0.4\textwidth]{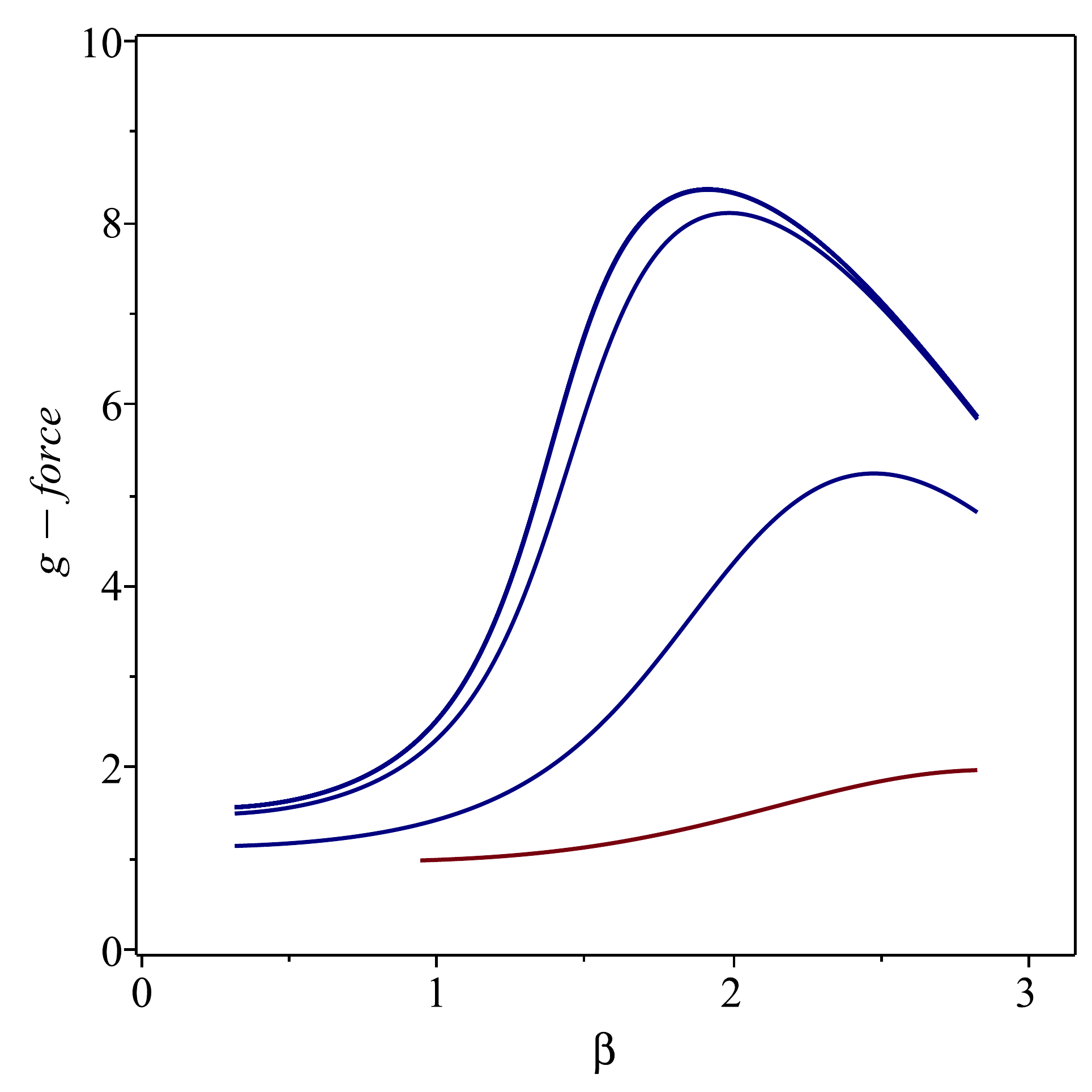}
\caption{The same as in figure \ref{fig-r1} but for the moderately steep slope ($\alpha=15^\circ$). }
\label{fig-r2}
\end{center}
\end{figure}

\begin{figure}
\begin{center}
\includegraphics[width=0.7\textwidth]{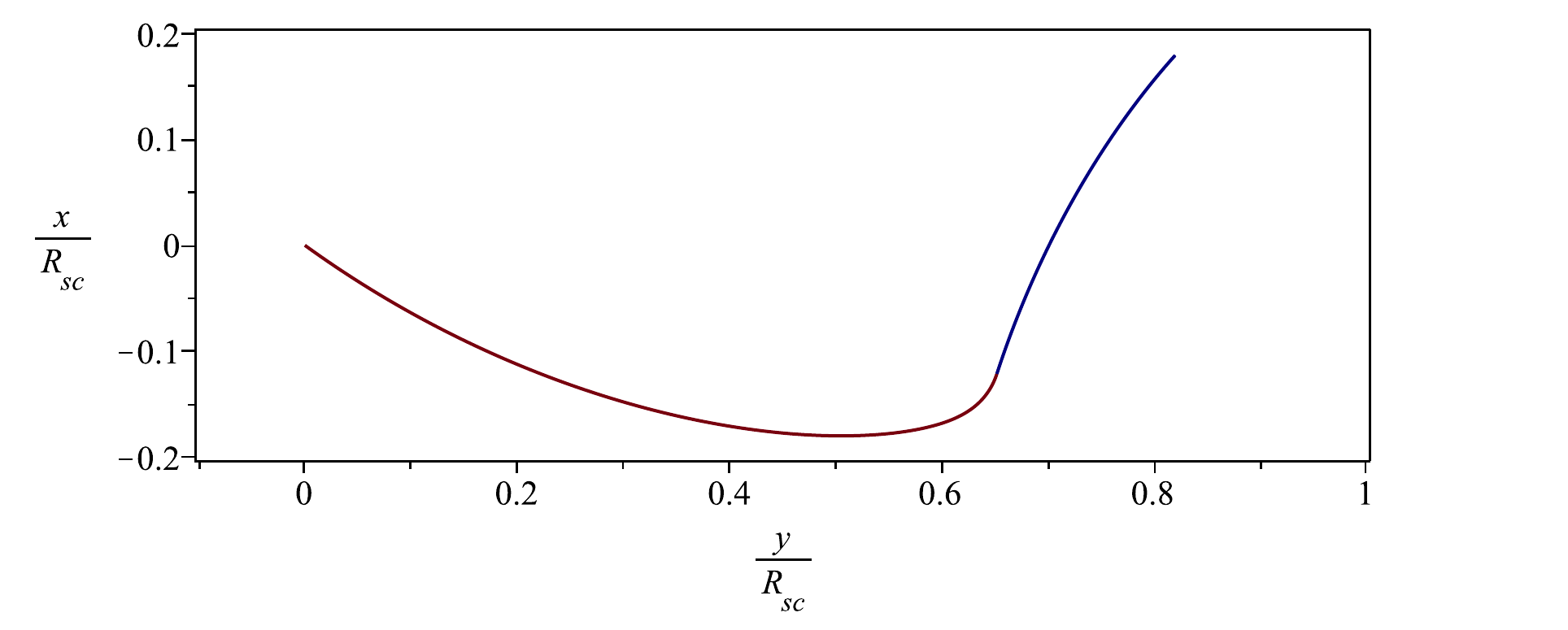}
\caption{ The trajectory of ideal carving run for the steep slope ($\alpha=35^\circ$).}
\label{fig-tr-r3}
\end{center}
\end{figure}
  
\begin{figure}
\begin{center}
\includegraphics[width=0.4\textwidth]{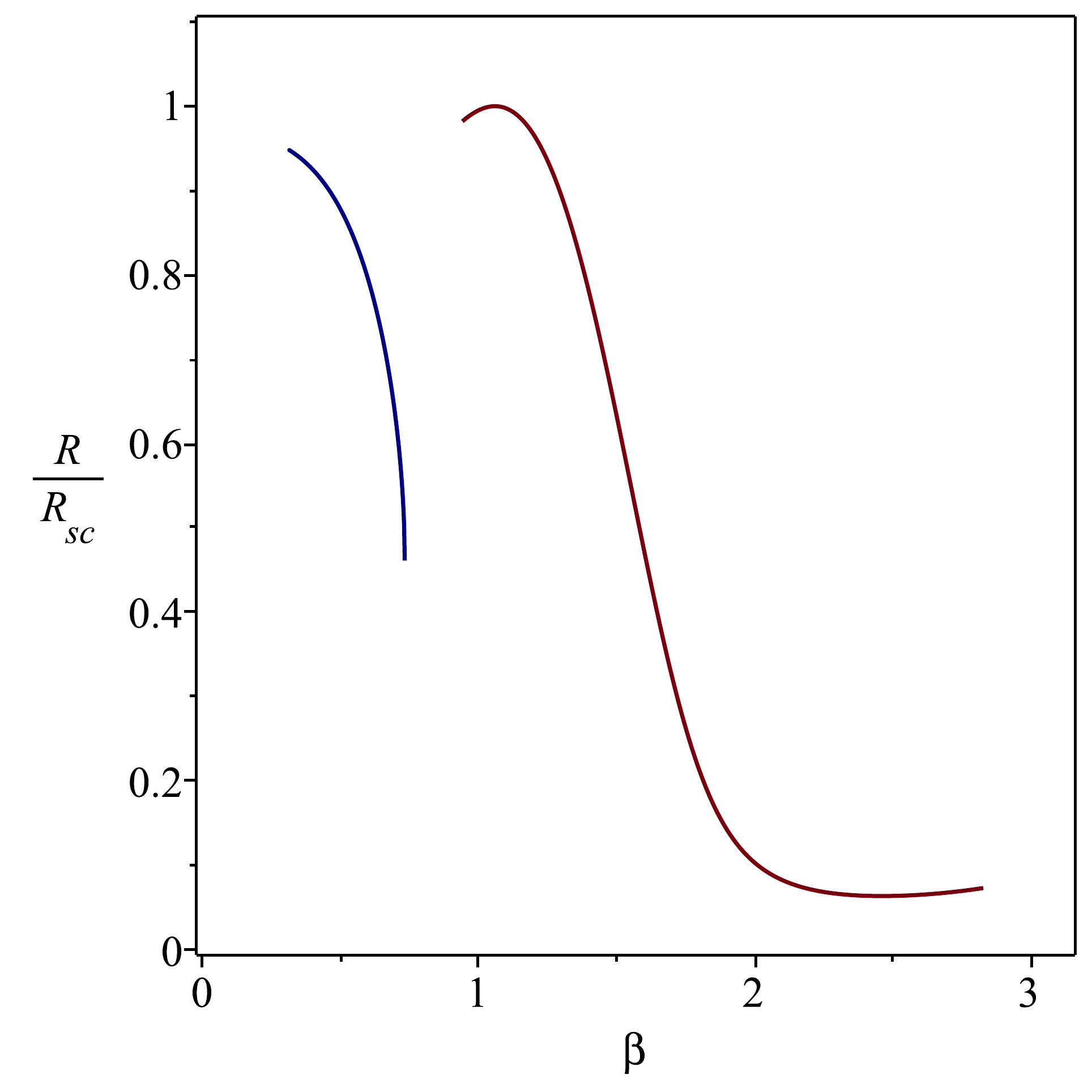}
\includegraphics[width=0.4\textwidth]{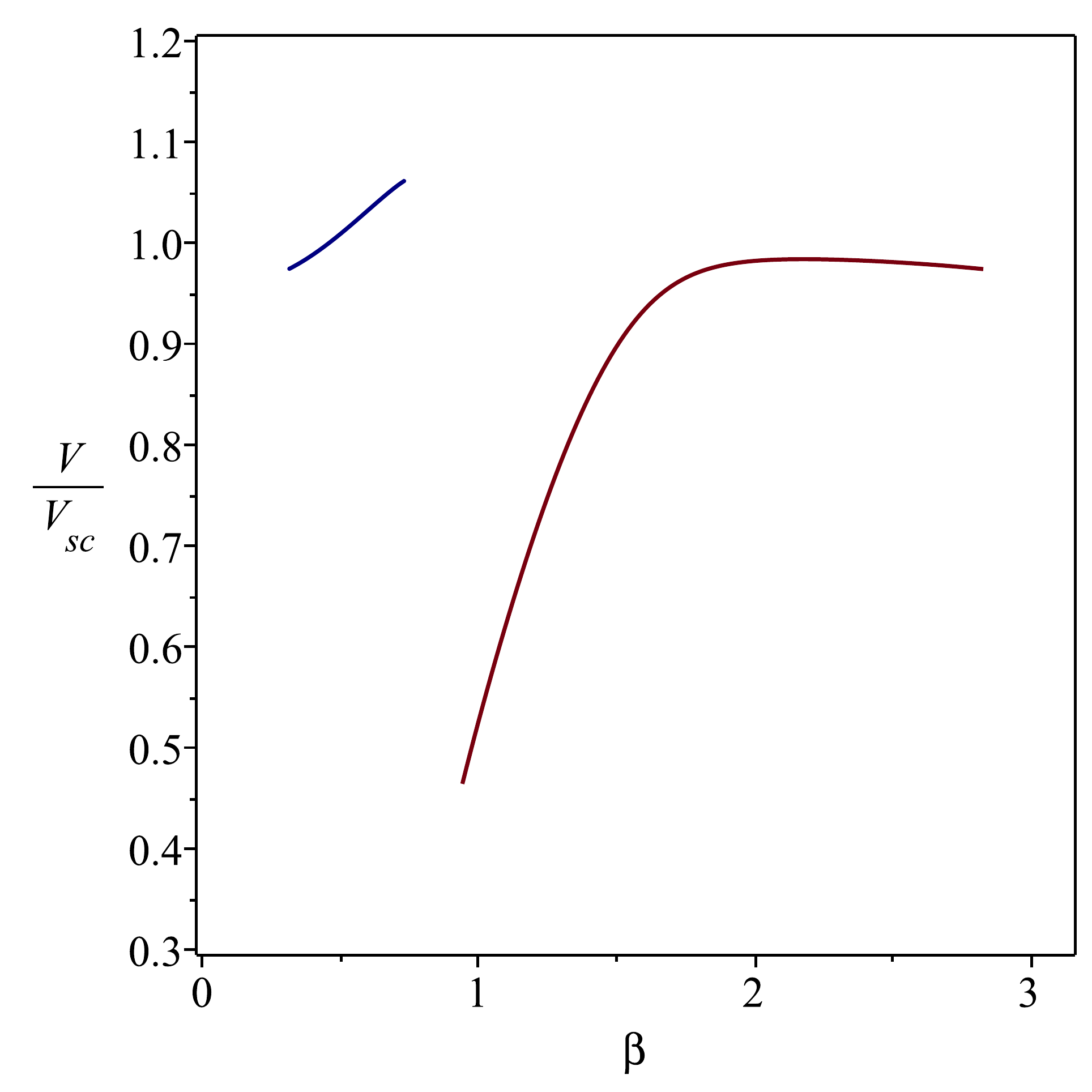}
\includegraphics[width=0.4\textwidth]{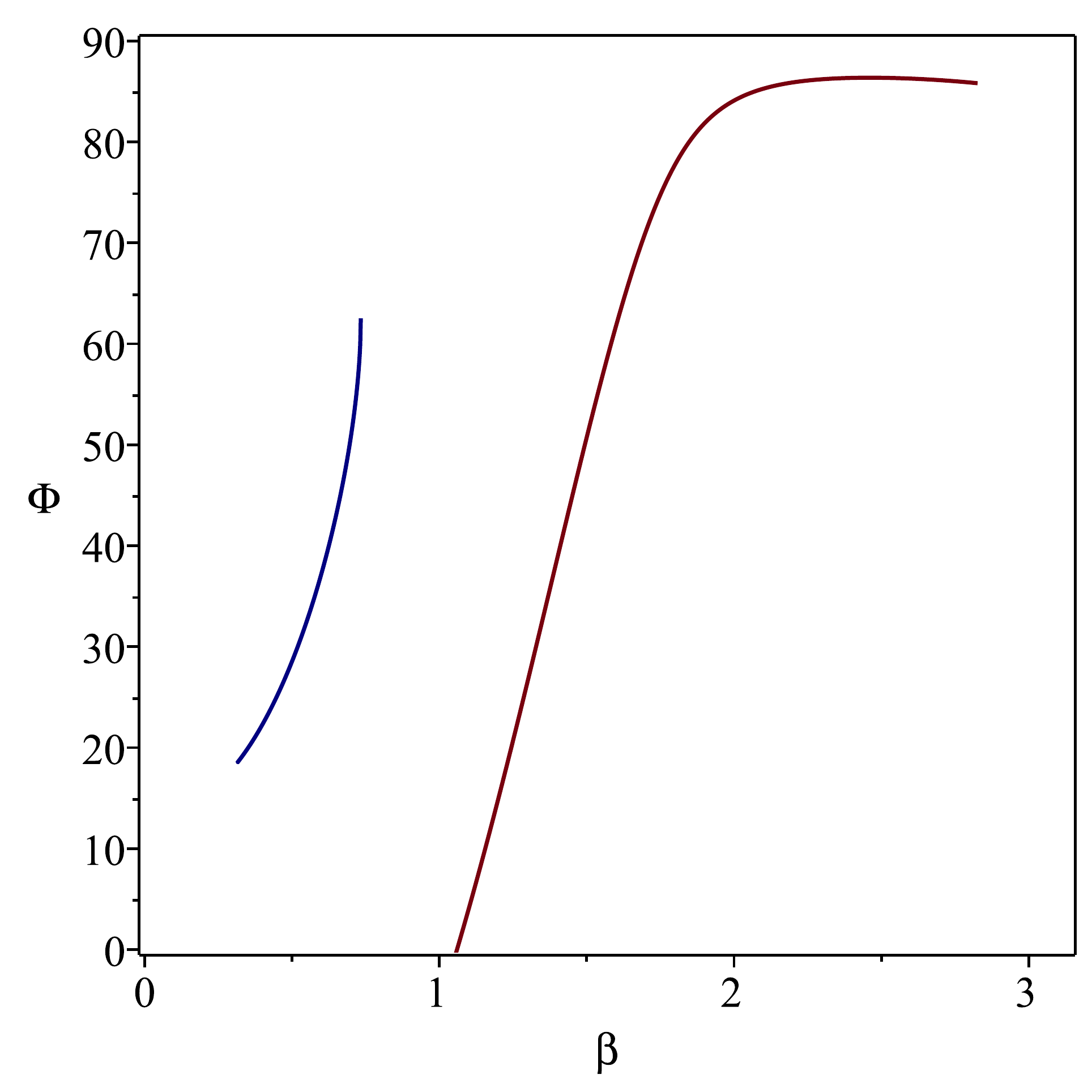}
\includegraphics[width=0.4\textwidth]{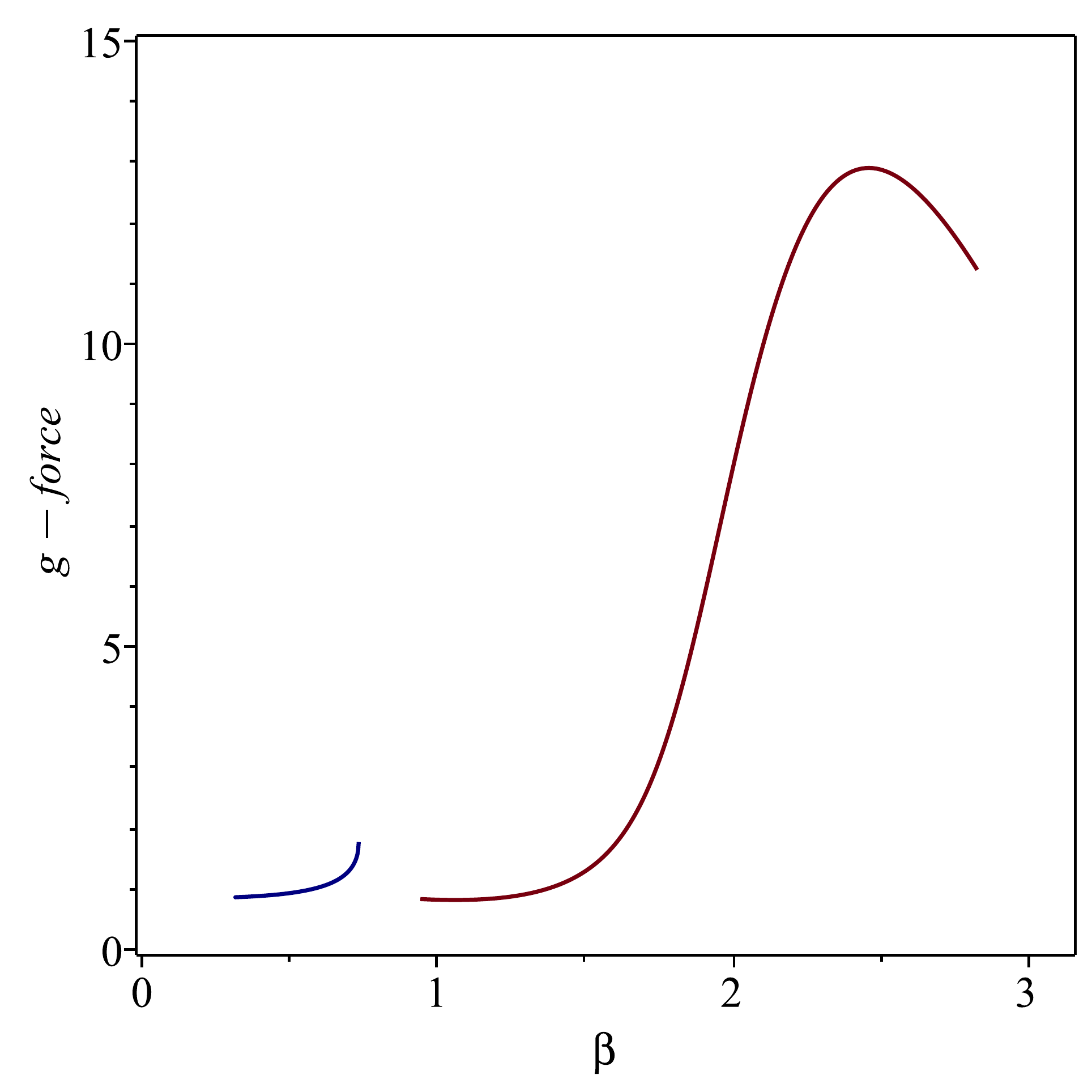}
\caption{The same as in figure \ref{fig-r1} but for the  steep slope ($\alpha=35^\circ$). }
\label{fig-r3}
\end{center}
\end{figure}

Figure \ref{fig-tr-r2} shows the turns of this run on the same section of the slope as in figure \ref{fig-tr-r1} for the flat run ($v\sub{ini}=0.57 V\sub{sc} = 0.2V\sub{g}$). Obviously, now the turns are much shorter and not so rounded, with the individual turn shape reminiscent of the letter ``J'', rather than ``C'' .  
Figure~\ref{fig-r2} shows the evolution of $R$, $v$, $\Phi$ and the g-force for the first 20 turns of this run.  One can see that like in the flat slope case the solution converges to an asymptotic one. Now this occurs very quickly --  in figure~\ref{fig-r2} the curves become indistinguishable beginning from the forth turn. The top-left panel of figure~\ref{fig-r2} confirms that on average the turn radius is significantly lower than for the flat run. Moreover it varies dramatically, from $R\approx 0.6 R\sub{sc}$ at the turn initiation down to $R\approx 0.12R\sub{sc}$ soon after the fall-line. The latter is approximately 1.6m (one SL ski length) when the solution is scaled to $R\sub{sc}=13$m.  

Surprisingly, the speed of the asymptotic solution remains just below $V\sub{sc}$ and well below $V\sub{g}$. The latter indicates that contrary to the expectations based on the analysis of fall-line gliding, in this slalom run the aerodynamic drag is not the dominant factor in determining the saturation speed. The reason is the extremely high effective weight and hence the friction force.  According to the data,  the aerodynamic term $\Kn v^2\approx 0.03$ whereas even for the lowest value of  $1/R \approx 1.6$ found at the start of the turn the friction term  $\mu/R\approx0.06$ is higher than this. Figure \ref{fig-r2} suggests $\langle 1/R\rangle \approx 5$  and  $\langle v \rangle \approx 1$. Hence in equation (\ref{s5})  the geometric term $\langle \sin\beta\rangle \tan\alpha=0.213$, the friction term $\langle \mu/R \rangle \approx 0.2$ and the aerodynamic drag term $\Kn \langle v^2 \rangle \approx 0.03$. Thus, the geometric and the friction terms almost cancel each other out and hence the solution becomes sensitive to the value of the friction coefficient. 
       
The corresponding dimensional value of the mean skier speed in the asymptotic solution is $v\approx 35$ km/h. This is indeed very close to the typical speed of slalom runs, indicating that racers ski very close to the carving limit.  However, the inclination angle of this run reaches very 
high values, $\Phi\approx 80^\circ$ in the lower-C. This and the extreme values of the g-force show that in this run we are likely to be already beyond what is achievable in practice.  In the sledge simulations for the slope of this gradient \cite{M08,M09},  carving in the upper-C of the turn followed by  skidding  in the Lower-C, with the mean turn radius exceeding the sidecut radius of the skis. The skidding was caused by the show fracturing under the increasing shear stress applied to the snow by the skis. However according to equation (\ref{Phi-cond}),  the snow hardness and critical shear stress used in the simulations  ($H=0.01\,$N$\,$mm$^{-3}$, $S\sub{c}=0.036\,$N$\,$mm$^{-2}$) implied the rather low critical inclination angle,  $\Phi\sub{c}\approx30^\circ$.        
  
\subsection{Steep slope}
\label{sslope}

The setup of this run differs from the previous ones by the higher slope gradient,  $\alpha=35^\circ$. For this gradient the difference between the aerodynamic and carving speed limits is even more dramatic than for the moderately steep slope, with
$V\sub{sc}=0.2V\sub{g}$.  In this run the initial speed is set to   $v\sub{ini}=0.46 V\sub{sc}$ ($0.1 V\sub{g})$.  
For these parameters, the solution becomes singular even before reaching the fall-line of the second turn and cannot be continued thereafter 
(see figure \ref{fig-tr-r3}).  Figure (\ref{fig-r3}) reveals the nature of the problem.   Already during the first turn the skier is forced into and extreme inclination and experiences overwhelming g-force. In real life, skier's legs would collapse at this point but in the model the run continues. Very soon after the transition into the second turn the speed shoots above the carving limit and the solution approaches the point where no equilibrium position for the skier can be found. Hence the run cannot be continued even theoretically.  As one can see, this occurs still in the Upper-C at the point where the turn radius remains relatively large and the skier inclination relatively small. We have discussed this possibility in Section \ref{interpretation}. 
 
\begin{figure}
\begin{center}
\includegraphics[width=0.9\textwidth]{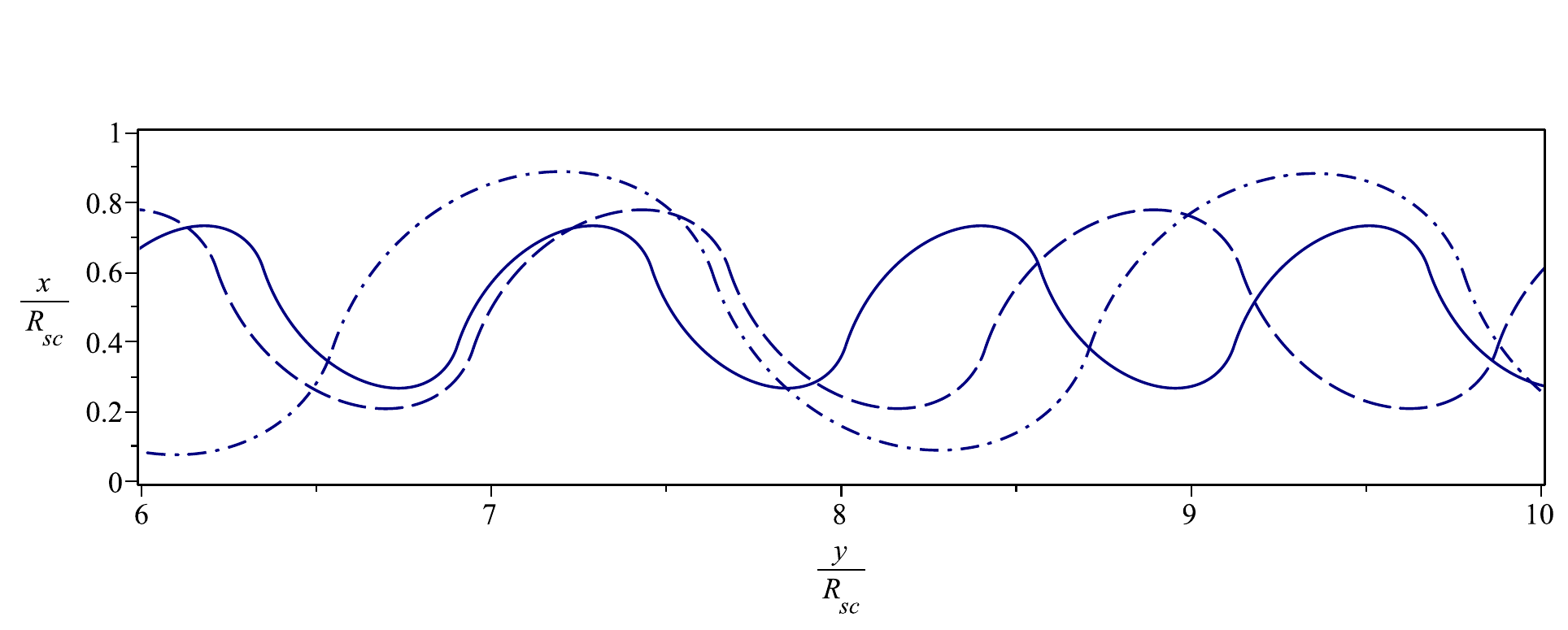}
\caption{ The trajectories of ideal carving runs on the $\alpha=13^\circ$ slope with SL (solid), GS (dashed) and DH (dash-dotted) skis.}
\label{fig-tr-comb}
\end{center}
\end{figure}
  
\begin{figure}
\begin{center}
\includegraphics[width=0.4\textwidth]{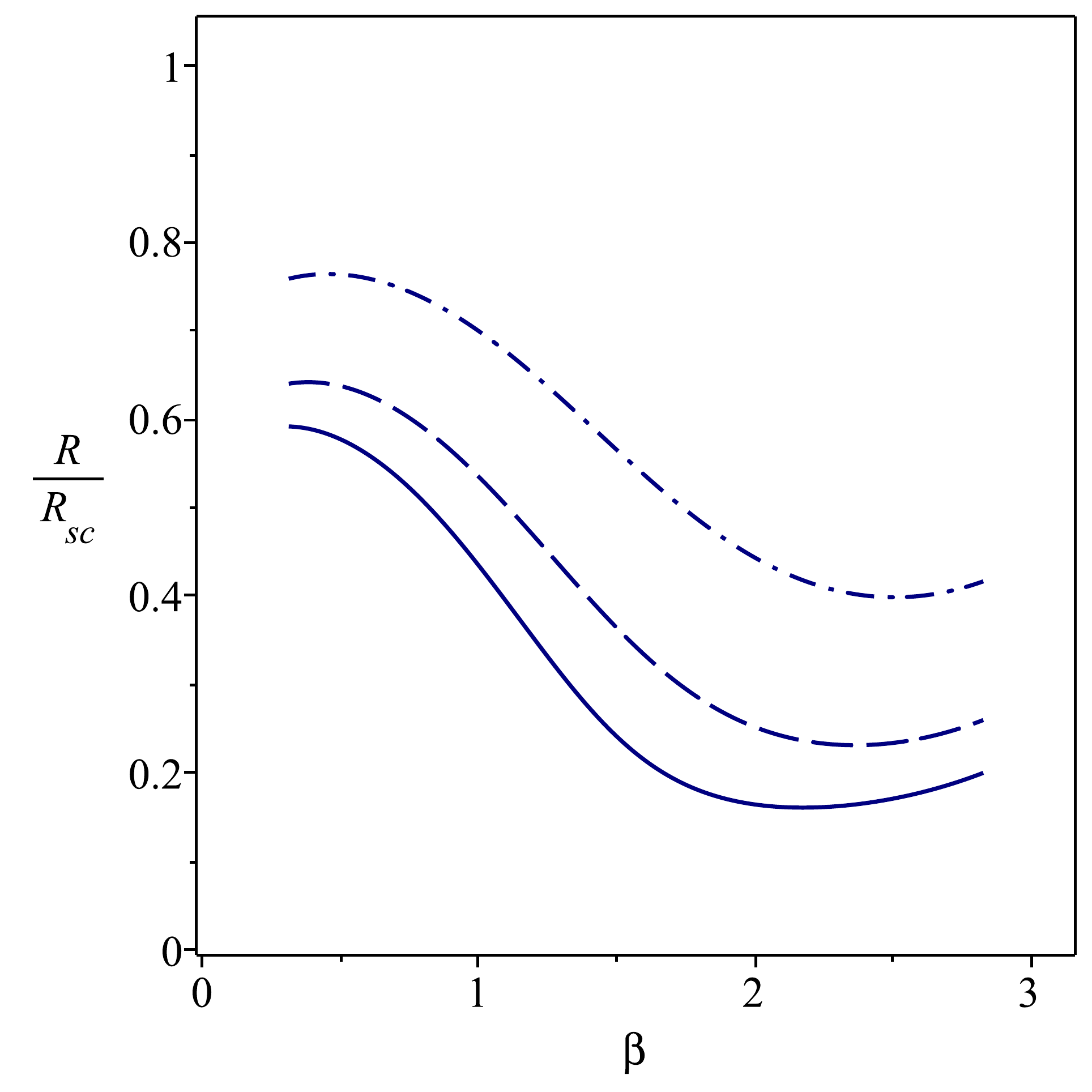}
\includegraphics[width=0.4\textwidth]{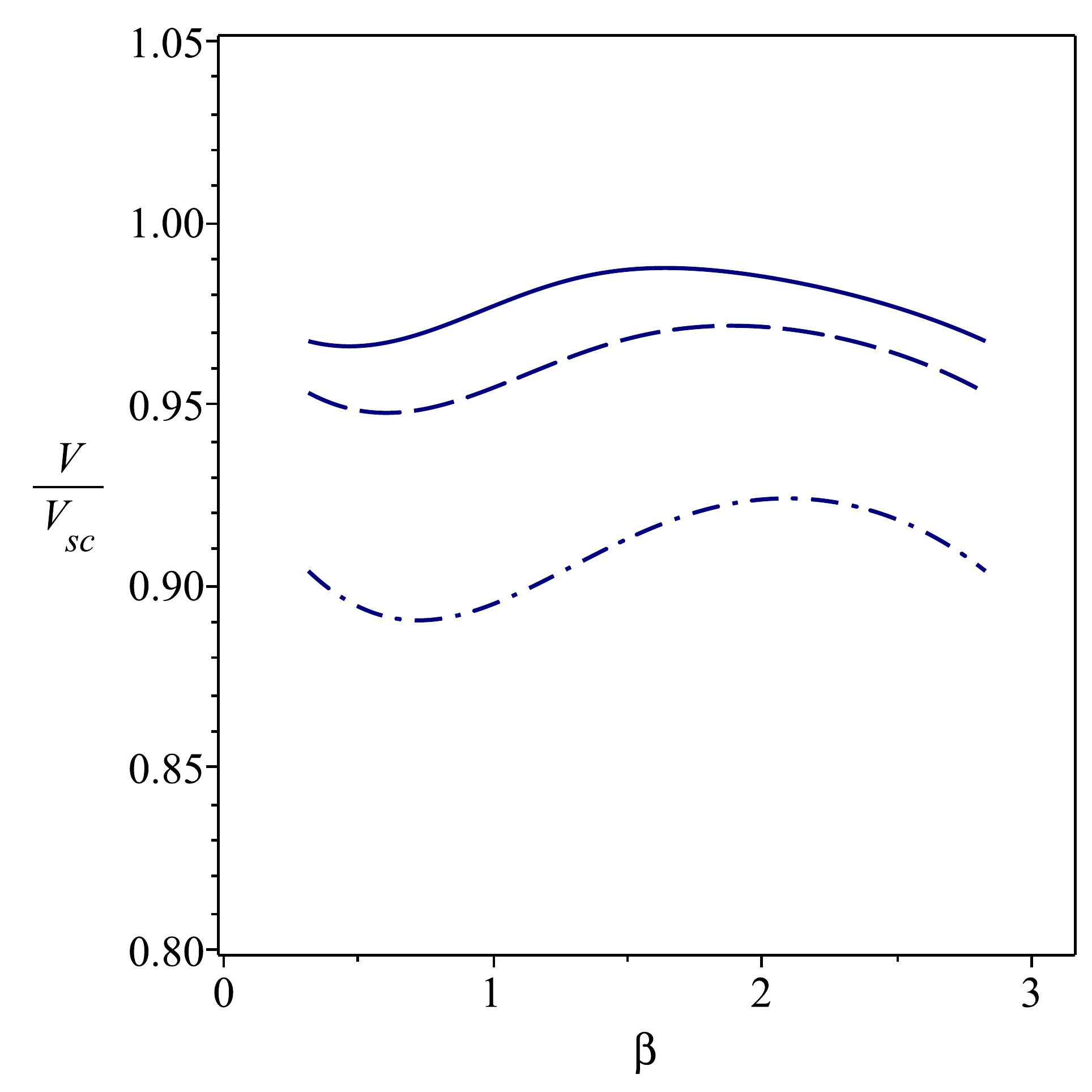}
\includegraphics[width=0.4\textwidth]{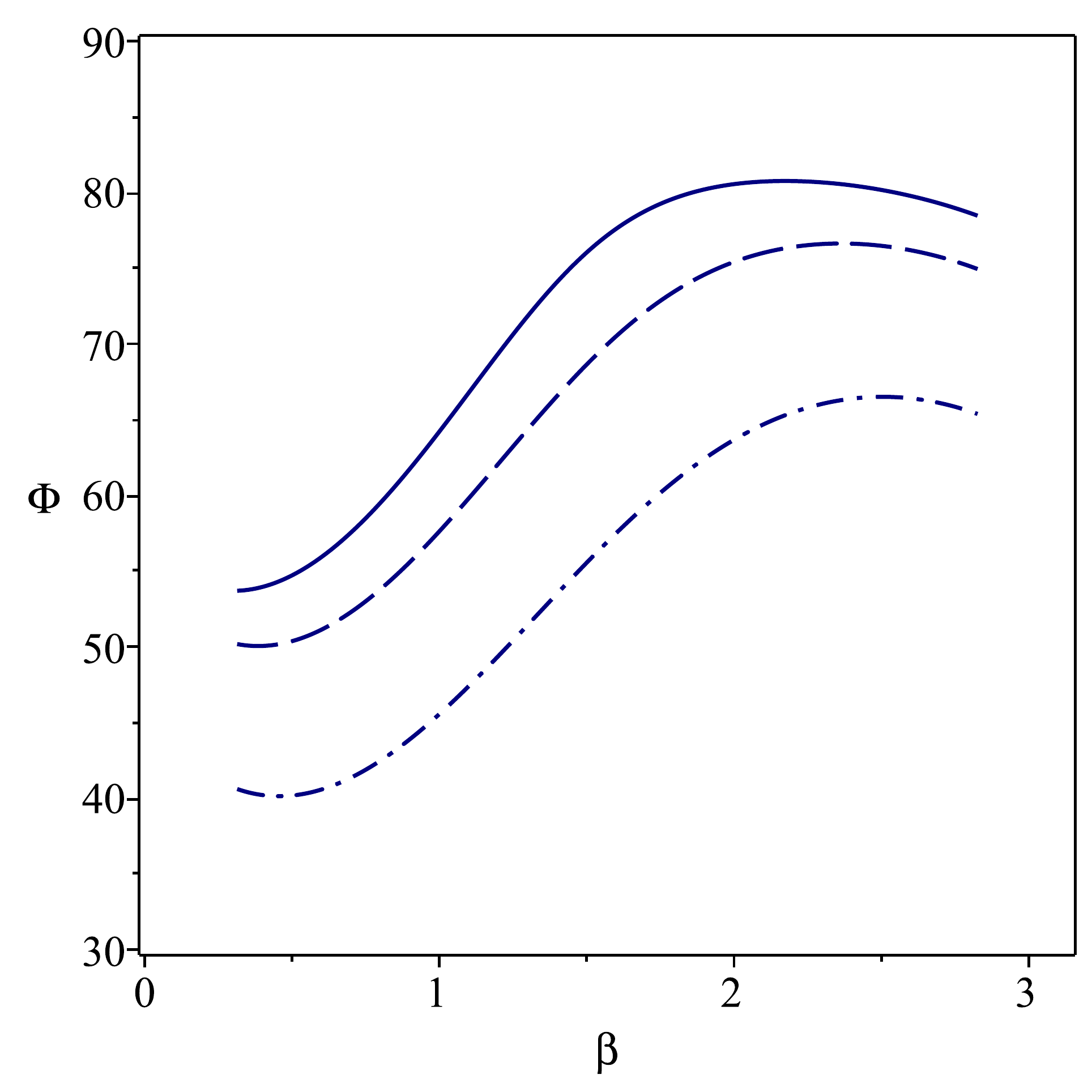}
\includegraphics[width=0.4\textwidth]{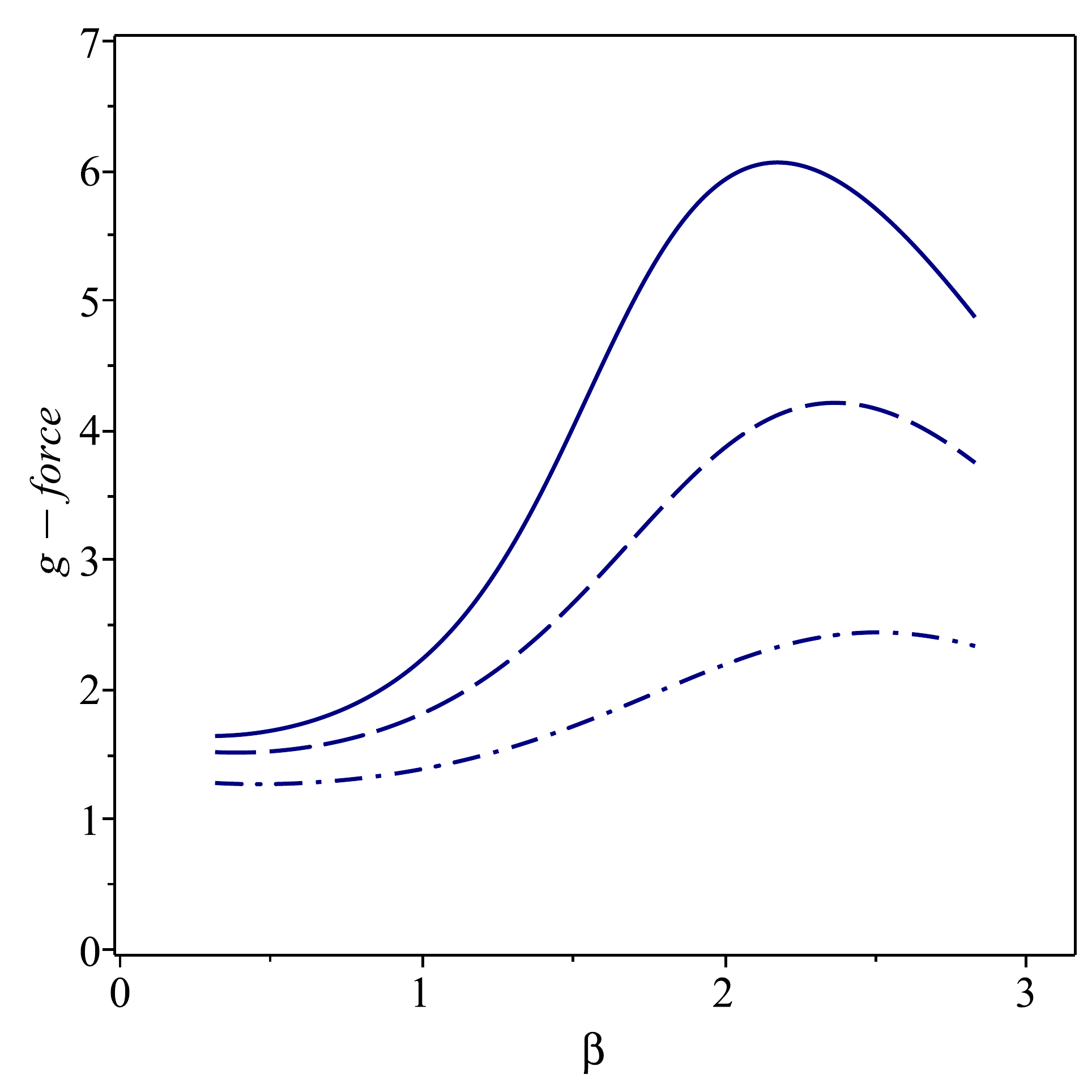}
\caption{The asymptotic turn parameters for ideal carving runs on the $\alpha=13^\circ$ slope with SL (solid), GS (dashed) and DH (dash-dotted) skis. }
\label{fig-comb}
\end{center}
\end{figure}

\section{The approximate similarity of carving runs}
\label{sidecut-effect}
 
How different are the carving turns in different alpine disciplines?   
The equations of carving turn (\ref{DVDB}--\ref{DICE}) have only three dimensionless parameters which modify these equations and hence their solutions -- the coefficient of dynamic friction $\mu$, the slope angle $\alpha$ and the dynamic sidecut parameter $\Kn={R\sub{sc}}/{L\sub{g}}$.  Out of the three only $\Kn$ depends on the ski sidecut radius. If we put $\Kn=0$ then the 
dependence of dimensionless solutions of the sidecut radius would vanish completely.  This means that for the same slope angle and coefficient of friction (the same snow conditions and wax), the slalom and downhill carving runs would be just scale versions of each other,  with the trajectory size scaling as $\propto R\sub{sc}$, skier's speed as $\propto \sqrt{R\sub{sc}}$ and both the g-force and the inclination angle being exactly the same.     

In reality, $\Kn\not=0$ and it grows with the sidecut radius. However, it remains small ($\,\Kn \ll 1\,$) even for DH skis. 
In the speed equation (\ref{DVDB}), $\Kn$ is a coefficient of the aerodynamic drag term $\Kn v^2$. Hence, what differentiates carving runs of different alpine disciplines is the relative importance of the aerodynamic drag.  Because the drag increases  with $\Kn$  we expect the speed of the downhill run to be slower than that of the slalom run, when both are measured in the units of their own $V\sub{sc}$. The smaller speed implies the larger turn radius (in the units of $R\sub{sc}$, see equation \ref{DICE}) and hence the lower g-force (equation \ref{Gforce-Fn}) and inclination angle (equation \ref{turn-r}).  To check this, we compared the runs on the same slope  ( $\alpha=13^\circ$ and 
$\mu=0.04$) with SL,  GS (giant slalom) and DH skis. Figure~\ref{fig-tr-comb}  shows the dimensionless trajectories of these runs in the asymptotic regime. Although the trajectories are relatively similar they are not exact copies of one another, with larger $\Kn$ yielding longer and smoother turns. Figure~\ref{fig-comb} shows the evolution of 
the main run variables during the turns and confirms that the larger sidecut radius yields turns with the smaller inclination angles and lower g-force.

We extended the study described in the previous section in order to determine the critical slope steepness $\alpha\sub{c,t}$ beyond which pure carving is impossible for SL, GS and DH skis even theoretically.  The result is  
\begin{equation}
   \alpha\sub{c,t}= \begin{cases}
     17^{\circ} &  \etext{for}  \Kn= 0.0325\,  (13\mbox{m})\,, \\ 
     19^{\circ} &  \etext{for}  \Kn=0.0875\, ( 30\mbox{m})\,, \\ 
     21^{\circ} &  \etext{for}  \Kn=0.1250\, ( 50\mbox{m})\,,  
  \end{cases} \label{alpha-c}
\end{equation}
where in the brackets we show the dimensional sidecut radius corresponding to $L\sub{g}=400$m. Taking into account the additional limitations based on the shear resistance of the snow, physical strength of the skiers and skis (see Sec.\ref{top-angle}), the practical critical gradients will be even smaller.  

\section{The snow friction and the critical gradient}
\label{sec:gradient}

The data presented in (\ref{alpha-c}) were obtained for the friction coefficient $\mu=0.04$. However, the coefficient depends on many factors and may vary significantly.  Since a higher friction coefficient $\mu$ implies a lower saturation speed of the run and hence a weaker centrifugal force, this may also allow carving on steeper slopes. Here we explore this avenue taking a little more pragmatic definition of executable carving turns. Following the discussion in Section~\ref{top-angle}, we will demand the inclination angle not to exceed the critical value $\Phi\sub{c}$. Realistically, it should be around $70^\circ$ as the higher values would lead to the g-force $> 3$.   

The Ideal Carving Equation (\ref{DICE}) can be written as  
$$
     v^2 = \sin\Phi + \cos\Phi \cos\beta\tan\alpha \,.  
$$   
Provided $\alpha$ is sufficiently small, the second term on the right side is small near $\Phi\sub{c}$ and hence  
$v^2 = \sin\Phi\sub{c}$.  
    
\begin{figure}
\begin{center}
\includegraphics[width=0.6\textwidth]{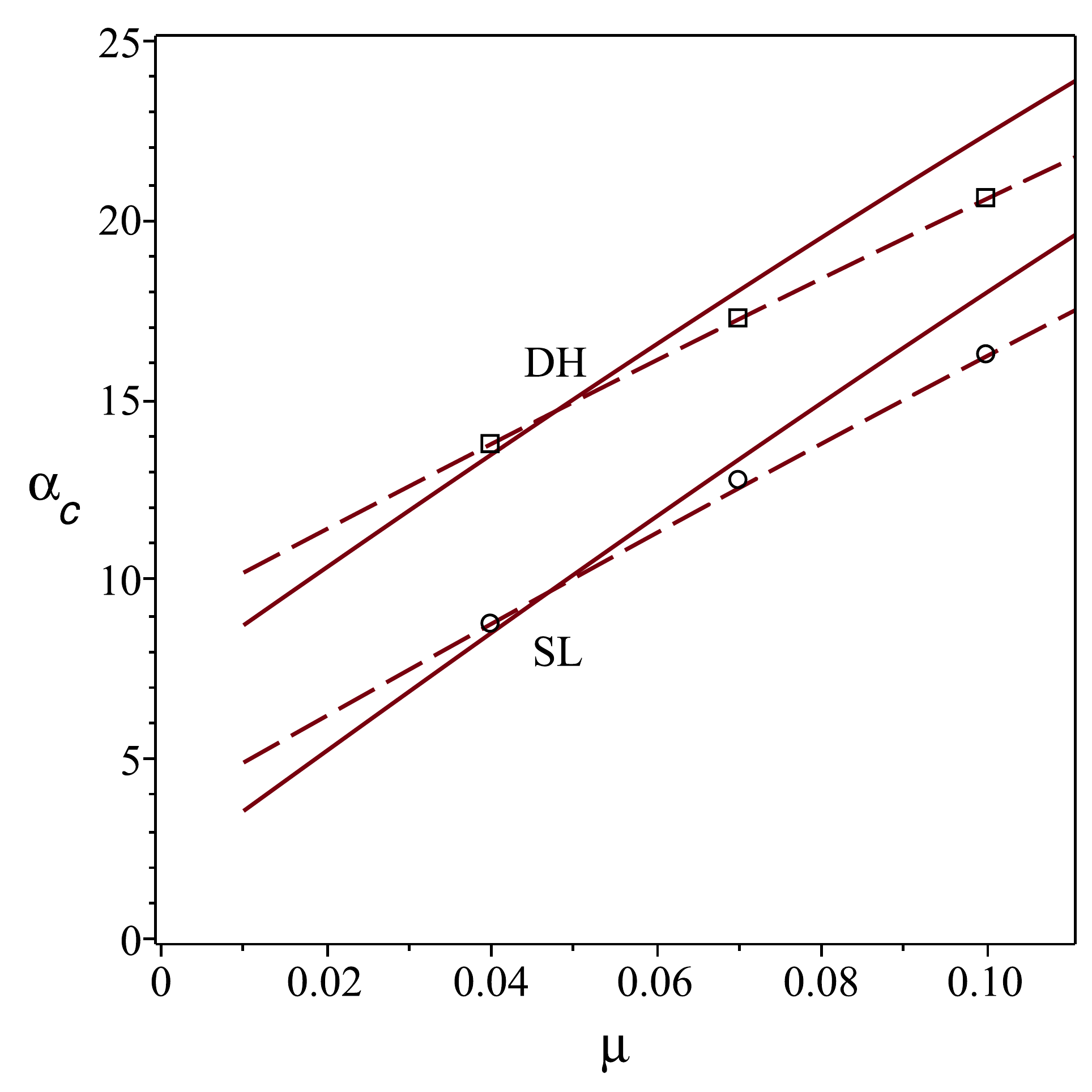}
\caption{The critical slope gradient as a function of the friction coefficient for $\Phi\sub{c}=70^\circ$. The solid lines show the analytic predictions based on equation (\ref{cr-angle}). The markers show the numerical data and the dashed lines their fit by the approximations (\ref{cr-angle-SL},\ref{cr-angle-DH}). } 
\label{fig-cgrad}
\end{center}
\end{figure}

In the asymptotic solution, the speed varies rather weakly and hence its derivative is closed to zero. According to equation (\ref{DVDB}), it vanishes exactly when    
$$
 \sin\beta\tan\alpha -\frac{\mu}{\cos\Phi} -\Kn v^2 =0 \,. 
$$
 Combining the last two equation we find the critical slope angle as a function of the friction coefficient 
 \beq
   \tan\alpha\sub{c} =  (\frac{\mu}{\cos\Phi} +\Kn)(\sin\beta-\Kn \cos\Phi \cos\beta)^{-1} \,.
 \label{cr-angle}
 \eeq  
 In the run where $\Phi$ peaks at $\Phi\sub{c}=70^\circ$ ($\mu=0.04$ and $K=0.0325$), we see that $v$ peaks somewhat before 
 $\Phi$, at the point with $\beta\approx 2$ and $\Psi\approx 66^\circ$. Substituting these values into equation (\ref{cr-angle}) we obtain the    
 equation of lines shown in Figure \ref{fig-cgrad}. To check this prediction we also run models with $\mu$=0.04, 0.07 and 0.10 where we varied the slope angle until $\Phi$ peaks within one percent of $\Phi\sub{c}$.  The results are not far away from the analytic prediction (see Figure \ref{fig-cgrad}).   They are well fitted by the linear equations 
 
 \beq
   \tan\alpha\sub{c} =  2.3 {\mu} + 0.06 \,.
 \label{cr-angle-SL}
 \eeq  
 for the SL skis ($\Kn=0.0325$) and 
 \beq
   \tan\alpha\sub{c} =  2.2 {\mu} + 0.16 \,.
 \label{cr-angle-DH}
 \eeq  
 for the DH skis ($\Kn=0.125$).

\section{The effect of angulation}
\label{sec:angulation}

\begin{figure}
\begin{center}
\includegraphics[width=0.58\textwidth]{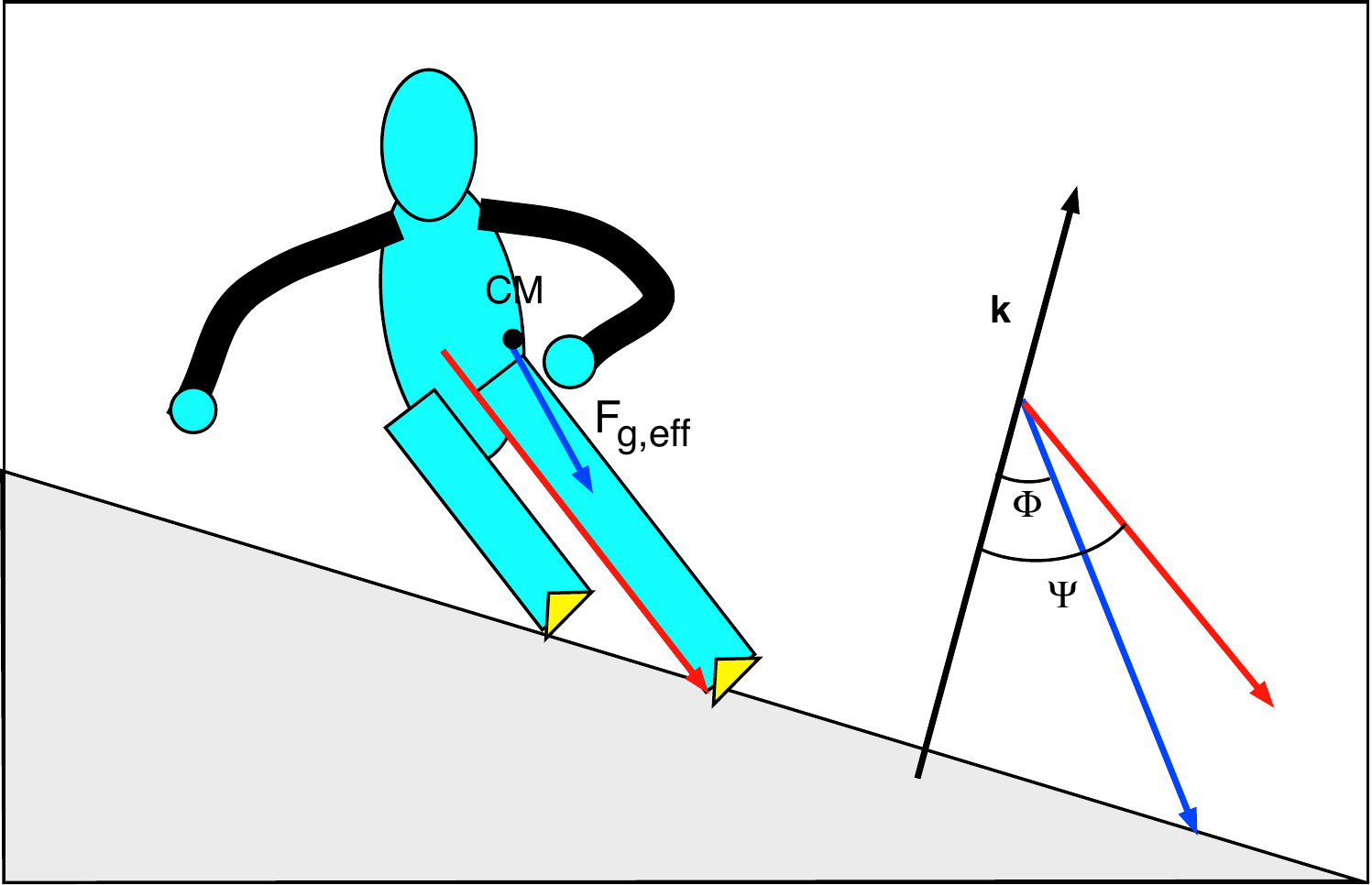}
\includegraphics[width=0.405\textwidth]{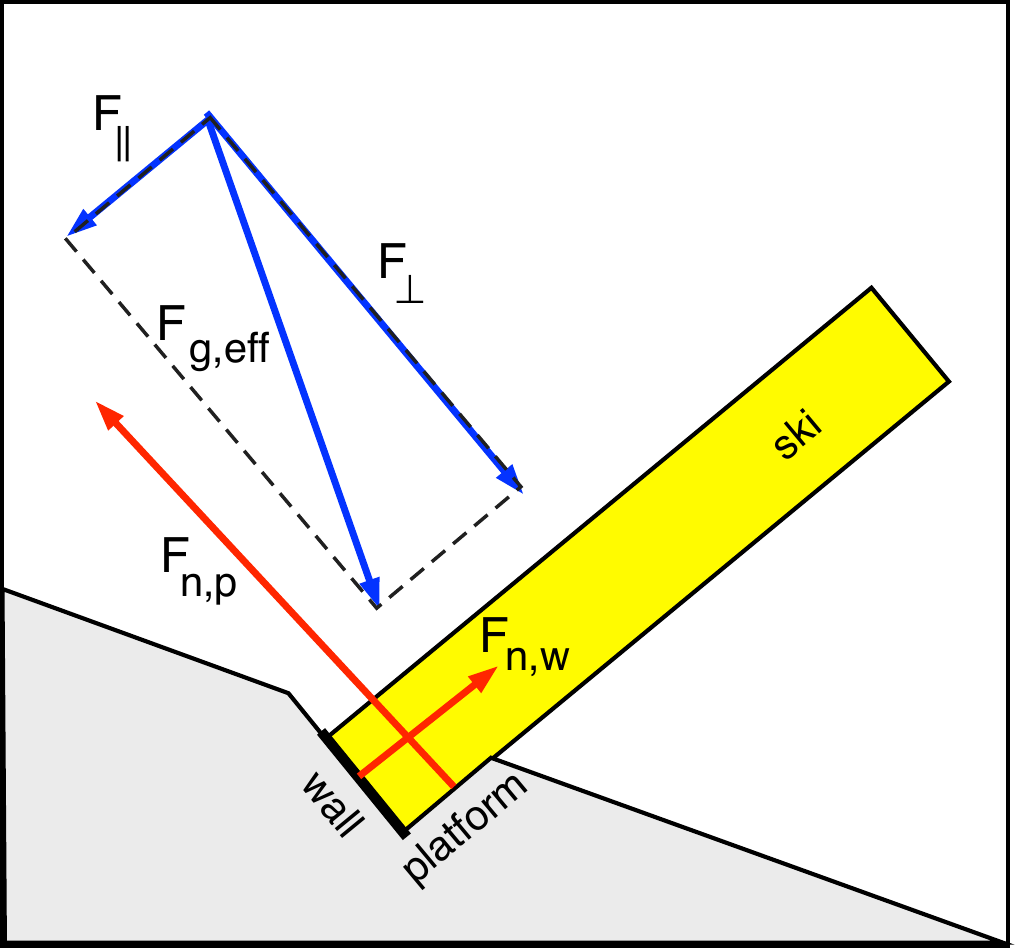}
\caption{ Inclination and angulation. Left panel: Because of the angulation at the hip, the inclination angle of the skis $\Psi$ is higher than the inclination angle of the skier $\Phi$. Right panel: Because $\Psi>\Phi$ the effective gravity has not only the component normal to the platform cut in the snow by the ski  ($\bF_\perp$) but also the component parallel to the platform and pushing the ski into the platform wall ($\bF_\parallel$). The wall reacts with the force $\bF\sub{n,w}$, balancing $\bF_\parallel$. }
\label{angulation}
\end{center}
\end{figure}

In the derivation of the ideal carving equation (Eq.\ref{ICE}) we assumed that the whole skier's body was aligned with the direction of the effective gravity and hence the inclination angle $\Psi$ of the skis was the same as the inclination angle $\Phi$ of the effective gravity force (and hence of the skier). However, in general $\Theta=\Psi-\Phi \not=0$ as skiers tend to bend (angulate) their body in the frontal plane ( see figure \ref{angulation}) ).  This is normally done by moving hip and to some degree knees to the inside of the turn so that $\Theta>0$ \cite{LM10}. There are at least two benefits of such angulation.  

Firstly,  it introduces a good safety margin against accidental side-slipping of the skis. Indeed, the case  of $\Theta=0$, which we have been focusing on so far, is only marginally stable against such a slipping because in this case the effective gravity has no component aligned with the platform that skis make in the snow. A small deviation from this equality towards the $\Theta<0$ state, e.g. due to the skier's hip unintentionally moving to the outside of the turn, will create such a component in the direction away from the slope. The snow reaction force will remain normal to the platform and hence unable to prevent ski from the side-slipping \cite{LM10}. 

If $\Theta>0$ then the platform-aligned component of effective gravity exists as well but it points towards the slope and presses the skis against the inside wall of the platform, making the side-slipping impossible. The wall reacts with its own reaction force (force $F\sub{n,w}$ in figure \ref{angulation}) and as the result, the total $\bF\sub{n}$ is no longer normal to the platform and hence to the ski base. This allows the total normal reaction force to balance the effective gravity force even if the latter is not normal to the platform.  Provided  $\Psi$ significantly exceeds $\Phi$, small variations of these angles due to various imperfections (perturbations) cannot affect the outcome.    

Secondly, it allows to vary the turn radius, thus introducing some control over the carving turn. 
Using the notation of Section~\ref{SL} 
$$
    \tan\Psi = (\xi^2-1)^{1/2} \,,
$$
whereas   
$$
    \tan\Phi = a\xi+b \,.
$$
For $0<\Phi<\Psi<90^\circ$, we have 
$$
\tan\Psi =\eta\tan\Phi\,,
$$ 
where $\eta >1$.  This yields the modified ideal carving equation
\beq
   (\xi^2-1)^{1/2} =  \eta(a\xi +b) \,.
  \label{ang-1}
\eeq   
A careful (and a bit tedious) analysis  of this equation shows that $d\xi/d\eta >0$ provided $\eta a<1$ and hence higher $\eta$ means lower turn radius $R$ (see Appendix \ref{angulation-turn-r}). 
In other words, by increasing the angulation the skier can tighten the arc and the other way around, which is well know to skiing experts 
(e.g. \cite{HH06}).

Replacing $a$ with $a'=\eta a >a$ and $b$ with $b'=\eta b >b$, we can write equation (\ref{ang-1}) in exactly the same form as the  original ICE (Eq.\ref{ICE}).  This immediately allow us to make a number of useful conclusions on the carving speed limits in the case of positive angulation.     Clearly,  $a+b>0$ if and only if $a'+b'>0$ and  hence the lower speed limit (\ref{cc2d},\ref{cc3d}) remains unchanged. However, the condition $a'<1$ now results in the constraint 
\beq
      v <  V\sub{sc}(\eta) \etext{where} V\sub{sc}(\eta) = \sqrt{\frac{g R\sub{sc}}{\eta} \cos\alpha}  \,.
  \label{upper-limit-ang}
\eeq  
Thus, for an angulated skier ($\eta>1$) the upper speed limit is lower that for a stacked one ($\eta=1$).

\begin{figure}
\begin{center}
\includegraphics[width=0.5\textwidth]{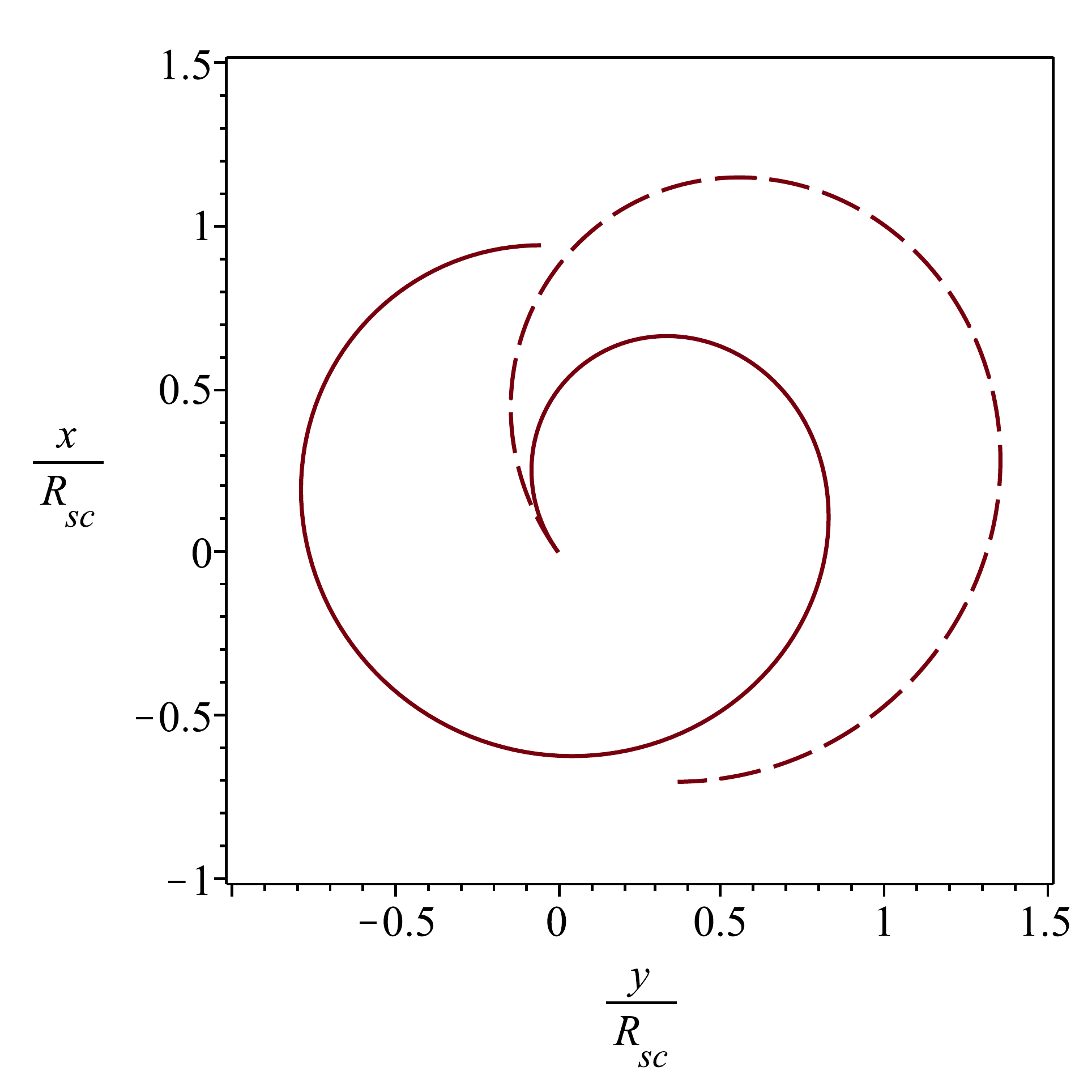}
\caption{ The trajectory of ideal carving run for the steep slope ($\alpha=20^\circ$) and steered turn initiation with $\Delta\beta=30^\circ$.}
\label{fig-tr-a}
\end{center}
\end{figure}

\begin{figure}
\begin{center}
\includegraphics[width=0.4\textwidth]{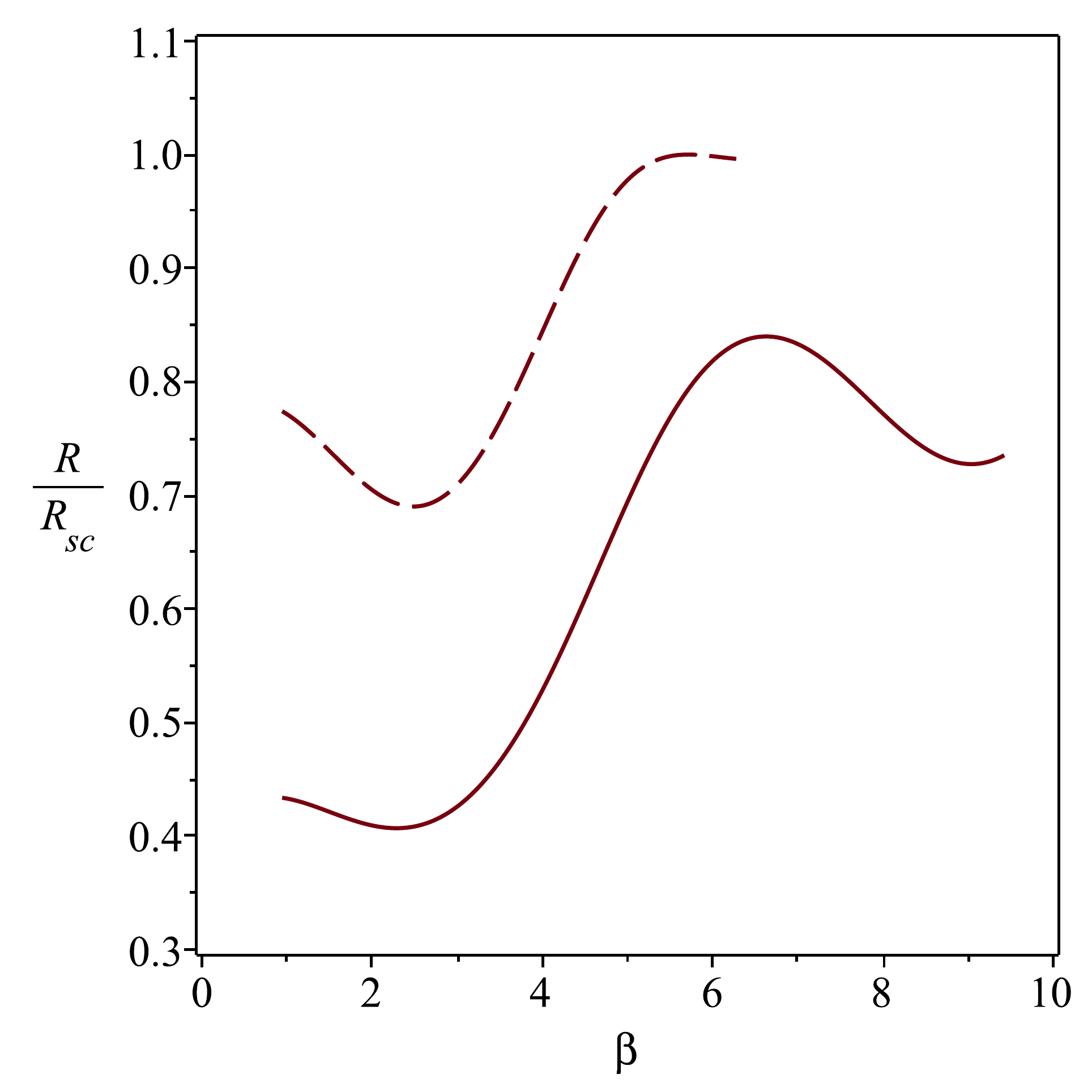}
\includegraphics[width=0.4\textwidth]{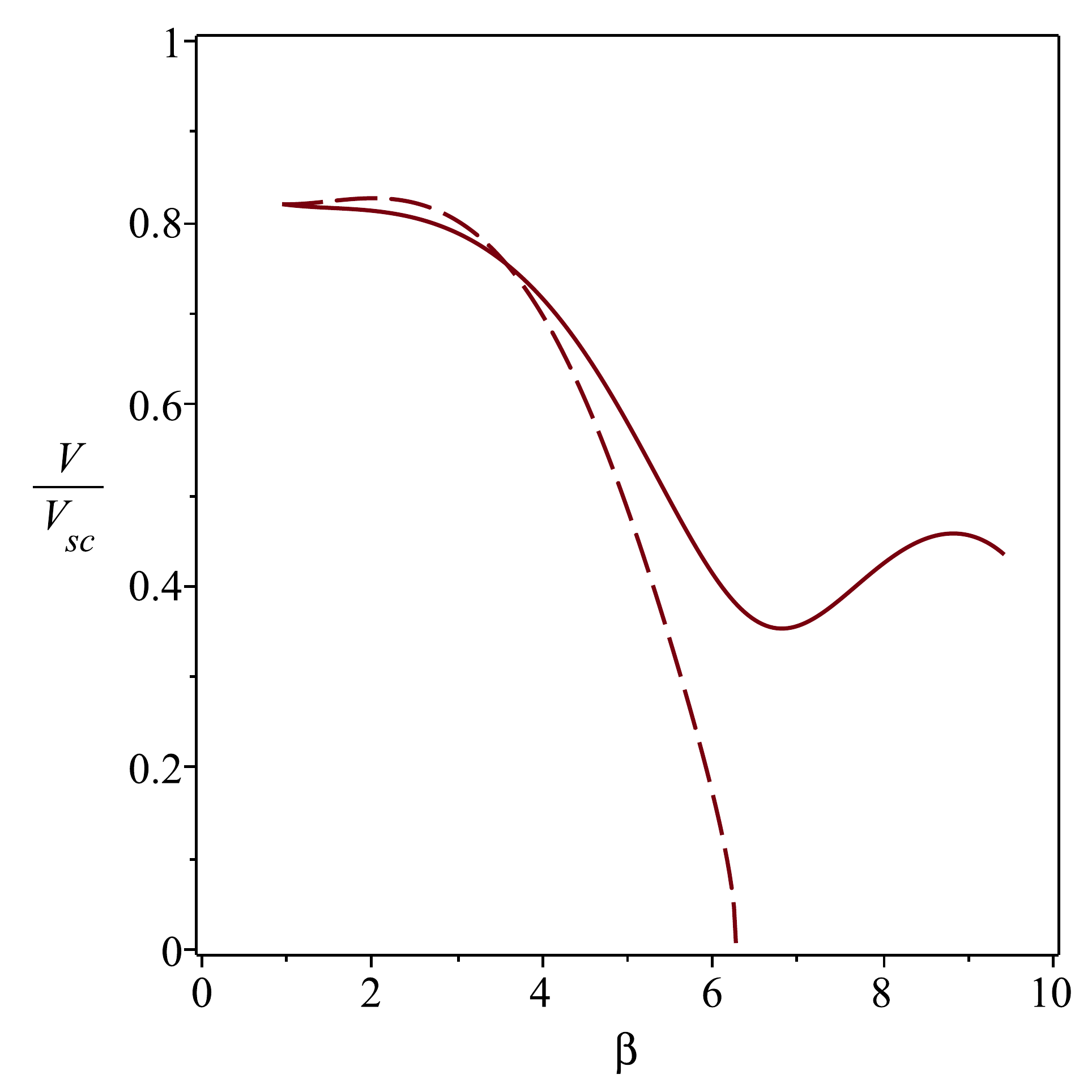}
\includegraphics[width=0.4\textwidth]{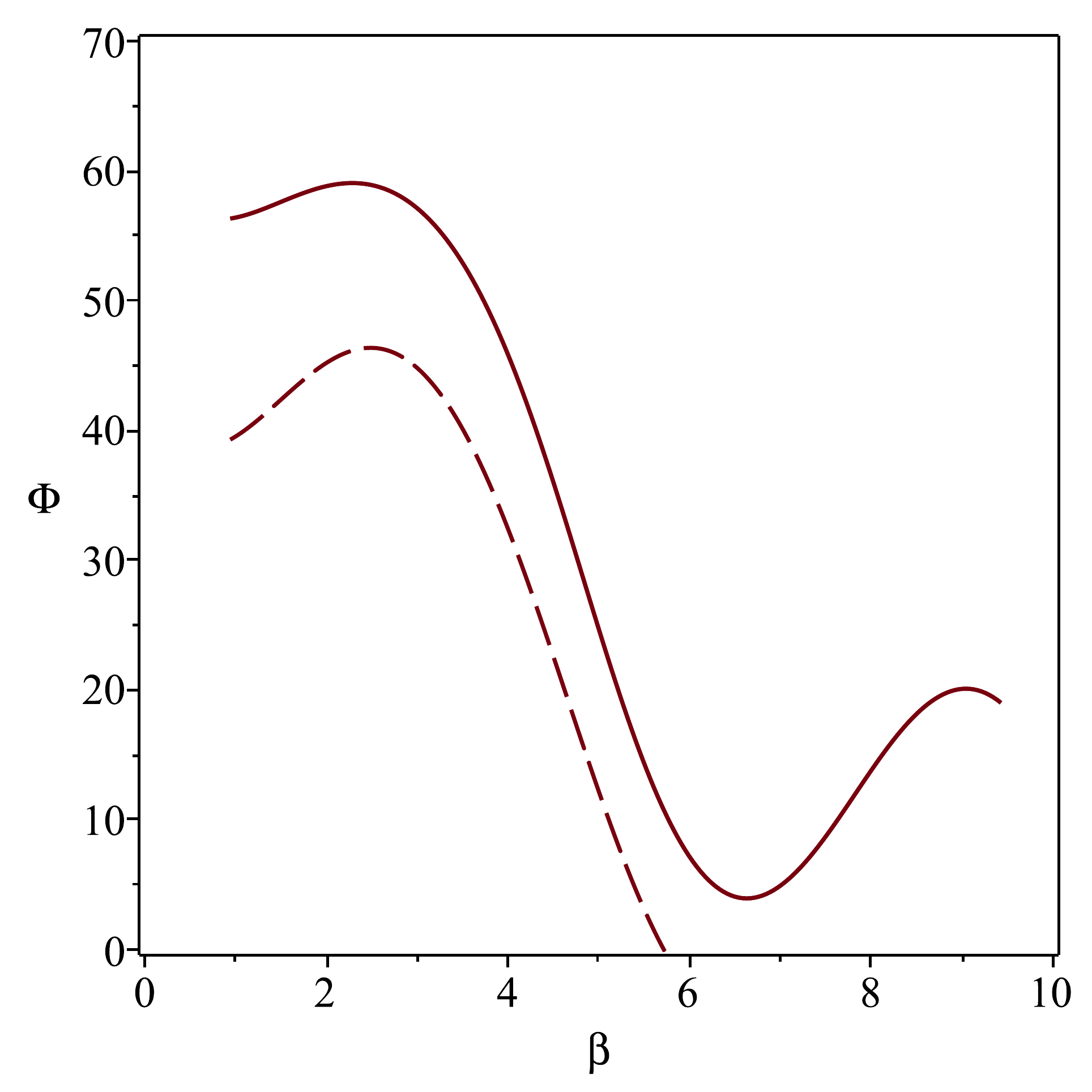}
\includegraphics[width=0.4\textwidth]{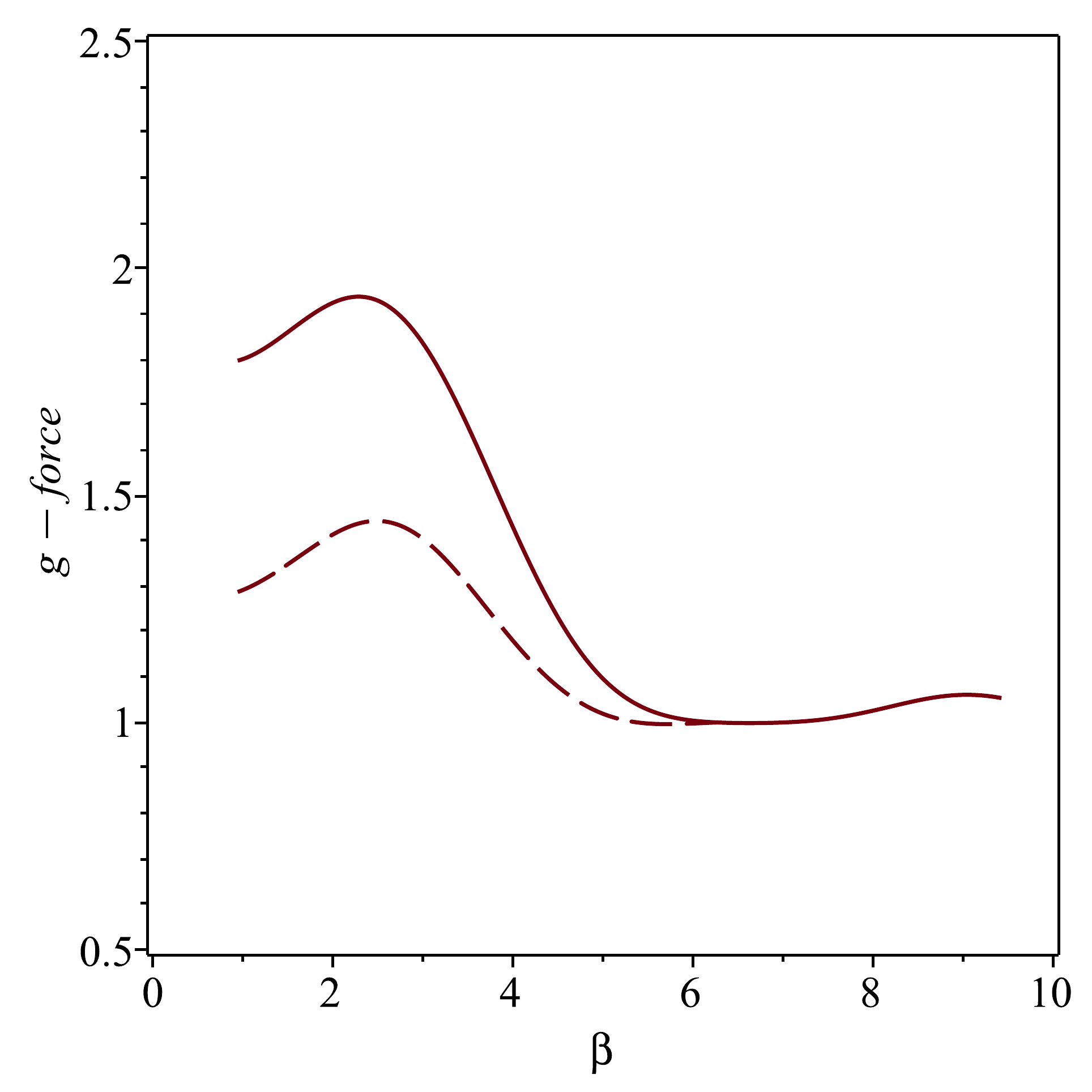}
\caption{The attempts at $360^\circ$ turn with (solid lines) and without (dashed lines) angulation. }
\label{fig-a}
\end{center}
\end{figure}

In order to illustrate the effect of angulation, we consider the so-called $360^\circ$ carving turn. 
By this we understand a carving turn which continues  in the clockwise (or counter-clockwise) direction all the way, first downhill then uphill and finally downhill again. In this example, we assume that the skier angulation depends only on the inclination angle,  $\Theta= \Theta(\Phi)$, as dictated by the equation     
\beq
   \tan\Psi = A + \tan\Phi \,, 
  \label{ang-2}
\eeq   
where $A$ is constant. It is easy to see that $A=\tan\Theta(0)$.  It is also easy to verify that $\Theta(\Phi)$ is a monotonically decreasing 
function which vanishes as $\Phi\to 90^\circ$, and that it yeilds $\Psi<90^\circ$ for all $0\le \Phi<90^\circ$.

Here we present the results for the slope angle $\alpha=5^\circ$, the initial skier speed $v\sub{ini}=0.5V\sub{g}$ and the initial angle 
of traverse $\beta\sub{ini}= 54^\circ$. Such a high speed could be gained at a steeper uphill section of the slope. 
Figure \ref{fig-a} shows the results for two runs: one with no angulation ($A=0$, dashed lines) and one with strong angulation ($A=\tan(30^\circ)$, solid lines).    
The run without angulation fells a bit short of success. It stops on approach to the summit as the skier's speed drops to zero.  

In the run with angulation, the turn is much tighter and the skier reaches the summit retaining a fair fraction of the initial speed. This allows them to continue and complete the $360^\circ$-turn.   Because of the similar initial speed but the lower turn radius, the centrifugal force and hence the total g-force of this run are higher.  This is also reflected in the higher inclination angle $\Phi$, which peaks at $\approx 67^\circ$ when $\beta\approx 130^\circ$. 
Significant body angulation is physically demanding also because the skeleton is no longer well stacked and hence bears a smaller fraction of the total skier's weight.   This results in a higher risk of injury.

\section{The effect of steering}

There are two obvious limitations associated with pure carving. First, the carving turn radius cannot be changed at will as it is almost uniquely 
determined by the speed, traverse angle and slope gradient. If the race course dictates a  smaller or larger turn radius some deviation from pure carving has to be introduced. Second, our experiments with ideal carving model show that on steep slopes this type of skiing leads to building up of excessive speed and eventual loss of grip. In order to enable runs on such slopes skiers must have additional means of speed control.  

These limitations can be overcome using the well-known technique of steering.  In this technique, the skis are set at some angle relative to the direction of motion. We will denote this angle, called the steering angle, as $\Delta\beta$. In such a position, skis displace a certain amount of snow ( snowploughing ), or scrape the top layer of ice, which leads to braking. The exact degree of braking depends on how the steering is executed. In the traditional fully steered turns, the steering is  continued all the way to the turn completion. In its modern variant, it is used mostly at the initiation phase of the turn. Once the speed is sufficiently reduced and/or the desired direction of the skis is reached, it is terminated  and the rest of the turn is continued as carving. In the so-called ``power-slide'', the return to carving is accompanied by forceful reduction of the steering angle via pivoting of the skis.            

In our modelling of such hybrid turns, we assume that at the turn initiation skis are rapidly pivoted to the desired steering angle and then they preserve their orientation relative to the fall line until carving is resumed. At the resumption point, the velocity component normal to the skis vanishes completely. Thus if $\beta\sub{fin}$ and  $v\sub{fin}$ are the angle of traverse and the speed at the completion of the previous turn respectively, the carving phase of the next turn starts with
\beq
  \beta\sub{ini} = (180^\circ - \beta\sub{fin}) + \Delta\beta 
  \label{str1}
\eeq   
and   
\beq
  v\sub{ini} = v\sub{fin} \cos(\Delta\beta) \,. 
  \label{str2}
\eeq   
The skier's kinetic energy is reduced by the factor $\cos^2\Delta\beta\approx1-(\Delta\beta)^2$ for small $\Delta\beta$. 
In the model we also ignore the finite duration of the steering phase and assume that it is negligibly short instead. 

To illustrate the effects of steering we made a run for a steep slope with $\alpha=20^\circ$ run and the steering angle
$\Delta\beta=45^\circ$. For this steering angle, exactly one half of skier's kinetic energy disappears at the transition between turns.   
This is just enough to avoid the singularity in the carving phase. The run is initiated with $x\sub{ini}=y\sub{ini}=0$, $\beta\sub{ini}=0.3\pi$ ($54^\circ$), and  $v\sub{ini}=(1/3) V\sub{sc}$ (=$0.1V\sub{g}$). Each turn terminates when the angle of traverse reaches $\beta\sub{fin}=0.9\pi$ ($162^\circ$). $\mu=0.04$ and $\Kn=0.0325$.

\begin{figure}
\begin{center}
\includegraphics[width=1.0\textwidth]{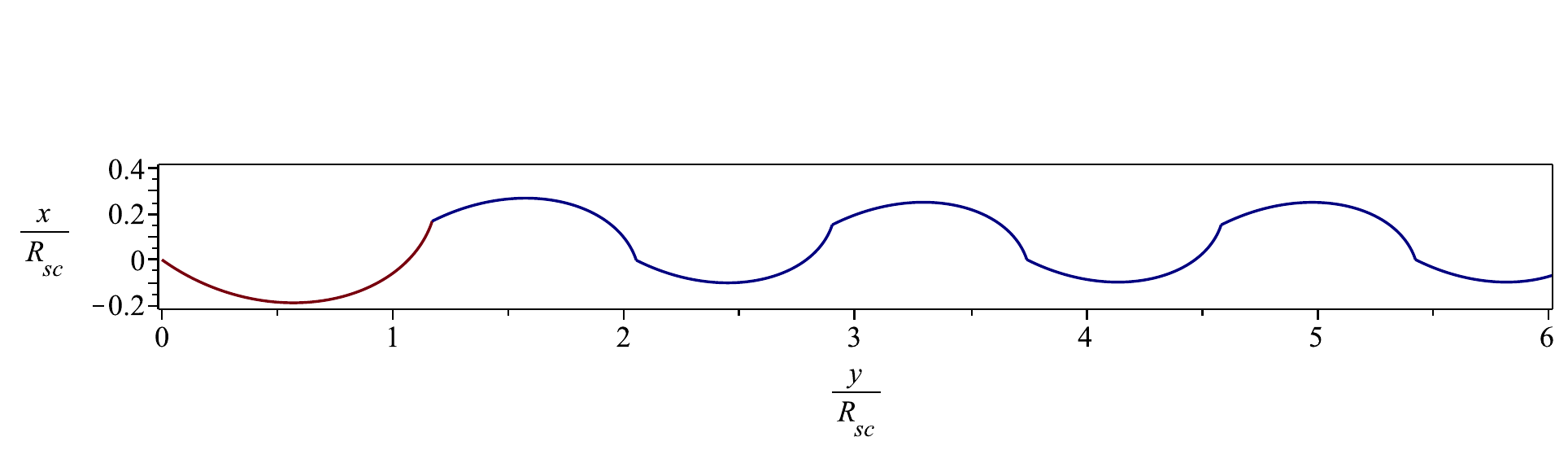}
\caption{ The trajectory of ideal carving run for the steep slope ($\alpha=20^\circ$) and steered turn initiation with $\Delta\beta=45^\circ$.}
\label{fig-tr-r4}
\end{center}
\end{figure}

\begin{figure}
\begin{center}
\includegraphics[width=0.4\textwidth]{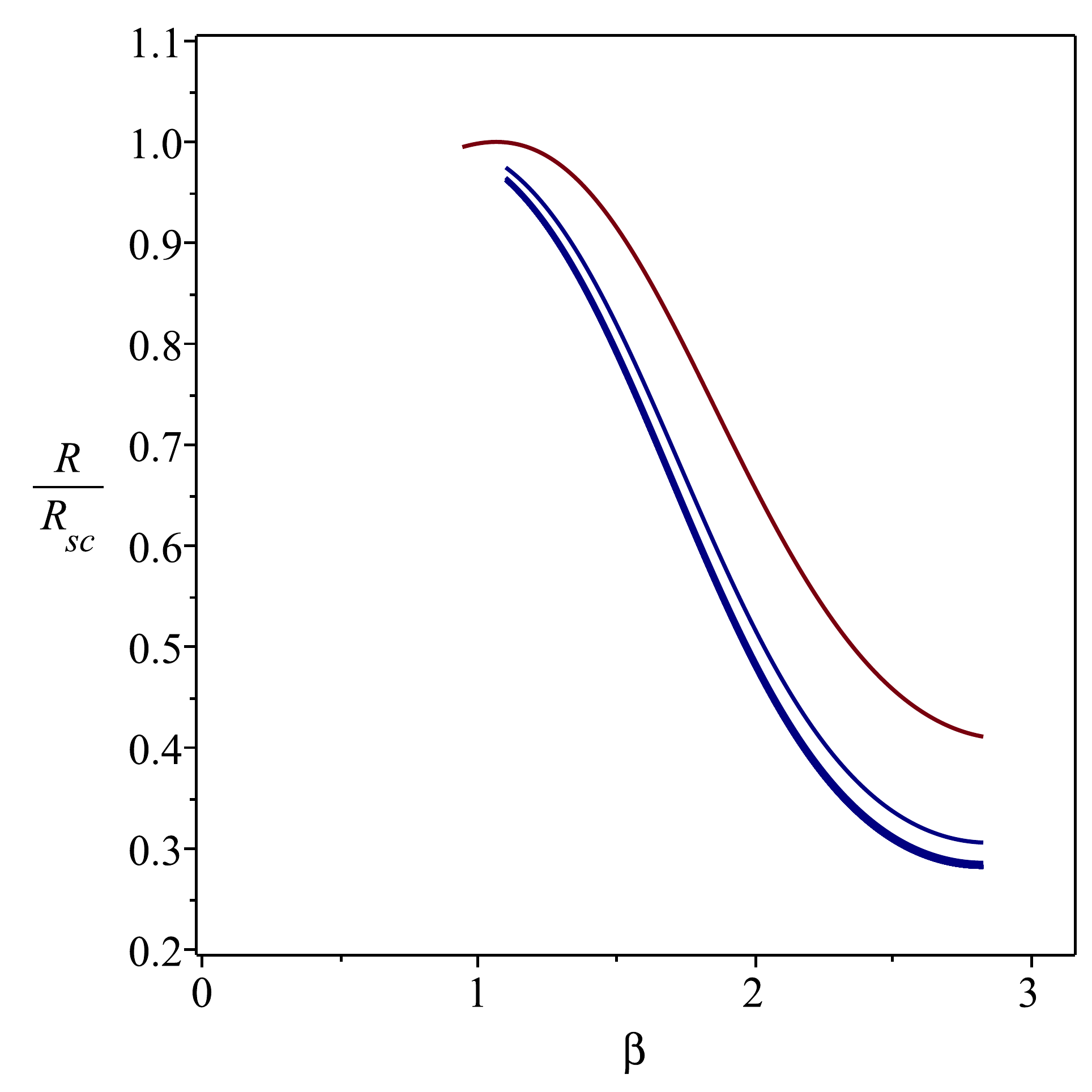}
\includegraphics[width=0.4\textwidth]{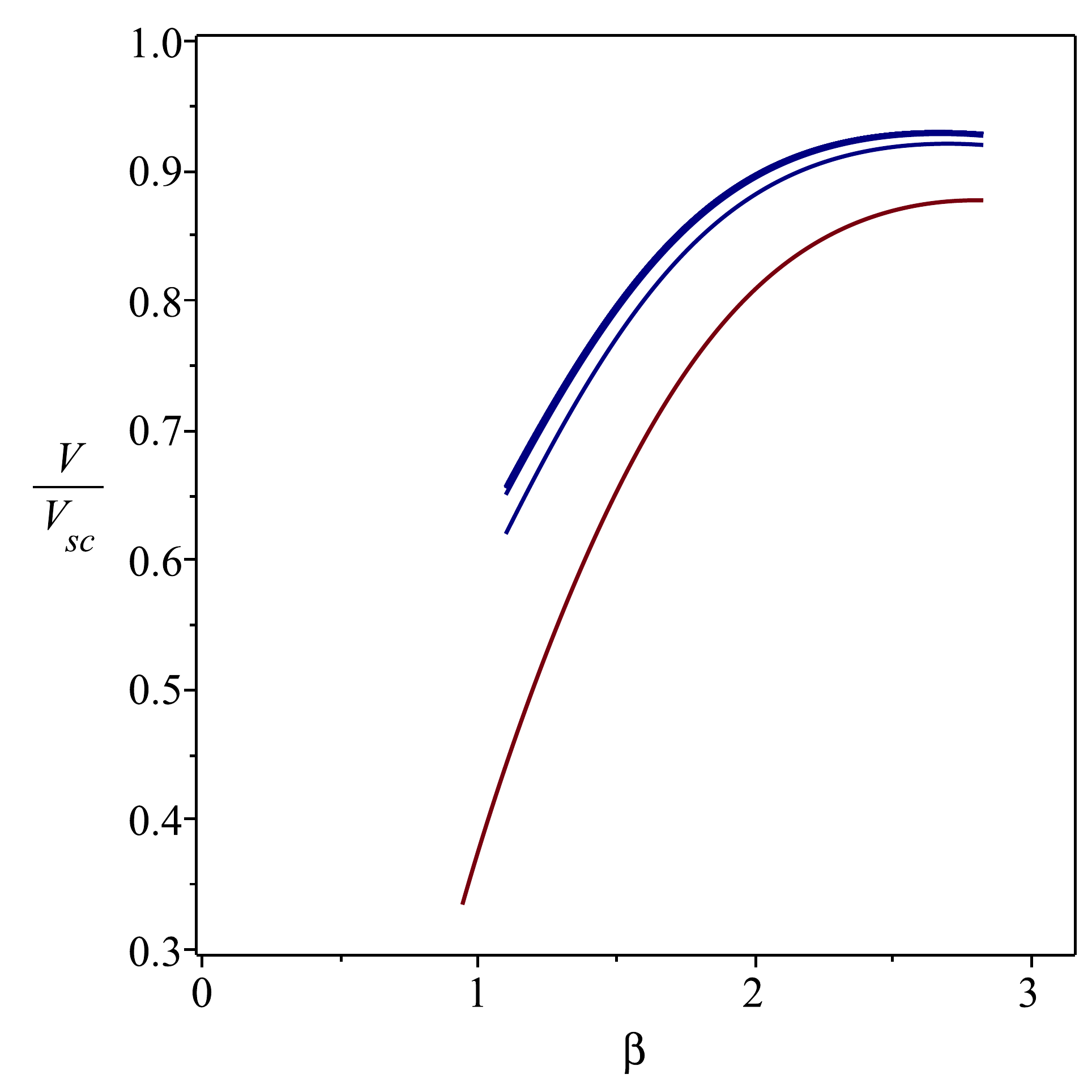}
\includegraphics[width=0.4\textwidth]{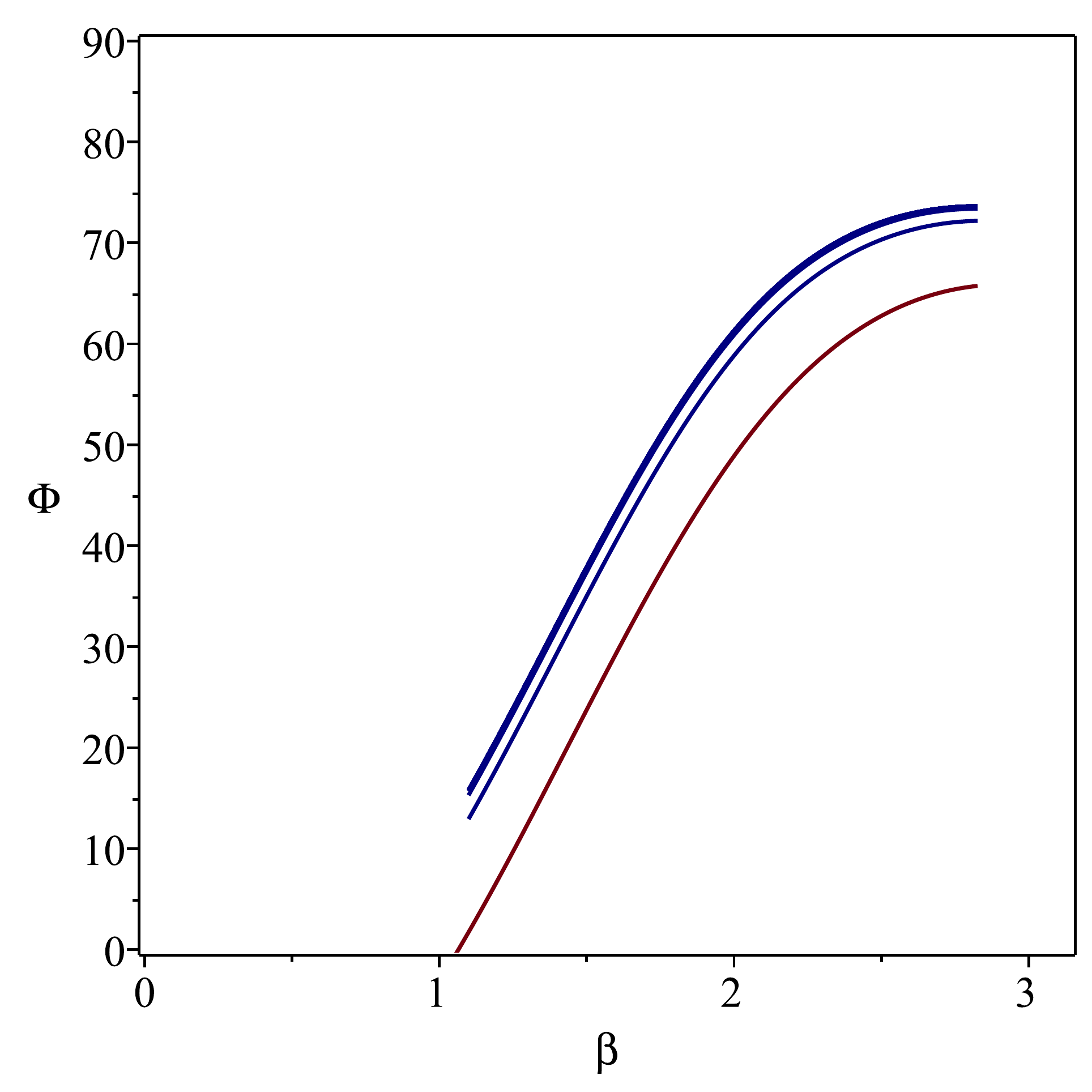}
\includegraphics[width=0.4\textwidth]{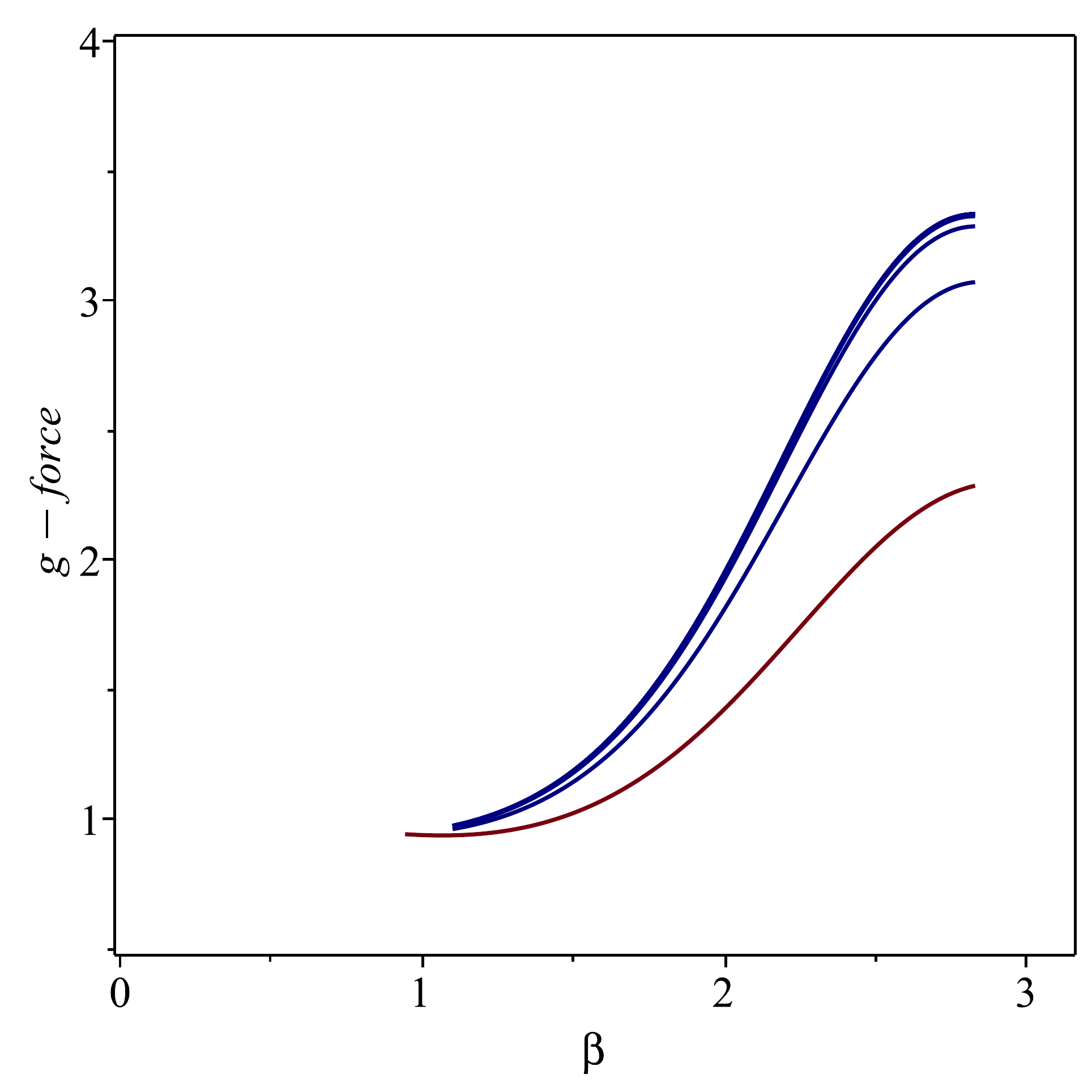}
\caption{The same as in figure \ref{fig-r1} but for the  steep slope ($\alpha=20^\circ$) and steered turn initiation with $\Delta\beta=45^\circ$. }
\label{fig-r4}
\end{center}
\end{figure}

Figure \ref{fig-tr-r4} shows the trajectory of the run and figure \ref{fig-r4} the evolution of its key variables. One can see that the trajectory is not smooth any more as each transition between turns introduces a break.  It would smooth out in a more accurate model where the transition is not instantaneous.   
As before, the solution quickly converges to a limiting one where each next turn is a copy of the previous one.  The local radius of curvature, inclination angle and g-force still vary dramatically throughout the turn but remain at more-or-less realistic levels. Obviously, a higher steering angle would moderate them furthermore.

\section{Conclusions}

In this paper we described a simple mathematical model of carving turns in alpine skiing, which can be applied to snowboarding as well.  The model combines a system of ordinary differential equations governing the CM motion with the algebraic Ideal Carving Equation (ICE), which emerges from the analysis of skier balance in their frontal plane.  ICE relates the local radius of the CM trajectory to skier's speed and direction of motion relative to the fall line and hence provides closure to the system.      

In the case of fall-line gliding, the skier speed grows until the gravity force is balanced by the aerodynamic drag and snow friction forces. Unless the slope is very flat, the snow friction is a minor factor and can be ignored, whereas  the balance between the gravity and the aerodynamic drag yields the saturation speed $V\sub{g}$. For this reason, $V\sub{g}$ and the distance $L\sub{g}$ required to reach it could be considered as the characteristic scales of alpine skiing.  However in the case of carving turns, ICE introduces another characteristic speed, the carving speed limit $V\sub{sc}$ \cite{JF04}. When the skier speed exceeds $V\sub{sc}$, carving becomes impossible in the whole Lower-C part of the turn.  In fact, at the fall line the turn radius $R\to 0$ and the g-force diverges  as $v\to V\sub{sc}$.   Because as a rule  $V\sub{sc}$ is lower than $V\sub{g}$, this speed scale is more relevant in the dynamics of carving turns.  Moreover, the radius of carving turns cannot exceed the ski sidecut radius $R\sub{sc}$, making the latter a natural length scale of this dynamics. These two scales yield the dimensionless equations of carving turn which have only three dimensionless control parameters: the slope gradient angle $\alpha$, the dynamic coefficient of friction $\mu$ and the dynamic sidecut parameter $\Kn=R\sub{sc}/L\sub{g}$, which determines the relative strength of the aerodynamic drag. 
  
In addition to the upper limit, IDE also sets a lower speed limit, which must be satisfied to enable carving in the upper-C. Below this limit the centrifugal force is too small and the balanced skier position is inconsistent with the ski edging required to start the turn.  

We used the model to explore ski runs composed of linked carving turns on a slope with constant gradient.  While in reality such turns are linked via a transition phase of finite duration, in our simulations the transitions are instantaneous and take place at a specified traverse angle (the angle between the skier trajectory and the fall line). At the transition, the skier speed and direction of motion remain invariant, whereas the inclination angle and hence the turn radius jump to the values corresponding to the next turn.  Under these conditions, the solution would either approach a limit cycle, where the turns become indistinguishable one from another, or terminate after hitting the speed limit. 
     
The results show the existence of a critical slope angle $\alpha\sub{c,t}$  above which the ideal carving run becomes a theoretical impossibility. Namely, as $\alpha\to\alpha\sub{c,t}$ the skier inclination angle approaches $90^\circ$ and the g-force diverges. The value of  $\alpha\sub{c,t}$ depends on the coefficient of friction and to a lesser degree on the sidecut parameter. For $\mu=0.04$ we find $\alpha\sub{c,t}= 17^\circ$ for SL skis,  $\alpha\sub{c,t} = 19^\circ$ for GS skis and $\alpha\sub{c,t} = 21^\circ$ for DH skis.  In practice, a number of factors, such as the strength of snow, skis and human body, come into play well before this theoretical limit and restrict pure carving to even flatter slopes. For example, demanding that the skier inclination angle remains below  $\Phi\sub{c}=70^\circ$ (and hence the g-force below three), we find that for SL skis the critical inclination angle $\alpha\sub{c} \approx 9^\circ$  if $\mu=0.04$ and  $\alpha\sub{c} \approx 16^\circ$ if  $\mu=0.1$.  For DH skis the corresponding values are  $\alpha\sub{c} \approx 14^\circ$ and $21^\circ$. Overall, we find that the critical gradient increases linearly with $\mu$.     

Slopes of sub-critical gradient can be roughly divided into the flat and moderately steep groups.  For flat slopes, the aerodynamic drag is not dominant over the snow friction even in the case of fall-line gliding. In carving runs, the role of the snow friction is even more important and the turn speed saturates well below both  $V\sub{g}$ and  $V\sub{sc}$. 
The carved arcs are nearly circular and their radius is only slightly below the sidecut radius of the skis.  The skier inclination angle and the g-force stay relatively small.    

On slopes of moderate gradient the saturation speed of fall-line gliding is very close to $V\sub{g}$, which is significantly above $V\sub{sc}$, but the speed of carving runs saturates near $V\sub{sc}$. The last condition makes the carving turns quite extreme. Their shape begins to deviate from the rounded shape of the letter C and remind the letter J instead,  with the local turn radius significantly decreasing on the approach to the fall line.  As the radius decreases,  the centrifugal force and hence the total g-force experienced by the skier grow. In order to stay in balance, they have to adopt large inclination to the slope.   The high effective gravity leads to high normal reaction from the snow and significantly increased friction, which is the reason why the speed stays well below $V\sub{g}$.  

On slopes of super-critical gradient ($\alpha>\alpha_c$) the speed quickly exceeds $V\sub{sc}$ after which the carving turn cannot be continued.  If practice this means a loss of balance, skidding and maybe even a crash.   

When given in the units of $R\sub{sc}$ and $V\sub{sc}$, the carving run solutions corresponding to $13\mbox{m} <R\sub{sc}<50\mbox{m}$ are rather similar. In fact, $R\sub{sc}$ enters the problem only via the dynamic sidecut parameter $\Kn=R\sub{sc}/L\sub{g}\ll 1$ which appears only in the speed equation where it defines the relative strength of the aerodynamic drag force.  The fact that $\Kn\ll1$  ensures that the dynamical importance of the drag term is small and hence the solutions are not very sensitive to the variations of $R\sub{sc}$.    Yet the drag term is not entirely negligible and the solutions show some mild variation with $R\sub{sc}$ within the studied range. In particular, turns corresponding to a larger sidecut radius are less extreme, with smaller inclination angles and weaker g-forces. This slight variation also explains the weak dependence of the critical gradient $\alpha\sub{c}$ on $R\sub{sc}$, as  described earlier.  

In the recent  field study of the forces experienced by top GS skiers, it was found that on average the aerodynamic drag 
was about 7 times below the snow friction \cite{S13}. The authors attributed this result to the fact that the participating racers could not perform pure carving and skidded instead.  This skidding could be understood as a sign of imperfect turn execution, implying that a better prepared skier could approach the aerodynamic speed limit $V\sub{g}$. Given the fact that this would involve the speed increase by more than twice, this seems highly unlikely.   Our study provides an alternative explanation. First of all, on steep slopes the perfect carving is theoretically impossible. On such slopes some efficient measures of speed control, of which  skidding seems to most available, must be employed to keep it below $V\sub{sc}$ and hence well below $V\sub{g}$. Secondly, in carving runs on slopes of subcritical gradient, the snow friction outperforms the aerodynamic drag force.  For example in our simulated carving run on a $15^\circ$ slope (see Sec.\ref{mss}), the friction dominates the drag on average by the factor of 3.
  
Skiing on super-critical slopes must involve speed control via skidding.  Watching runs made by the best athletes, including those competing on the World Cup circuit, one can notice that they execute pure carving turns only on the flat sections of the race track whereas on the steep sections they perform hybrid turns which combine pivoting and skidding early in the turn with carving later on. In order to probe the potential of this technique,  we made a simple modification of our model via introduction of the ski pivot and the corresponding reduction of the skier speed at the transition between carving turns.  The solutions obtained with this model confirm the expectations.   Since the radius of a carving turn is not well controlled by the skier but is largely dictated by their speed and the parameters of the slope, the ski pivoting is also helpful when the skier has to execute a tighter turn than the one possible with pure carving.   

Some degree of control over the radius of carving turns can also be reclaimed via the skier angulation.
We derived a slightly modified version of ICE which takes the angulation into account and analysed its implications. First, the angulation does not allow to remove the carving speed limit $V\sub{sc}$. On the contrary, it becomes even stricter. Secondly, given the same speed and slope parameters, the turn radius decreases with angulation.  To demonstrate this, we described a little experiment which demonstrates how the angulation helps to execute the full 360$^\circ$ carving turn. The angulation also increases the skier stability and helps to prevent the side slipping of their skis, reducing the importance of ski boot canting.  On the other hand, the shorter radius means the higher g-force and hence makes skiing more physically demanding. This is particularly unwelcome as the angulation reduces the ability of the body's skeletal structure to bear the skier's weight and puts more stress of the body's muscular system.   Loading of the inside ski may also help to control the turn radius as it allows to increase the ski inclination angle as well.  However this approach also becomes less effective at speed where large inclination angles imply that the inside leg is highly flexed and hence much less capable of supporting weight than the extended outside leg.      

We studied the stability of the lateral balance set by the original ICE, where the skier is balancing on a single ski. Although we find that it is unstable we do not think that this instability has important implications.   In particular, skiers can control it using the angulation of their body. Moreover, the skier weight is usually distributed over both skis which is obviously a stabilising factor as well.   Nowadays the top skiers are beginning to employ another way of the stabilisation -- by pushing the inside hand against the snow. This additional support may allow significantly lower inclination angles than those dictated by the Ideal Carving Equation, which is based on the assumption that the skier balances entirely on the outside ski. 

Obviously, our model of carving turns is rather simplified and ignores many details. In fact, it is much simpler than other models explored in the past, particularly those with the multi-segment representation of the skis and skiers \cite{M06,M08,M09,M10}.  However, we believe that it captures the basic dynamics of  carving turns quite well and that its simplicity allows to enhance our understanding of  this dynamics.   It may be worthwhile to develop this model a bit further.   For example, one can think of a more realistic description of the transition between turns in a ski run.  The role of angulation can be explored more thoroughly and the way it is used in various phases of the turn can be based more closely on the runs performed by actual athletes.  The model of hybrid turns can be improved via more realistic treatment of the skidding phase. Real slopes vary in steepness, pushing skiers further against the snow or sending them into the air. Moreover, skiers actively move their CM up and down to handle the terrain. Finally, field experiments can be used to test the model against the real carving runs. 

\appendix
\section{Edge radius of flexed skis} 
\label{RBS}
   
\begin{figure}
\begin{center}
\includegraphics[width=0.455\textwidth]{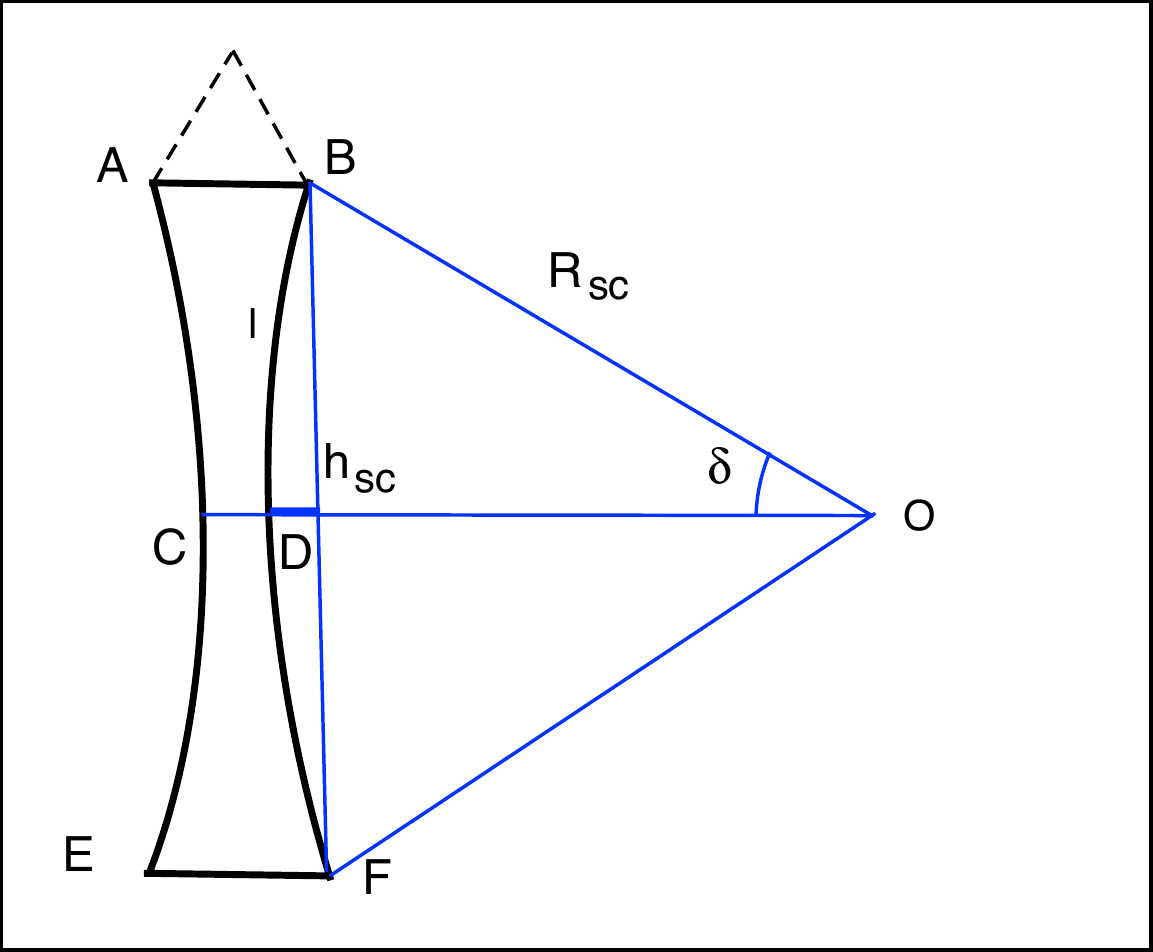}
\includegraphics[width=0.455\textwidth,viewport=0mm 2.5mm 11.7cm 10cm]{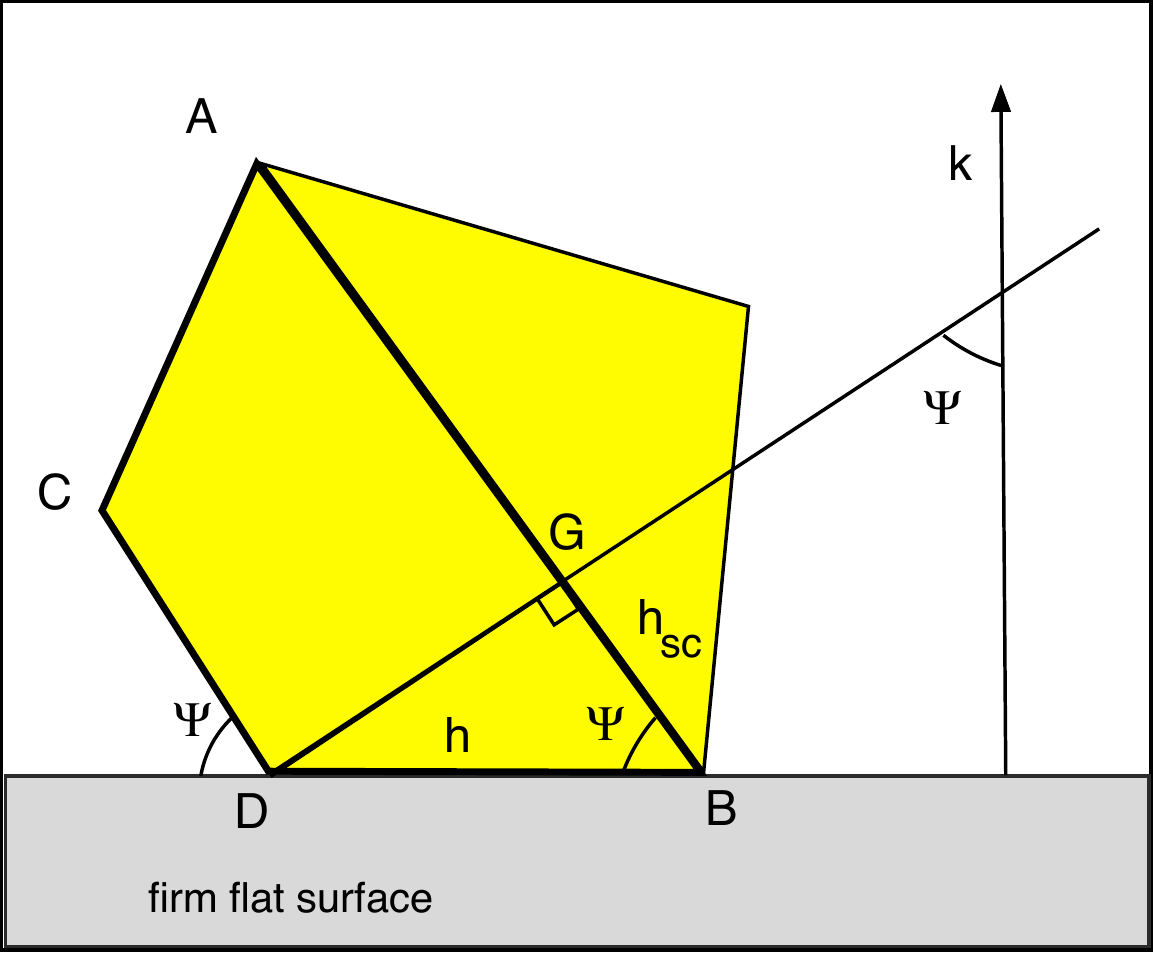}
\caption{ {\it Left panel}: Shaped ski, its sidecut $h\sub{sc}$ and sidecut radius $R\sub{sc}$. The dashed line shows the ski tip which is not important for determining  $R\sub{sc}$.   {\it Right panel}: The same ski with the edge BDF pressed against a hard flat surface at the inclination angle $\Psi$ as seen in projection on the plane normal to the ski at its waist. In this projection, DGB is a right angle triangle. }
\label{figure5}
\end{center}
\end{figure}

Here we analyse the geometry of a carving ski and how it changes when placed at the inclination angle  $\Psi$ to a flat surface, 
which we assume to be hard and hence not changed in the process apart from a small cut possibly made in it by the ski's sharp steel edges.   
We start with the case where the ski lays flat as shown in the left panel of figure \ref{figure5}.  In the figure, the ski is highly symmetric 
with no difference between its nose and tail sections.  Although the real skis are wider at the nose this does not matter as long as their
running  edges can be approximated as circular arcs of radius $R\sub{sc}$, called the sidecut radius. Denote as D the point in the 
middle of the edge BF and as $l$ the distance between B and D (or A and C) along the edge. As seen in this figure, $l=R\sub{sc}\delta$, 
where $\delta$ is the angular size of the edge DB as seen from its centre of curvature. This angle is normally rather small. For an SL ski of length $l\sub{ski}\approx 2l=1.65\,$m and $R\sub{sc}=12.7\,$m we have $\delta\approx 0.065$ $(3.\!\!^\circ7)$.  For other kinds of racing skis, it is even smaller.   The sidecut depth $h\sub{sc}$ is defined as the distance between D and the straight line BF connecting the opposite 
ends of the edge. Obviously,   
\beq
   h\sub{sc} = R\sub{sc}(1-\cos\delta)  \,.
  \label{r1}
\eeq   
Using the first two terms of the Maclauren expansion for $\cos\delta$ 
$$
     \cos\delta = 1 -\frac{1}{2}\delta^2 + O(\delta^4)
$$  
and then substituting $\delta=l/R\sub{sc}$ one finds the approximation 
\beq
   h\sub{sc} \simeq \frac{l^2}{2R\sub{sc}} \,.
  \label{r2}
\eeq   
%
  
Now suppose that this ski is kept at the angle $\Psi$ to a firm flat surface and that it is pressed in the middle until its lower edge comes 
into the contact with the surface along its whole length (excluding the tip).  
In this position the edge can still be approximated as an arc but a different one.  Denote its radius as $R$ and its ``sidecut'' depth as 
$h$ (see the right panel of figure \ref{figure5}).   Obviously $R$ and $h$ are connected in the same way as $R\sub{sc}$ and 
$h\sub{sc}$ 
\beq
  h = R(1-\cos\delta)  \, \,.
  \label{r2a}
\eeq   
where now $\delta = l/R$.  When $\delta\ll1$, this is approximated as 
\beq
   h \simeq \frac{l^2}{2R} 
  \label{r3}
\eeq   
and hence 
\beq
   \frac{R}{R\sub{sc}} = \frac{h\sub{sc}}{h} \,.
  \label{r4}
\eeq   
Analysing the right angle triangle GDB of the right panel in figure \ref{figure5}, one finds that $h\sub{sc}=h\cos\Psi$ and hence 
\beq
   \frac{R}{R\sub{sc}} = \cos\Psi \,.
  \label{turn-rr}
\eeq   
This is equation (3.7) in \cite{LS04}.  

\section{Loading of the inside ski} 
\label{LIS}

In the main paper, we focus on the case where the skier balances entirely on the inner edge of the ski which is located on the outside of the turn's arc. Indeed, one of the first things learned in ski lessons is keeping most of the load on the outside ski. However, some loading of the inside ski is needed to gain better stability, turn control and reduction of stress on the outside leg.  
Here we analyse the effect of the partial loading of the inside ski. Denote as $\bF\sub{n,i}$ and $\bF\sub{n,o}$ the normal reaction forces from the inside and the outside skis respectively and as $\br\sub{i}$ and $\br\sub{o}$ the position vectors connecting the skier's CM with the inner edges of the skis in the transverse plane. The force balance then reads 
\beq
      \bF\sub{g,eff} + \bF\sub{n,i}+\bF\sub{n,o} = 0 \,,
      \label{ap1}
\eeq 
whereas the torque balance in the transverse plane is 
\beq
      \br\sub{i}\times\bF\sub{n,i} + \br\sub{o}\times\bF\sub{n,o} = 0 \,.
       \label{ap2}
\eeq 
Provided the shanks of both legs are parallel to each other,  equation (\ref{ap1}) implies 
$$
     \bF\sub{n,i} = -a \bF\sub{g,eff} \etext{and} \bF\sub{n,o} = (a-1) \bF\sub{g,eff} \,,
$$
where $0\leq a \leq 1$. Substituting these into equation (\ref{ap2}), we obtain 
$$
   \bF\sub{g,eff}\times(a\br\sub{i}+(1-a) \br\sub{o}) = 0
$$
and hence 
$$
    \bF\sub{g,eff} = A ( \br\sub{o}+a(\br\sub{i}-\br\sub{o})) \,. 
$$
Thus  $\bF\sub{g,eff}$ points to the inner edge of the outside ski when $a=0$,  to the inner edge of the inside ski when $a=1$ and to somewhere in between when $0<a<1$.

\section{The effect of angulation on the turn radius}
\label{angulation-turn-r}
The ideal carving equation for an angulated skier is 
\beq
   (\xi^2-1)^{1/2} =  \eta(a'\xi +b') \,.
  \label{atr-1}
\eeq   
where $\xi=R\sub{sc}/R\ge 1$, $a'=\eta a$, $b'=\eta b$ and $\eta= \tan\Psi/\tan\Phi\ge1$ is the angulation parameter (see equation \ref{ang-1}). Differentiation of this equation with respect to $\eta$ yields 
\beq
  \oder{\xi}{\eta} = \frac{\eta (a'\xi +b')^2}{S(\xi)} \,,
  \label{atr-2}
\eeq 
where
$$
S(\xi) = (1-a'^2) \xi -a'b' = \xi -a'(\xi^2-1)^{1/2}\,.
$$
It is easy to see that $S(\xi)>0$ for any $\xi\ge 1$ provided $0<a'<1$ and so is $d\xi/d\eta$. Hence 
$$
\oder{R}{\eta}  <0 \etext{if} 0<a'<1 \,.
$$

\section{Lateral stability}
\label{stability}

Here we analyse the stability of skier's balanced position in the transverse plane, focusing on the simplified case of loading only the outside 
ski.    

\subsection{Stacked position}

We first consider the stability of a stacked skier. Suppose that in the equilibrium position the inclinations angles of the ski (skier) and effective gravity 
are $\Psi_0$ and $\Phi_0=\Psi_0$ respectively. Using the notation of Sec.\ref{SL}, 

\beq
    \tan\Psi_0=(\xi_0^2-1)^{1/2}\,, \quad \tan\Phi_0 = a\xi_0 +b \,,
\label{ls-0}
\eeq
where $\xi_0$ stands for the equilibrium turn radius.  Consider a perturbation which changes the ski angulation position but keeps the 
skier velocity unchanged (and hence $a$ and $b$ as well). However, the turn radius changes and so does the effective gravity. We need to determine if the 
modified effective gravity is a restoring force or ti pushes the system further away from the equilibrium.      
In the perturbed state $\Psi=\Psi_0+\delta\Psi$, $\Psi=\Phi_0+\delta\Phi$ and  $\xi=\xi_0+\delta\xi$,  where $\delta A$ stands for the perturbation of $A$.        
It is clear that when $\delta\Psi >0$ the instability condition reads $\Psi>\Phi$ or  $\delta\Psi>\delta\Phi$, where as for $\delta\Psi >0$ it is 
 $\delta\Psi<\delta\Phi$. Both this cases are captured in the instability condition 
\beq
   \frac{\delta\Phi}{\delta\Psi} <1 \,.
  \label{ls-1}
\eeq   

Using equation (\ref{ls-0}) we find 
\beq
   \delta(\tan\Psi) = \frac{\xi_0}{\tan\Psi_0} \delta \xi \,,
  \label{ls-2}
\eeq   
and 
\beq
   \delta(\tan\Phi) = a \delta \xi \,.
  \label{ls-3}
\eeq   
Hence 
\beq
   \frac{\tan\delta\Phi}{\tan\delta\Psi} = \frac{a\tan\Psi_0}{\xi_0} = 
       a \sin\Psi_0  \,.
  \label{ls-4}
\eeq   
According to the condition (\ref{cc1d}), in carving turns $a<1$ and hence equation (\ref{ls-4}) immediately yields 
${\tan\delta\Phi}/{\tan\delta\Psi} <1$. This implies  ${\delta\Phi}/{\delta\Psi} <1$ and therefore we conclude the lateral equilibrium is 
unstable.

\subsection{Angulated position} 
	
While the balance of an angulated skier is still unstable, in this case there is an additional way of controlling the instability, namely by 
a suitable change of the angulation.  If $\delta\Psi$  is the perturbation of the ski inclination angle and $\delta\Phi$ is the corresponding 
perturbation of the effective gravity angle that the skier can restore their balance via changing their angulation by the amount 
\beq
         \delta\psi = \delta\Psi-\delta\Phi \,.
\eeq

\section*{Acknowledgments} 
In order to solve the equations of ideal carving, we used the software package {\it Maple} (Maple is a trademark of Waterloo Maple Inc.).


\end{document}